\documentclass[aps,pra,twocolumn,superscriptaddress,longbibliography,footinbib]{revtex4-1}  %
\usepackage{graphicx} 
\usepackage{float} 
\usepackage{dcolumn} 
\usepackage{bm,color}
\usepackage{amsmath,amssymb,dsfont,amstext,amsfonts}
\usepackage{extarrows}
\usepackage{xcolor}
\usepackage{wasysym}
\usepackage{mathtools}
\usepackage{bbold}
\usepackage[export]{adjustbox}
\usepackage{mathdots}
\usepackage{siunitx}
\usepackage[colorlinks=true,linkcolor=blue,urlcolor=blue,citecolor=blue]{hyperref}

\usepackage[normalem]{ulem} 
\usepackage{color}

\def\mod{\:\mathrm{mod}\:}

\begin{document}
\title{High-chirality \textcolor{black}{and multi-quaternion Weyl nodes} in hexagonal ReO$_3$}
\author{Siyu Chen}
\email{sc2090@cam.ac.uk}
\affiliation{TCM Group, Cavendish Laboratory, University of Cambridge,
J. J. Thomson Avenue, Cambridge CB3 0HE, United Kingdom}
\affiliation{Department of Materials Science and Metallurgy, University of Cambridge, 27 Charles Babbage Road, Cambridge CB3 0FS, United Kingdom}
\author{Robert-Jan Slager}
\email{rjs269@cam.ac.uk}
\affiliation{TCM Group, Cavendish Laboratory, University of Cambridge,
J. J. Thomson Avenue, Cambridge CB3 0HE, United Kingdom}
\author{Bartomeu Monserrat}
\email{bm418@cam.ac.uk}
\affiliation{TCM Group, Cavendish Laboratory, University of Cambridge,
J. J. Thomson Avenue, Cambridge CB3 0HE, United Kingdom}
\affiliation{Department of Materials Science and Metallurgy, University of Cambridge, 27 Charles Babbage Road, Cambridge CB3 0FS, United Kingdom}
\author{Adrien Bouhon}
\email{adrien.bouhon@su.se}
\affiliation{TCM Group, Cavendish Laboratory, University of Cambridge,
J. J. Thomson Avenue, Cambridge CB3 0HE, United Kingdom}
\affiliation{Nordita, Stockholm University and KTH Royal Institute of Technology, Hannes Alfv{\'e}ns v{\"a}g 12, SE-106 91 Stockholm, Sweden}

\begin{abstract}
\textcolor{black}{The formation of two-band nodal points in gapless topological phases, referred to as conventional Weyl nodes, relies solely on translational symmetry. However, when coupled with other spatial and spatio-temporal symmetries, unconventional Weyl nodes with high degeneracy, pronounced chirality, \textcolor{black}{and complementary quaternion charges} can manifest.} In this work, we identify ReO$_3$ as an ideal unconventional Weyl semimetal in which rotation and screw symmetries as well as their combination with time-reversal symmetry play a crucial role. To show this, we first revisit in detail the algebraic determination of the chirality of Weyl nodes from the spinful irreducible representations of the occupied and unoccupied bands, and then combine it with the complementary $C_{2}T$-symmetry-protected patch Euler class \textcolor{black}{and non-Abelian frame charges} that indicate the pinning of the Weyl nodes on $C_{2}T$-invariant planes. Notably, we find one of the Weyl nodes in ReO$_3$ as the first example of a node with multiple nontrivial quaternion charges. Supporting our findings with first-principles calculations, we furthermore reveal very clear Fermi arc signatures of the high-chirality Weyl nodes at the Fermi level for different surface orientations. We finally investigate the effect of strain upon which the \textcolor{black}{robustness of Weyl nodes} clearly demonstrates their Chern (\textit{i.e.} chirality conservation) and quaternionic (\textit{i.e.} symmetry-plane pinning) topological nature. 
\end{abstract}

\maketitle


\section{Introduction}
Topological insulators and semimetals\,\cite{Rmp1,Rmp2, Armitage2018} have taken a prominent role in condensed matter physics and materials science both theoretically and experimentally. Shortly after the discovery of topological materials, it was realized that crystalline symmetries, in addition to the ``tenfold way'' symmetries~\cite{Kitaev,SchnyderClass}, also play a fundamental role in the characterization of these materials, culminating in a rich landscape of different phases and phenomena\,\cite{Clas1, Clas2, HolAlex_Bloch_Oscillations, Clas3, Wi2, Clas4, Clas5, Codefects2, UnifiedBBc,Codefects1,ShiozakiSatoGomiK, Chenprb2012, mSI, mtqc, magenticpaper, wiedersemi, subdimhigh,Ft1, Axion3, songtcq, bouhon2019wilson, Kemp_2022,Montambaux_2018, vmechn_2022, Sjoqviist_2004}. Using general gluing constraints~\cite{Clas3} \textcolor{black}{along with low-energy $k \cdot p$ effective models~\cite{tang2021}, classification algorithms~\cite{Clas4, Clas5} have been already formulated to map out a large fraction of topological insulators and semimetals, yielding a catalogue of topological materials accompanied by the emergent quasi-particles that they host~\cite{zhang2019, vergniory2019, tang2019, PhysRevB.102.045130, tang2022, yu2022, zhang2022, liu2022}}. 

Beyond this paradigm of symmetry-indicated phases, a new direction in the form of multi-gap Euler topological phases was recently discovered\,\cite{2023quantumgeo, morris2023andreev, zhao2022observation, jiang2021observation, bouhonGeometric2020, Eulerdrive, Jiang_meron, chen2021manipulation, AnEuler, BJY_nielsen, bouhon2019nonabelian, Guo1Dexp}. In such systems, symmetries that ensure reality conditions for the associated wave functions endow the band crossings residing between different gaps with non-Abelian frame charges\,\cite{Wu1273, BJY_nielsen, bouhon2019nonabelian}. Braiding of these band crossings can further induce topological phase transitions where isolated two-band subspaces acquire nontrivial multi-gap topological invariants\,\cite{bouhonGeometric2020,bouhon2022geo2}. These new phases can be characterized both globally, via the Euler class\,\cite{BJY_nielsen, BJY_linking,zhao2017PT}, or locally, via the patch Euler class~\cite{bouhon2019nonabelian}, and they are increasingly being explored in a variety of physical systems\,\cite{morris2023andreev, zhao2022observation, Lange2022, jiang2021observation, AnEuler, Eulerdrive, Jiang_meron, Jankowskidisorder, wahl2024exactprojectedentangledpair,chen2021manipulation, Guo1Dexp}. 

Overall, band crossings and symmetry conditions are intricately related, and both concepts form a rich interplay that results in new physical phases. 
Among the diverse crystalline systems, noncentrosymmetric crystals with spin-orbit coupling and time-reversal symmetry (\textit{i.e.} the class $\mathsf{AII}$ of the tenfold way\,\cite{Kitaev,Schnyder08}) have been shown to exhibit stable band crossings\,\footnote{Hereafter, following the widely used terminology, we refer to the nodal points of these band crossings as Weyl nodes.}\,\cite{Volovik,zhao2013topFS}: each time-reversal invariant momentum point (TRIMP) hosts (minimally) a Weyl node, and crystalline symmetries can protect further band degeneracies at specific high-symmetry regions of the Brillouin zone (BZ). In particular, rotation symmetry protects a Weyl node on the rotation-invariant line whenever two bands with distinct rotation-symmetry eigenvalues are permuted in energy order from one point to another of the line~\cite{BradCrack}, where the symmetry eigenvalues are associated with irreducible representations (IRREPs) of the point group of the system. It has been shown that $C_n$ rotation (\textit{i.e.} $2\pi/n$ rotation) symmetries for $n=4$ and $6$ alone can enforce high-chirality Weyl nodes~\cite{multi-weylfang, Vanderbilt_screw, chang2017, zhang2018hexagonal, gonzalez2020hexagonal}. Remarkably, the chirality of such a Weyl node can be readily read from the characters of IRREPs ~\cite{multi-weylfang}. 

Beyond the degeneracies directly indicated by the IRREPs of the point group, nonsymmorphic crystalline symmetries, which combine a point group symmetry with a translation by a faction of the primitive Bravais vectors, such as screw and glide symmetries, can also enforce further band degeneracies. One approach to study them is to consider the projective representations associated with a specific high-symmetry region of the BZ boundary that can lead to the additional pairing of bands~\cite{BradCrack}. More generally, nonsymmorphic symmetries impose the permutation in energy order of bands with distinct point-symmetry eigenvalues from one momentum to another momentum distant by a full reciprocal primitive lattice vector, thus enforcing the crossing of bands protected by the point group symmetry\,\cite{Wi2, BBS_nodal_lines, young2015dirac, zhao2016nonsym}. The combination of the band permutations in different directions can lead to the existence of additional band crossings away from the BZ boundary which can be read out from the IRREPs of occupied bands over the whole BZ, thus providing an assessment of the global topology of the band structure~\cite{Wi2}. Also, nonsymmorphic symmetries typically lead to a number of bands that must be connected by symmetry-protected band degeneracies larger than two\,\cite{Vishwanath_filling_gapless}, that is two being the minimal number of connected bands, via Weyl nodes at the TRIMPs, when there is only time-reversal symmetry. Such higher minimal connectivity of bands, when considered globally over the whole BZ, has been the subject of many works in the last few years\,\cite{Vishwanath_filling_gapless, barry2016, Wi2, BBS_nodal_lines, geilhufe2017three, geilhufe2017data, Thomas_line, Wieder_magnetic_Dirac, FURUSAKI2017788, hourglass, Furusaki_line, zhang2018hexagonal, gonzalez2020hexagonal, wieder2016spin, Kane_2D_SOC}. 

Finally, it is worth mentioning that Zak and Michel have long pointed out that the above irreducible connectivity of bands is related to the minimal dimensionality of elementary band representations\,\cite{Zak1, Zak2, Zak3, Zak4, michel1999, michel2001}. The observation that not all elementary band representations must be gapless then led to a general framework, coined topological quantum chemistry, to assess nontrivial band topology from the IRREPs\,\cite{Clas5, Clas3, HolAlex_Bloch_Oscillations, Ft1,bouhon2019wilson}.
    
In this paper, we explore the interplay of band degeneracies, symmetry, and topology, using ReO$_3$ as a concrete material platform. In particular, we identify ReO$_3$ as an ideal unconventional Weyl semimetal in which the spatial symmetries (\textit{i.e.} rotation and screw) and the $C_2T$ spatio-temporal symmetries (\textit{i.e.} combining a $C_2$ rotation with time reversal) all play a crucial role. For this, we revisit in detail the algebraic determination of the chirality of Weyl nodes from spinful IRREPs, starting from the derivation of all their characters. The analysis of band topology of Chern topology from the IRREPs over the whole BZ, combined with the Nielsen-Ninomiya theorem~\cite{NIELSEN1981219}, allows us to account for all the Weyl nodes relevant at the Fermi level, \textit{i.e.} those doubly degenerate ones that are indicated straightforwardly by IRREPs and a special triply degenerate one that arises ``accidentally'' from a crystal symmetry point of view. \textcolor{black}{Very interestingly, we find for the first time that the triply degenerate Weyl nodes with high chirality appearing precisely at the Fermi level further carry multiple nontrivial quaternion charges protected by distinct $C_2 T$ symmetries.} The study of the bulk topology is completed with a detailed bulk-boundary correspondence that predicts clear signatures of high-chirality Weyl nodes in the form of Fermi arc multiplets within surface spectra that are all observable at the Fermi level. \textcolor{black}{Finally, the analysis culminates with the simulation of strain providing a direct probe of the robustness of the Weyl nodes upon breaking the hexagonal rotation symmetry.} 

The paper is organized as follows. In Sec.\,\ref{sec_material}, we address via first-principles calculations the crystal structure and the electronic band structure of ReO$_3$, a material that is experimentally known. There, we show the presence of high-chirality Weyl nodes in close proximity to the Fermi level. In Sec.\,\ref{Sec: theory}, we expose an algebraic approach, based on symmetry-constrained Wilson loops, for determining the chirality of Weyl nodes imposed by rotation symmetries, using the space group of ReO$_3$ as a case study. We provide detailed explanations of the methodology, highlighting its simplicity and universality for all space groups of noncentrosymmetric systems in the Altand-Zirnbauer class $\mathsf{AII}$. In Sec.\,\ref{BCvsIRREPS}, we describe the complementarity of the algebraic method of Sec.\,\ref{Sec: theory} to the numerical method of Sec.\,\ref{sec_material} in determining the bulk topology of the system. \textcolor{black}{In Sec.\,\ref{sec_Euler}, we introduce and evaluate the topologies associated with a reality condition protected by the antiunitary $C_2T$ symmetries, allowing us to attribute complementary patch Euler classes to the doubly degenerate Weyl nodes and multiple quaternion charges to the triply degenerate Weyl node.} We then derive the symmetry-indicated bulk-boundary correspondence at the Fermi level in Sec.\,\ref{sec_BBC}, and follow in Sec.\,\ref{sec_surface} with the surface spectrum obtained for two distinct surface orientations exhibiting Fermi arcs clearly visible at the Fermi level. These provide a clear-cut signature of the high chirality of Weyl nodes. In Sec.\,\ref{sec_strain}, we finally discuss the effect of strain on the bulk configuration of Weyl nodes, relating to their high chiralities, patch Euler classes, \textcolor{black}{and their multiple quaternion charges}. Finally, we summarize the contributions and findings of our research in Sec.\,\ref{sec_conclusion}.

\section{R\lowercase{e}O$_3$ as a high chirality Weyl semimetal}
\label{sec_material}
\subsection{Structure}
\begin{figure}
\begin{tabular}{cc}
\includegraphics[width=0.48\linewidth]{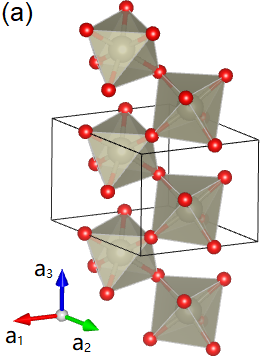} & 
\includegraphics[width=0.43\linewidth]{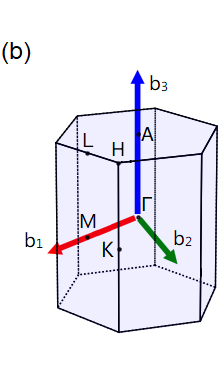}
\end{tabular}
\caption{(a) Crystal structure of hexagonal ReO$_3$ and (b) its BZ. The gray and red balls represent the rhenium and oxygen atoms respectively. The six-fold screw axis is along the $\boldsymbol{a}_3$- and $\boldsymbol{b}_3$-directions.}
\label{struture}
\end{figure}

Through X-ray experiments, ReO$_3$ has been shown to crystallize in a nonsymmorphic hexagonal space group, $\mathsf{P}6_322$, at $9$\,GPa and $500^{\circ}$C, and this phase can be recovered to ambient conditions\,\cite{dyuzheva1987}. The corresponding crystal structure is visualized in Fig.\,\ref{struture}(a). The primitive cell contains two rhenium atoms and six oxygen atoms, where the rhenium atoms occupy the Wyckoff position 2c $(1/3, 2/3, 1/4)$ and each bond with six equivalent oxygen atoms with the Wyckoff position 6g $(x, 0, 0)$, where $x = 0.374$. This arrangement results in a slightly distorted ReO$_6$ octahedron, and the ReO$_6$ octahedra are connected by sharing a corner oxygen atom, forming a chain along the $\boldsymbol{a}_3$-axis with six-fold screw symmetry. 

It is worth highlighting that, although hexagonal ReO$_3$ is reported to be synthesized under high pressure and temperature conditions, it is observed to be metastable under ambient conditions. We have confirmed the theoretical dynamical stability of ReO$_3$ under ambient pressure by calculating its phonon dispersion~\cite{Lloyd2015,chen2022} and demonstrating the absence of imaginary vibrational modes~\cite{S1}. 

\subsection{Computational details}\label{sec_comp}
We perform first-principles calculations for ReO$_3$ using density functional theory (DFT)~\cite{DFT-Hohenberg-Kohn, DFT-Kohn-Sham} as implemented in the Vienna \textit{ab initio} Simulation Package ({\sc vasp})~\cite{VASP-Original-Paper}. The exchange-correlation functional is treated using the generalized gradient approximation parametrized by Perdew-Burke-Ernzerhoff (PBE)~\cite{PBE-exchange-correlation}. The interaction between ions and electrons is modeled through pseudopotentials in the projector-augmented wave formalism~\cite{VASP-PAW-One, VASP-PAW-Two}, where the valence electrons of rhenium are $5d^56s^2$ and the valence electrons of oxygen are $2s^2 2p^4$. The plane-wave expansion is truncated at an energy cutoff of $550$\,eV, and the BZ integration is evaluated using a $15\times15\times11$ Monkhorst-Pack $\boldsymbol{k}$-point grid~\cite{monkhorst1976}. For the band structure analysis, we invoke {\sc irreps}~\cite{iraola2022} to determine the symmetry eigenvalues and IRREPs of the band structure. 

To numerically calculate the chirality of Weyl nodes (as a validation of the method introduced in Sec.\,\ref{Sec: theory}), we construct an effective tight-binding model to reproduce the 12 bands near the Fermi level (which constitute an ideal isolated set of bands). The model Hamiltonian is obtained through {\sc wannier90}~\cite{wannier90} and subsequently symmetrized according to the crystallographic symmetry of ReO$_3$, where the initial projection functions are chosen as random Gaussian functions and are optimized through the procedure of maximal localization~\cite{marzari2012}. Furthermore, we use the iterative Green's function method, as implemented in {\sc wanniertools}~\cite{wu2018}, to calculate the surface states of ReO$_3$ (in Sec.\,\ref{sec_surface}) based on the obtained model Hamiltonian.

\subsection{Electronic structure from first principles}
\label{sec:electronic}

\begin{figure*}
\begin{tabular}{cc}
\includegraphics[width=0.403333\linewidth]{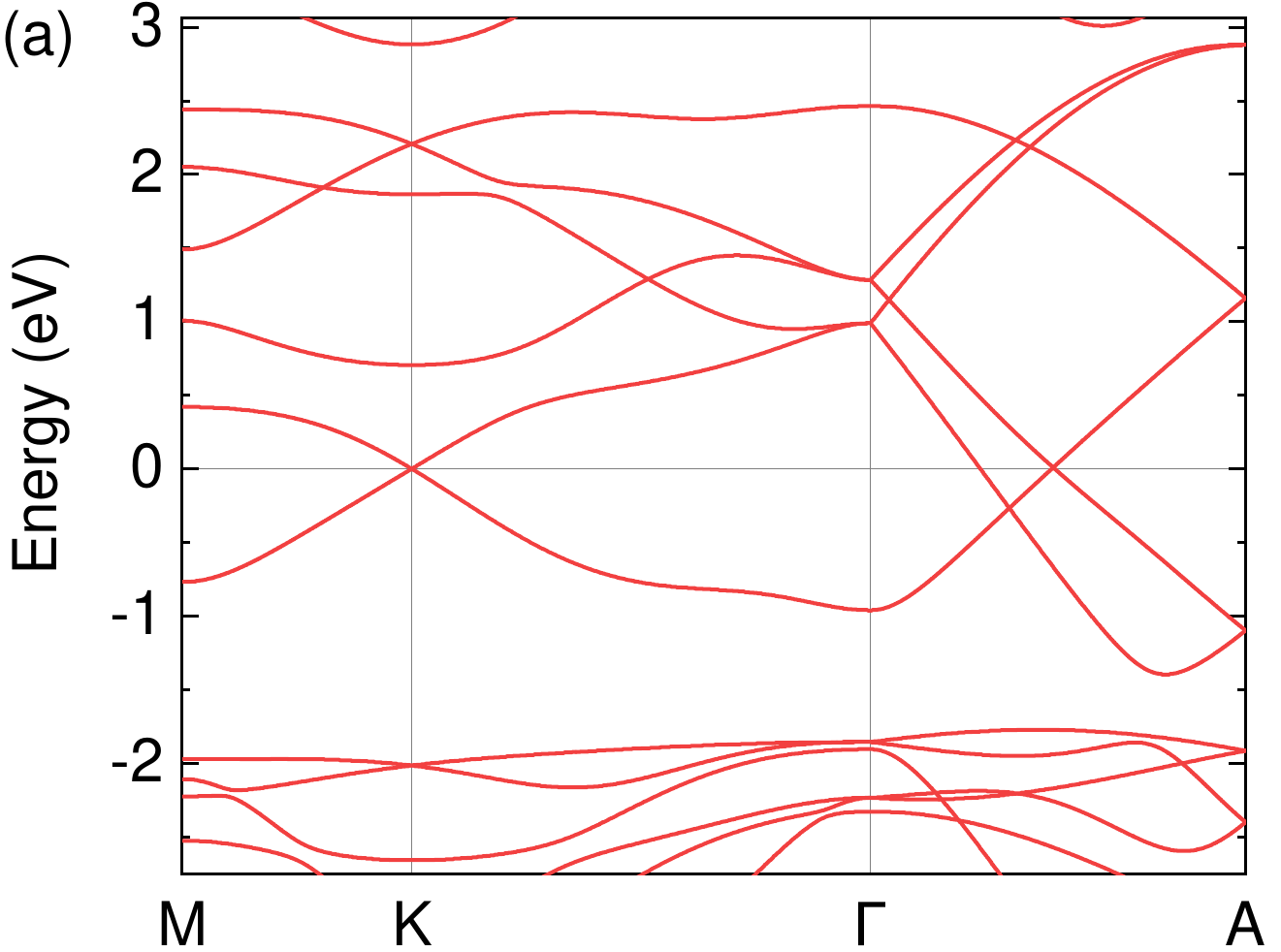}  \quad \quad \quad
\includegraphics[width=0.403333\linewidth]{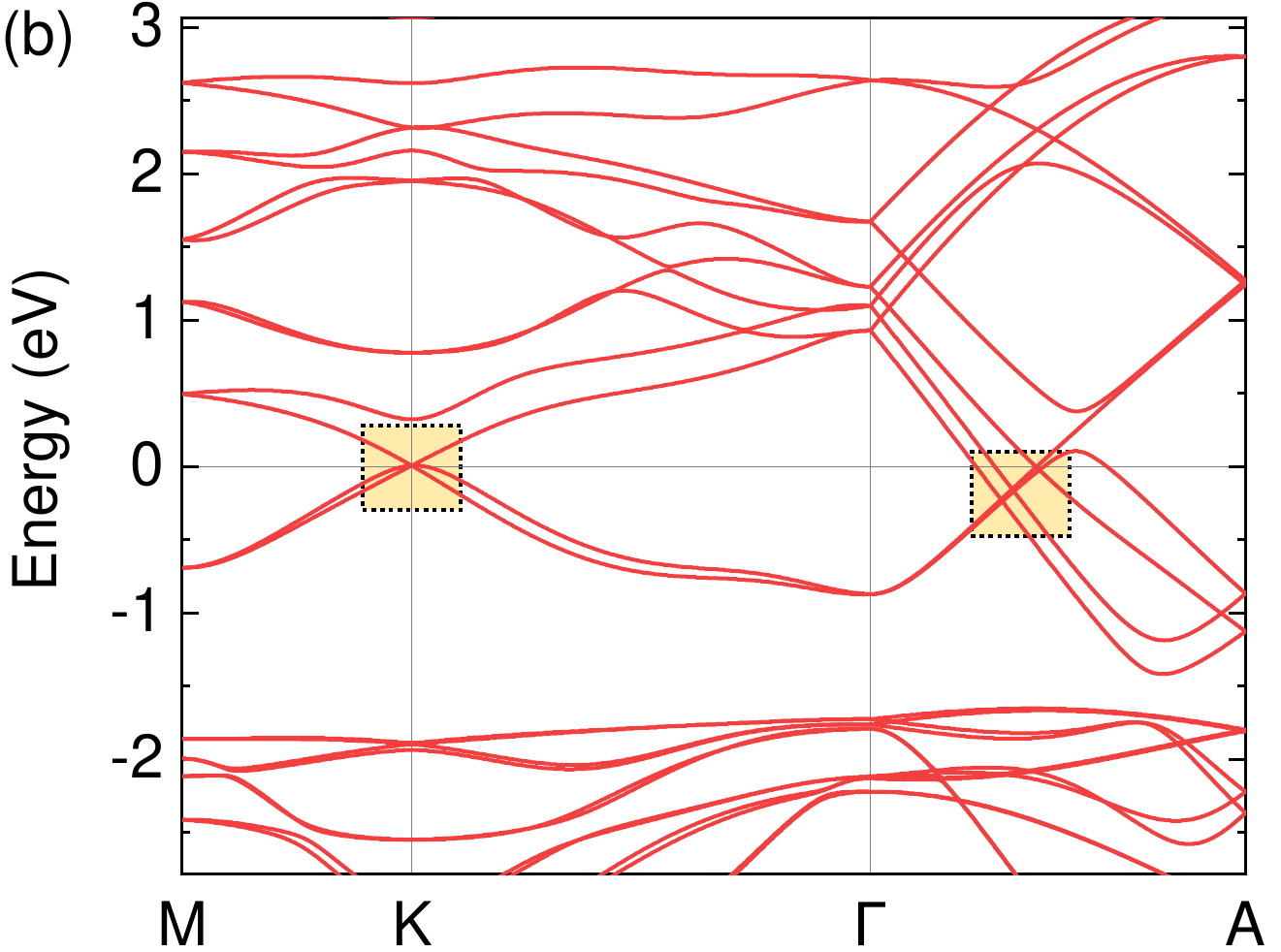} \\  

\includegraphics[width=0.403333\linewidth]{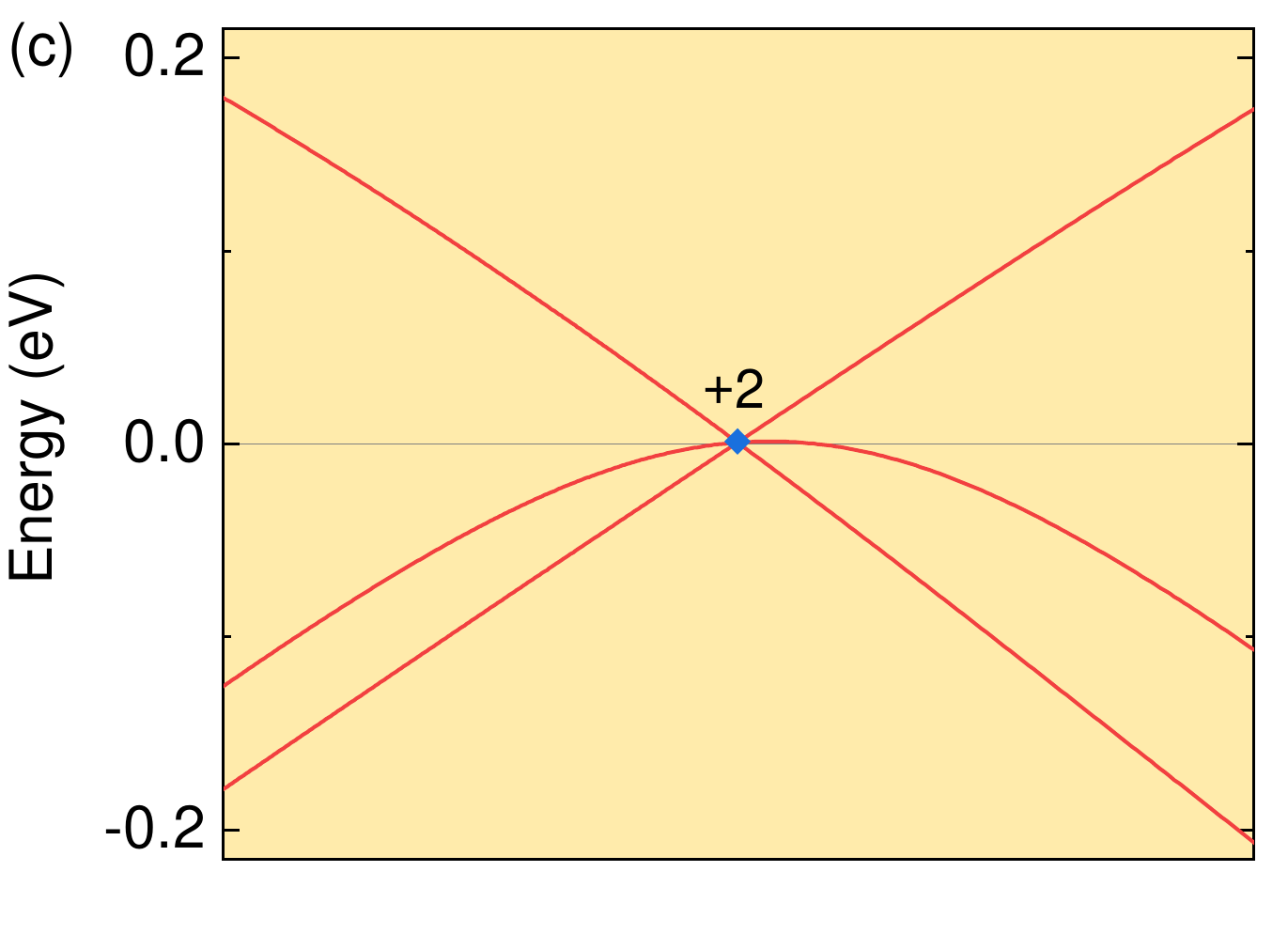}  \quad \quad \quad
\includegraphics[width=0.403333\linewidth]{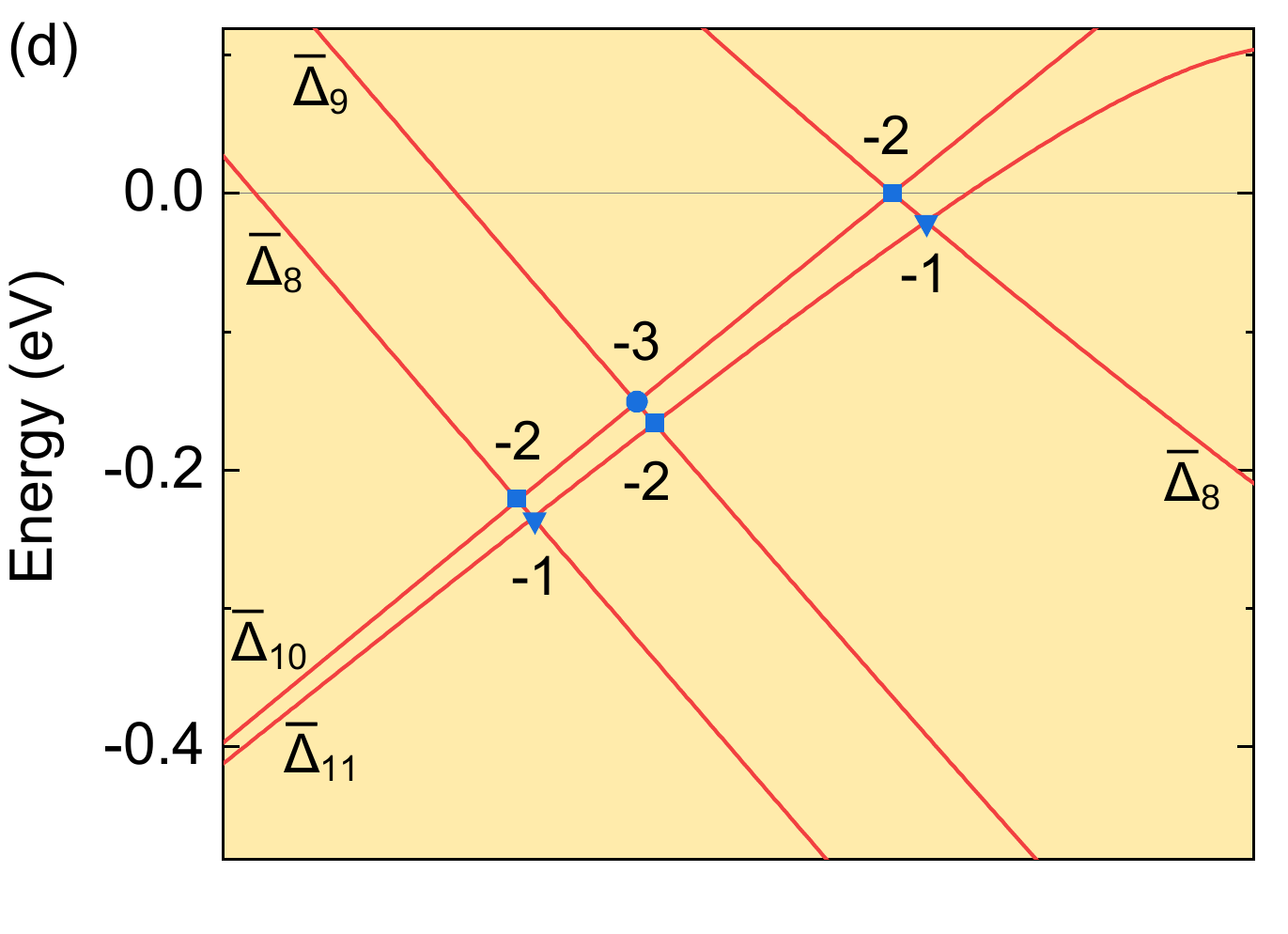}  

\end{tabular}
\caption{Band structures of ReO$_3$ (a) without SOC and (b) with SOC. The details near the $\text{K}$ and $\Gamma$ points in the case of SOC, indicated by the yellow shaded boxes, are shown in (c) and (d) respectively~\cite{S2}.}
\label{fig:band}
\end{figure*}

{\renewcommand{\arraystretch}{1.25}
\begin{table}
\begin{tabular}{cccccc}
\hline\hline
   M & K & $\Gamma$ & L & H & A \\
\hline
  $\overline{\text{M}}_5(2)$ & $\overline{\text{K}}_5(1)\oplus \overline{\text{K}}_6(2)$  & $\overline{\Gamma}_9(2)$ & $\overline{\text{L}}_3\overline{\text{L}}_5(2)$  & $\overline{\text{H}}_6(2)$    & $\overline{\text{A}}_4\overline{\text{A}}_5(2)$ \\
  $\overline{\text{M}}_5(2)$ & $\overline{\text{K}}_4(1)$    & $\overline{\Gamma}_7(2)$ & $\overline{\text{L}}_2\overline{\text{L}}_4(2)$ & $\overline{\text{H}}_6(2)$    & $\overline{\text{A}}_9(2)$    \\
  $\overline{\text{M}}_5(2)$ & $\overline{\text{K}}_6(2)$    & $\overline{\Gamma}_8(2)$ & $\overline{\text{L}}_2\overline{\text{L}}_4(2)$ & $\overline{\text{H}}_6(2)$    & $\overline{\text{A}}_8(2)$    \\
  $\overline{\text{M}}_5(2)$ & $\overline{\text{K}}_6(2)$    & $\overline{\Gamma}_7(2)$ & $\overline{\text{L}}_3\overline{\text{L}}_5(2)$ & $\overline{\text{H}}_4\overline{\text{H}}_5(2)$  & $\overline{\text{A}}_9(2)$    \\
  $\overline{\text{M}}_5(2)$ & $\overline{\text{K}}_4(1)$    & $\overline{\Gamma}_9(2)$ & $\overline{\text{L}}_2\overline{\text{L}}_4(2)$ & $\overline{\text{H}}_4\overline{\text{H}}_5(2)$ & $\overline{\text{A}}_6\overline{\text{A}}_7(2)$ \\
  $\overline{\text{M}}_5(2)$ & $\overline{\text{K}}_6(2)$    & $\overline{\Gamma}_8(2)$ & $\overline{\text{L}}_3\overline{\text{L}}_5(2)$ & $\overline{\text{H}}_6(2)$    & $\overline{\text{A}}_8(2)$    \\
     & $\overline{\text{K}}_5(1)$    &     &       &       &      \\
\hline\hline
\end{tabular}
\caption{Spinful IRREPs (and dimensionality) of the 12 bands near the Fermi level, ordered from low to high energy, following the labeling convention of Bilbao Crystallographic Server~\cite{Bilbao}.}
\label{tab:irreps}
\end{table}
}

Figures\,\ref{fig:band} (a-b) show the band structure of ReO$_3$ calculated without and with spin-orbit coupling (SOC), where the corresponding BZ has been given in Fig.\,\ref{struture}(b). It can be seen that the strong SOC interaction induced by the heavy element rhenium significantly splits the energy bands near the Fermi level. In the SOC case, the band structure displays a well-separated group of $12$ bands in the vicinity of the Fermi level with numerous crossing points. We have also applied a Hubbard correction to the $5d$-orbitals of rhenium to obtain a more accurate band structure~\cite{Liechtenstein1995, dudarev1998}, and it shows that there is no significant difference compared to Fig.\,\ref{fig:band}(b)~\cite{S2}. 

Focusing on the energy window within $E_\text{F} \pm 0.5$ eV, which is typically accessible in angle-resolved photoemission spectroscopy (ARPES) experiments, we have discovered two Weyl nodes that reside precisely on the Fermi level, as highlighted in Figs.\,\ref{fig:band} (c-d). One is located at the high symmetry point $\text{K}$, while the other is pinned on the high symmetry line $\Delta\equiv \overline{\Gamma\text{A}}$.

The Weyl node at the $\text{K}$ point is a triply degenerate point formed by two linear bands and one quadratic band. It is worth noting that this is distinct from previously studied unconventional Weyl nodes that arise from three-dimensional IRREPs of chiral point groups, $\mathsf{O}$ or $\mathsf{T}$~\cite{manes2012, barry2016, tian2021,multi-weylfang}. The point group of ReO$_3$, $\mathsf{D}_6$, has no spinful IRREP of more than two dimensions. \textcolor{black}{Instead, this triply degenerate Weyl node found here is a consequence of an accidental crossing between two legitimate IRREPs of the system. A careful DFT calculation reveals that the energy gap between $\overline{\text{K}}_5$ and $\overline{\text{K}}_6$ is less than 0.01\,meV.} The numerical integration of the Berry curvature around this triply degenerate Weyl node yields a high chirality of $+2$.

\begin{figure}
\centering
\includegraphics[width=0.85\linewidth]{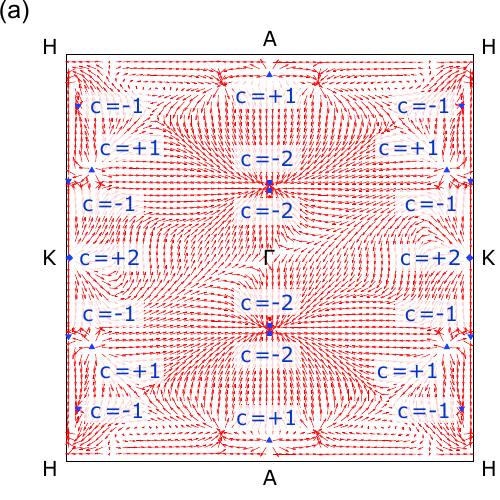}
\includegraphics[width=0.85\linewidth]{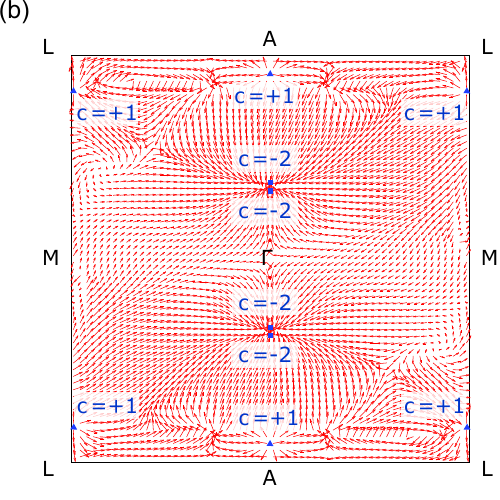}
\includegraphics[width=0.85\linewidth]{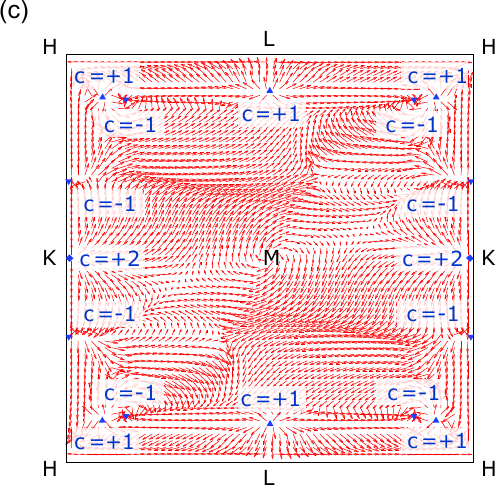}
\caption{Berry curvatures of the two lowest bands on the (a) $4\times\overline{\mathrm{K\Gamma AH}}$, (b) $4\times\overline{\mathrm{M\Gamma AL}}$, (c) $4\times\overline{\mathrm{KMLH}}$ planes. The Weyl nodes are labeled with their respective chirality $c$.}
\label{berry}
\end{figure}

The Weyl nodes pinned on the $\Delta$-line are formed by a set of accordion-like bands. Similar band structures have been observed in some other strong spin-orbit coupling materials with nonsymmorphic space groups, \textit{e.g.} AuF$_3$ crystallizing in the space group $\mathsf{P}6_122$~\cite{zhang2018}. However, in most nonsymmorphic crystals, these Weyl nodes pinned on the screw axis are far from the Fermi level. In stark contrast, ReO$_3$ exhibits six Weyl nodes in a narrow energy window within $0.25$\,eV of the Fermi level, making it an ideal candidate for experimental observation. We show in Sec.\,\ref{Sec: theory} that these Weyl nodes are symmetry-enforced by the screw rotations and demonstrate that the chiralities of the Weyl nodes along the $\Delta$-line can be inferred from symmetry alone. To establish this foundation, here we first calculate the IRREPs of the $12$ bands near the Fermi level at the high symmetry points (see Table\,\ref{tab:irreps}) and summarize the total chirality of Weyl nodes on three vertical irreducible BZ boundary planes (see Table\,\ref{tab:net_C}). Furthermore, Fig.\,\ref{berry} shows the Berry curvature of the two lowest bands out of the $12$ bands on the three high-symmetry planes (denoted by $4\times\overline{\mathrm{K\Gamma AH}}$, $4\times\overline{\mathrm{M\Gamma AL}}$, and $4\times\overline{\mathrm{KMLH}}$), and also the chirality of each Weyl node on the plane as calculated by integrating the Berry curvature around each node. One can clearly see that the doubly degenerate Weyl nodes on the $\Delta$-line act as a drain of Berry curvature, and the triply degenerate Weyl nodes located at the $\text{K}$ and $\text{K}^\prime$ points act as a source of Berry curvature.

{\renewcommand{\arraystretch}{1.5}
\begin{table*}
\begin{tabular}{cccccccc}
\hline\hline
   \quad\quad K(2) \quad\quad & \quad\quad $\overline{\Gamma \text{A}}$(2) \quad\quad & \quad\quad $\overline{\text{ML}}$(6) \quad\quad & \quad\quad $\overline{\text{KH}}$(4) \quad\quad & \quad\quad $\overline{\text{K} \Gamma \text{AH}}$(12) \quad\quad & \quad\quad $\overline{\text{M} \Gamma \text{AL}}$(12) \quad\quad & \quad\quad $\overline{\text{KMLH}}$(12) \quad\quad  \\
\hline
  $+2$ & $-3=-2-2+1$  & $+1$ & $-1$  & $0=-1+1$    & $0$     & $0=-1+1$  \\
\hline\hline
\end{tabular}
\caption{Net chirality on high symmetry points, lines, and planes. The numbers in parentheses indicate the symmetry multiplicity.}
\label{tab:net_C}
\end{table*}
}

\section{IRREPs indicated high-chirality Weyl nodes}
\label{Sec: theory}

We now explore the Weyl nodes from a deeper theoretical perspective, invoking the symmetries present in the underlying space group $\mathsf{P}6_322$ with point group $\mathsf{D}_6$. 

To facilitate the following discussion, we introduce the concept of the filling number $\nu$, ranging from $1$ to $12$, of the bands in the vicinity of the Fermi level, contained in the energy window from $-1.5\,\text{eV}$ to $3.5\,\text{eV}$, and ordered in increasing energy, such that the Bloch energy eigenvalues satisfy $E_{\nu_1}(\boldsymbol{k})\leq E_{\nu_2}(\boldsymbol{k})$ for all $\boldsymbol{k}$ and $\nu_1<\nu_2$ with $\nu_1,\nu_2\in\{1,\dots,12\}$. We use $\nu=n$ to formally separate all the bands below, say $\{E_{1}(\boldsymbol{k}),\dots,E_{n}(\boldsymbol{k})\}$, and above, $\{E_{n+1}(\boldsymbol{k}),\dots,E_{12}(\boldsymbol{k})\}$, a given Weyl node at some quasi-momentum $\boldsymbol{k}^*$, \textit{i.e.} with $E_{n}(\boldsymbol{k}^*)=E_{n+1}(\boldsymbol{k}^*)$. We show in Sec.\,\ref{sec_BBC} below that only the Weyl nodes with the filling number $\nu=2$ are relevant for the experimental observation of Fermi arcs at the Fermi level. 

\subsection{Screw symmetry eigenvalues}\label{sec_theory_A}
The screw symmetry $6_3\equiv\{C_{6}\vert \boldsymbol{\tau}\}$, made of a $C_6$ rotation around the $z$-axis combined with the fractional shift $\boldsymbol{\tau}=(3/6)\boldsymbol{a}_3=\boldsymbol{a}_3/2$ with $\boldsymbol{a}_3 = c(0,0,1)$ the primitive Bravais vector in the $z$-direction, protects Weyl nodes on the $\Delta$-line along the $k_z$-axis of the BZ. A generic point of the $\Delta$-line is invariant under the point subgroup $\mathsf{C}_6$, such that only the IRREPs of $\mathsf{C}_6$ are relevant to characterize the stability and topology of the Weyl nodes \footnote{Since the Bloch Hamiltonian is generically complex, \textit{i.e.} it belongs to the Altland-Zirnbauer symmetry class $\mathsf{A}$~\cite{Kitaev,SchnyderClass}, band crossings have codimension $3$ in the four-dimensional parameter space $(\omega,\boldsymbol{k})$ of one-body Green's functions, such that stable band crossings are {\it points} (zero-dimensional) ~\cite{Volovik,zhao2013topFS}.}. We list in Table \ref{table_C6} the $C_6$-symmetry eigenvalues (equivalently, the characters) for the spinless IRREPs of $\mathsf{C}_6$ which we label by $\Delta_j$ with $j=1,\dots,6$. It is worth noting that the $C_6$-eigenvalues are all of the form 
\begin{equation}
    \chi^l_{6} = \mathrm{e}^{\text{i} \frac{2\pi}{6}l}\;,~l\in\{0,1,\dots,5\}\equiv\mathbb{Z}_6\,,  
\end{equation}
that is the sextic roots of unity.

{\renewcommand{\arraystretch}{1.3}
\begin{table}[]
    \centering
    \begin{tabular}{c|cccccc}
    \hline\hline
     $\mathsf{C}_6$ & $\Gamma_1$ & 
     $\Gamma_2$ & $\Gamma_3$ & $\Gamma_4$ &
     $\Gamma_5$ & $\Gamma_6$  \\
     \hline 
      $l$ & $0$ & $3$ &
     $4$ & $1$ & $2$ & $5$ \\
     \hline
      $\chi^{l}_6\equiv \mathrm{e}^{\text{i}\frac{2\pi}{6}l}$ & $ 1$ & $ -1$ & $ \mathrm{e}^{-\text{i} \frac{2\pi}{3}}$
      & $ \mathrm{e}^{\text{i} \frac{\pi}{3}}$ & $ \mathrm{e}^{\text{i} \frac{2\pi}{3}}$ & $ \mathrm{e}^{-\text{i} \frac{\pi}{3}}$ \\
         \hline\hline 
    \end{tabular}
    \caption{
    $C_6$-symmetry eigenvalues (equivalently, the characters) of the spinless IRREPs of the point group $\mathsf{C}_6$ following the labeling convention of the Bilbao Crystallographic Server~\cite{Bilbao}. 
    }
    \label{table_C6}
\end{table}
}

{\renewcommand{\arraystretch}{1.6}
\begin{table}[]
    \centering
    \begin{tabular}{c|cccccc}
    \hline\hline
     $\mathsf{G}^{\Delta}$ & $\overline{\Delta}_7$ & 
     $\overline{\Delta}_8$ & $\overline{\Delta}_9$ & $\overline{\Delta}_{10}$ &
     $\overline{\Delta}_{11}$ & $\overline{\Delta}_{12}$  \\
     \hline 
     $(l,s)_{l\in\mathbb{Z}_6,s\in\{\pm1\}}$ &
     $\substack{(4,-1)\\ (5,+1)}$  &
      $\substack{(1,+1)\\ (2,-1)} $ &
      $\substack{(2,-1)\\ (3,+1)}$ &
      $\substack{(0,+1)\\ (5,-1)} $ &
      $ \substack{(0,-1)\\ (1,+1)} $ &
      $\substack{(3,-1)\\ (4,+1)} $ \\
      \hline
      \hline 
      $\zeta^{\overline{\Delta}_j}_k(6_3)= \mathrm{e}^{\text{i}\frac{\pi}{6} (2l-s)} $ & $-\text{i}$ & $\text{i}$ & $\mathrm{e}^{\text{i} \frac{5\pi}{6}} $ & $\mathrm{e}^{-\text{i} \frac{\pi}{6}} $ &
      $\mathrm{e}^{\text{i} \frac{\pi}{6}} $ &
      $\mathrm{e}^{-\text{i} \frac{5\pi}{6}}$\\
      \hline 
      $
      \begin{array}{c}
           \chi^{\overline{\Delta}_j}_{k}(6_3) 
      =\zeta^{\overline{\Delta}_j}_{k}(6_3) \kappa  \\
           {\scriptstyle (\kappa = \mathrm{e}^{-\text{i}k\pi})} 
      \end{array}$ & $-\text{i}\kappa$ & $\text{i}\kappa$ & $\mathrm{e}^{\text{i} \frac{5\pi}{6}}\kappa$ & $\mathrm{e}^{-\text{i} \frac{\pi}{6}} \kappa$ &
      $\mathrm{e}^{\text{i} \frac{\pi}{6}} \kappa$ &
      $\mathrm{e}^{-\text{i} \frac{5\pi}{6}} \kappa$  \\
         \hline\hline 
    \end{tabular}
    \caption{
    $6_3$-symmetry eigenvalues (equivalently, the characters) of the spinful IRREPs of the little co-group $\mathsf{G}^{\Delta}\cong \mathsf{C}_6$ of the $\Delta$-line (see derivation in Appendix \ref{ap_wannier_basis} and \ref{sec_reps_6}), following the labeling convention of the Bilbao Crystallographic Server~\cite{Bilbao}. Here $\zeta^{\overline{\Delta}_j}_{k}(6_3)$ gives the $6_3$-symmetry eigenvalues in the basis of the cell-periodic Bloch eigenstates. 
    }
    \label{table_C6_3}
\end{table}
}

We write a generic point of the $\Delta$-line, while avoiding the high-symmetry points $\Gamma$ and A, as 
\begin{equation}
    \boldsymbol{k}^{\Delta} = k\, \boldsymbol{b}_3~\text{for}~ k\in(0,\frac{1}{2})~\text{or}~k\in(\frac{1}{2},1)\,,
\end{equation}
with $\boldsymbol{b}_3$ the primitive reciprocal lattice vector dual to $\boldsymbol{a}_3$. In this setting, the quasi-momentum $\boldsymbol{k}^{\Delta}$ and the shift $\boldsymbol{\tau}$ are invariant, \textit{i.e.} $D_6\boldsymbol{k}^{\Delta}=\boldsymbol{k}^{\Delta}$ and $D_6^{-1}\boldsymbol{\tau}=\boldsymbol{\tau}$, where $D_6$ is the matrix representation of the action of $C_6$ on $\boldsymbol{k}$ (see Appendix \ref{sec_reps_6}). Taking into account the spin-$1/2$ degrees of freedom of the electronic system and the effect of the fractional shift $\boldsymbol{\tau}$ of the screw symmetry $6_3$, numbering the bands from lower to higher energy (\textit{i.e.} $n=1,2,\dots$), and assuming there is no accidental degeneracy between the Bloch eigenstates, we find that the eigenvalues of the $6_3$-screw symmetry are given by
\begin{equation}
\left\{
\begin{aligned}
    \langle \psi_{n} , k \vert {^{\{C_6\vert \boldsymbol{\tau}\}}} \vert \psi_{n} , k \rangle  &=  \chi_k^{\overline{\Delta}_{j_n}} (6_3)\,,\\
    \chi_k^{\overline{\Delta}_{j_n}}(6_3) &= \zeta_k^{\overline{\Delta}_{j_n}}(6_3)\,  \mathrm{e}^{-\text{i} k\pi} \,, \\
    \zeta_k^{\overline{\Delta}_{j_n}}(6_3) &\in\left\{ \mathrm{e}^{\text{i} (2l - s)\frac{\pi}{6}} \right\}_{l\in\mathbb{Z}_6,s\in\{\pm1\}} \,,
\end{aligned}\right.
\end{equation}
where $\zeta_k^{\overline{\Delta}_{j_n}}(6_3)$ is the eigenvalue of the $6_3$-screw symmetry in the basis of the cell-periodic Bloch eigenstates, \textit{i.e.}
\begin{equation}
\label{eq_zeta_six}
    \zeta_k^{\overline{\Delta}_{j_n}}(6_3) = \langle u_{n} ,  k \vert {^{\{C_6\vert \boldsymbol{\tau}\}}} \vert u_{n} , k \rangle \,,
\end{equation}
(see Appendix \ref{ap_wannier_basis} and \ref{sec_reps_6} for a detailed derivation).  The symmetry eigenvalues $\chi_k^{\overline{\Delta}_{j_n}}(6_3)$ correspond to the characters of the spinful IRREPs of the $6_3$-screw symmetry of the little co-group of the $\Delta$-line~\cite{BradCrack}
\begin{equation}
    \mathsf{G}^{\Delta} = \bigcup_{l\in \mathbb{Z}_6} \{C_6\vert\boldsymbol{\tau}\}^l \cong \mathsf{C}_6 \,,
\end{equation}
with $\{C_6\vert\boldsymbol{\tau}\}^0\equiv \{E\vert \boldsymbol{0}\}$. There is thus a two-to-one correspondence between the set of all combinatorial $(l,s)$-pairs and the set of spinful IRREPs,\textit{ i.e.}
\begin{equation}
    \left\{(l,s)\right\}_{l,s}\mapsto 
    \left\{\overline{\Delta}_j\right\}_{j\in\{7,8,9,10,11,12\}}\,.
\end{equation}
Here we follow the same labeling notation as the Bilbao Crystallographic Server~\cite{Bilbao}. The specific results are listed in Table \ref{table_C6_3}, where the symmetry representation $\{\zeta_k^{\overline{\Delta}_{j_{n}}}\}_n$ will be used below in the derivation of symmetry-indicated chiralities. 

\subsection{Screw symmetry protected Weyl nodes}\label{sec_theory_B}

After obtaining the screw symmetry eigenvalues, we are now able to evaluate the Chern number in an algebraic manner. This is essentially an application of the method first presented in Refs\,\cite{Wi2} and \cite{BBS_nodal_lines} (see also Ref.\,\cite{magenticpaper} for a tetragonal example) which relies on Wilson loop techniques developed in Refs.\,\cite{InvTIBernevig,Chenprb2012,hourglass,Alex_BerryPhase}. Contrary to the approaches in Refs.\,\cite{multi-weylfang,Vanderbilt_screw}, it does not require the derivation of a symmetry-constrained $\boldsymbol{k}\cdot\boldsymbol{p}$ Bloch Hamiltonian. In addition, we note that Ref.\,\cite{uribe2021chiralities} provides another approach based on equivariant cohomology considerations \textcolor{black}{and Ref.\,\cite{alpin2023} presents a similar argument as this work in terms of relating the Chern number and the exchange of IRREPs.} 

We begin by deriving all possible chiralities of Weyl nodes on the $\Delta$-line. First, it is worth bearing in mind that the crossing on the $\Delta$-line between any pair of bands, say bands $E_{n}(\boldsymbol{k})$ and $E_{n+1}(\boldsymbol{k})$, belonging to distinct IRREPs is protected by the screw symmetry $6_3$. Furthermore, such a crossing defines a twofold separation of the spectrum with all the bands $\{\dots,E_{n}(\boldsymbol{k})\}$, which we refer to as {\it occupied}, and the other bands $\{E_{n+1}(\boldsymbol{k}),\dots\}$, which we refer to as {\it unoccupied}. We emphasize that this separation is only formal and does not need to correspond to the Fermi level. 

{\renewcommand{\arraystretch}{1.25}
\begin{table*}[]
    \centering
    \begin{tabular}{c|ccccccccccccccc}
    \hline\hline
    $(j,j')$ & 
    $(7,8)$ &
     $(7,9)$ &
     $(7,10)$ &
     $(7,11)$ &
     $(7,12)$ &
     $(8,9)$ &
     $(8,10)$ &
     $(8,11)$ &
     $(8,12)$ &
     $(9,10)$ &
     $(9,11)$ &
     $(9,12)$ &
     $(10,11)$ &
     $(10,12)$ &
     $(11,12)$ 
     \\
     \hline
     $\begin{array}{c}c^{(j,j')}\\ c^{(j',j)} \end{array}$ & 
     $\begin{array}{r} -3 \\ 3 \end{array}$ & 
     $\begin{array}{r} 2 \\ -2 \end{array}$ & 
     $\begin{array}{r} -1 \\ 1 \end{array}$ & 
     $\begin{array}{r} -2 \\ 2 \end{array}$ & 
     $\begin{array}{r} 1 \\ -1 \end{array}$ &
     $\begin{array}{r} -1 \\ 1 \end{array}$ & 
     $\begin{array}{r} 2 \\ -2 \end{array}$ & 
     $\begin{array}{r} 1 \\ -1 \end{array}$ & 
     $\begin{array}{r} -2 \\ 2 \end{array}$ &
     $\begin{array}{r} -3 \\ 3 \end{array}$ & 
     $\begin{array}{r} 2 \\ -2 \end{array}$ & 
     $\begin{array}{r} -1 \\ 1 \end{array}$ & 
     $\begin{array}{r} -1 \\ 1 \end{array}$ & 
     $\begin{array}{r} 2 \\ -2 \end{array}$ & 
     $\begin{array}{r} -3 \\ 3 \end{array}$ \\
     \hline\hline 
    \end{tabular}
    \caption{
    Symmetry-indicated chirality $c[\mathcal{s}]=c^{(j,j')} \mod 6$ [Eq.\,(\ref{eq_chern_6})] of Weyl nodes corresponding to the crossing points at the filling number $n$ of any two bands of IRREPs $\overline{\Delta}_j$ and $\overline{\Delta}_{j'}$ on the $\Delta$-line, with $\mathcal{s}$ the surface wrapping the Weyl node with $k_0$ ($k_1$) the south (north) pole ($k_0<k_1$), and such that we assume the configuration of the $n$-th and $n+1$-th bands as $(\overline{\Delta}(n,k_0),\overline{\Delta}(n,k_1))=(\overline{\Delta}(n+1,k_1),\overline{\Delta}(n+1,k_0))=(\overline{\Delta}_{j},\overline{\Delta}_{j'})$. Permuting the IRREPs of the two bands, $(j,j')\mapsto (j',j)$, leads to a reversed chirality. We remark that one cannot distinguish $c^{(j,j')} = 3$ and $c^{(j,j')} = -3$ since the symmetry-indicated chirality $c[\mathcal{s}]$ is only defined modulo 6. In such a case, one must deduce the sign from the global consistency of all symmetry-indicated Weyl nodes in the BZ, namely the total chirality at a constant filling number must vanish over the whole BZ. (See the discussion in Sec.\,\ref{sec_other_WP}.) 
    }
    \label{table_chern}
\end{table*}
}

\begin{figure}
\begin{tabular}{ll}
(a) & (b)\\
\includegraphics[width=0.57\linewidth]{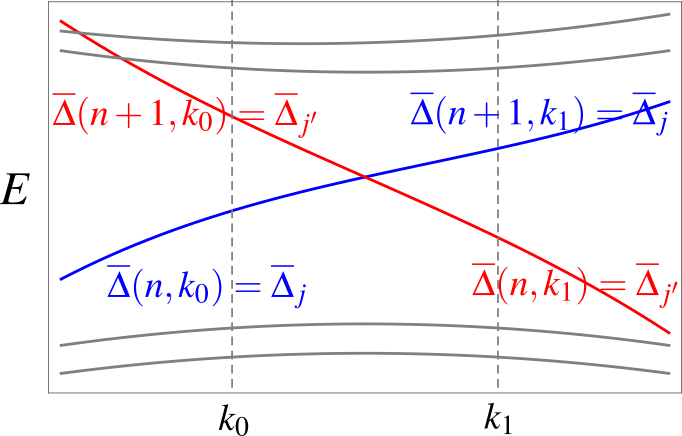} & 
\includegraphics[width=0.43\linewidth]{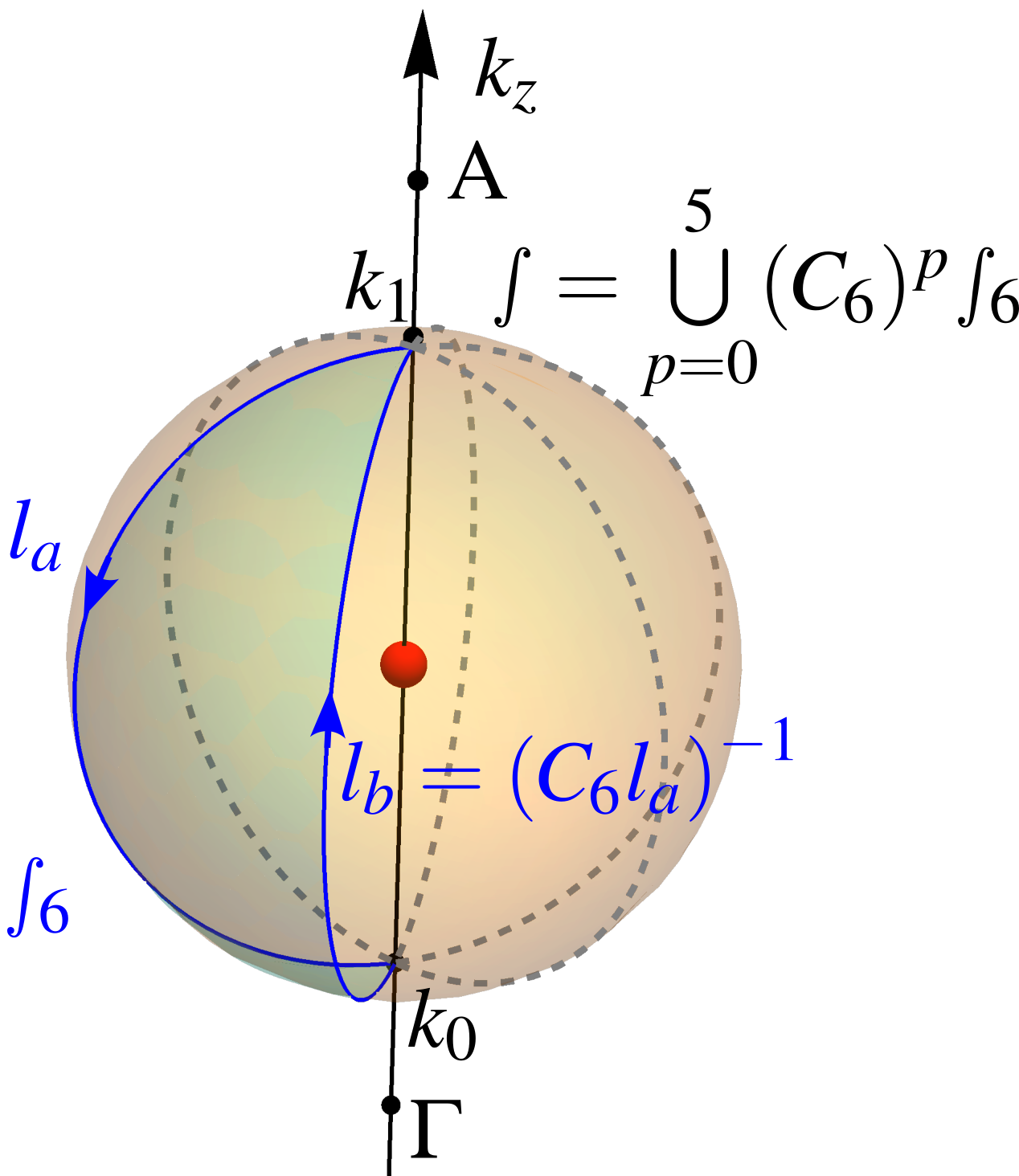}
\end{tabular}
\caption{(a) Energy dispersion in the vicinity of a $6_3$-symmetry protected band crossing between the IRREPs $\overline{\Delta}_j$ and $\overline{\Delta}_{j'}$ for a filling number $\nu=n$. (b) $C_6$-symmetry generated surface $\mathcal{s}$ wrapping the band crossing on the $\Delta$-line. One sixth slice, $\mathcal{s}_6$, is bounded by the loop $\partial \mathcal{s}_6 = l_b \circ l_a$ where the arc $l_a$ connects $k_1$ and $k_0$, and $l_b=(C_6 l_a)^{-1}$.}
\label{fig_ball}
\end{figure}

Our approach to computing the chirality of a given Weyl node is to use the quantization of the Wilson loop spectrum evaluated over all occupied bands for a filling number of a given band crossing and integrated along a specific base loop that is dictated by the screw symmetry $6_3$, where the Wilson loop is defined as
\begin{equation}
\left\{
\begin{aligned}
    \mathcal{W}[l] &= \langle \boldsymbol{u},k_0\vert \hat{W}_l \vert \boldsymbol{u},k_0\rangle \,,\\
    \hat{W}_l &= \prod\limits_{k}^{k_0\leftarrow k_1\leftarrow k_0}\vert \boldsymbol{u},k\rangle \langle \boldsymbol{u},k\vert \,,
\end{aligned}\right.
\end{equation}
with the base loop $l$ passing through the points $\{k_0,k_1\}$ and the $n$-tuple vector of occupied cell-periodic Bloch eigenstates $\vert \boldsymbol{u},k\rangle = (\vert u_1,k\rangle~\dots~\vert u_n,k\rangle)$ (see Appendix \ref{ap_wannier_basis}). More precisely, let us consider a band crossing between two bands of IRREPs $\overline{\Delta}_{j}$ and $\overline{\Delta}_{j'}$ at the filling number $n$, located at a generic point $k^*$ of the $\Delta$-line---still avoiding $\Gamma$ and A, such that $k^*\in(0,1/2)$ or $k^*\in(1/2,1)$. We then define two points $\{k_0,k_1\}$ on the same half of the $\Delta$-line surrounding the crossing point, \textit{i.e.} $0<k_0<k^*<k_1<1/2$ or $1/2<k_0<k^*<k_1<1$, and such that the band crossing under consideration is the only one in the segment $[k_0,k_1]$ for the same filling number. As an example, Fig.\,\ref{fig_ball}(a) shows a band crossing for the band configuration 
\begin{equation}
\left\{
\begin{aligned}
    \overline{\Delta}(n,k_0)&=\overline{\Delta}(n+1,k_1)=\overline{\Delta}_{j} \,,\\
    \overline{\Delta}(n,k_1)&=\overline{\Delta}(n+1,k_0)=\overline{\Delta}_{j'} \,.
\end{aligned}
\right.    
\end{equation} 
We now define a loop $l_{101}=l_b \circ l_{a}$, composed of two arcs: $l_a$ starting at $k_1$ and ending at $k_0$, and $l_b$ obtained as the reversal of the $C_6$ image of $l_a$, \textit{i.e.} $l_b=(C_6l_a)^{-1}$, starting at $k_0$ and ending at $k_1$, as shown in Fig.\,\ref{fig_ball}(b). The rank-$n$ Wilson loop over the base loop $l_{101}$ thus reads
\begin{equation}
\begin{aligned}
    \mathcal{W}_{l_{101}} &= \mathcal{W}_{l_b} \cdot \mathcal{W}_{l_a} = \mathcal{W}_{(C_6l_a)^{-1}} \cdot 
    \mathcal{W}_{l_a} \,,\\
    &= \left(\mathcal{W}_{C_6l_a} \right)^{-1} \cdot 
    \langle \boldsymbol{u},k_0\vert   \hat{W}_{l_a}  \vert \boldsymbol{u},k_1\rangle \,,\\
    &= \left(
    \langle \boldsymbol{u},k_0\vert   \hat{W}_{C_6l_a}  \vert \boldsymbol{u},k_1\rangle \right)^{-1} \cdot 
    \mathcal{W}_{l_a}  \,,\\
    &= \left(\mathcal{R}_{k_0} \langle \boldsymbol{u},k_0\vert^{\{C_6\vert\boldsymbol{\tau}\}}   \hat{W}_{C_6l_a}{^{\{C_6\vert\boldsymbol{\tau}\}}}\vert \boldsymbol{u},k_1\rangle   \mathcal{R}_{k_1}^{-1} \right)^{-1} \cdot  
     \mathcal{W}_{l_a} \,,\\
     &= \mathcal{R}_{k_1} 
     \langle \boldsymbol{u},k_0\vert   \hat{W}_{l_a}\vert \boldsymbol{u},k_1\rangle^{-1}   \mathcal{R}_{k_0}^{-1}  \cdot  \mathcal{W}_{l_a} \,,\\
     &= \mathcal{R}_{k_1} \cdot 
     \mathcal{W}_{l_a}^{-1} \cdot 
     \mathcal{R}_{k_0}^{-1}\cdot \mathcal{W}_{l_a}\,,
\end{aligned}
\end{equation}
where we have used the matrix form of Eq.\,(\ref{eq_zeta_six})
\begin{equation}
\begin{aligned}
     \mathcal{R}_{k}(\{C_6\vert\boldsymbol{\tau}\}) &= \langle \boldsymbol{u} ,  k \vert {^{\{C_6\vert \boldsymbol{\tau}\}}} \vert \boldsymbol{u} , k \rangle  \,,\\
      &= \text{diag}\left( \zeta^{\overline{\Delta}_{j_1}}_{k} (6_3), \dots,
      \zeta^{\overline{\Delta}_{j_n}}_{k}(6_3) 
    \right)\,.
\end{aligned}
\end{equation}

We obtain the Berry phase over $l_{101}$ from the determinant of the Wilson loop, \textit{i.e.}
\begin{equation}
\label{eq_berry_6}
\begin{aligned}
    \mathrm{e}^{-\text{i}\gamma^{(n)}_B[l_{101}]} &= \det \mathcal{W}[l_{101}]=  \frac{\det \left[\mathcal{R}_{k_1}(\{C_{6z}\vert\boldsymbol{\tau}\})\right]}{\det \left[\mathcal{R}_{k_0}(\{C_{6z}\vert\boldsymbol{\tau}\})\right] } \\
& = \left(\prod\limits_{a=1}^{n-1} \dfrac{\zeta^{\overline{\Delta}_{j_a}}_{k_1}}{\zeta^{\overline{\Delta}_{j_a}}_{k_0}}\right)  \dfrac{\zeta^{\overline{\Delta}(n,k_1)}_{k_1}}{\zeta^{\overline{\Delta}(n,k_0)}_{k_0}}
= 1\cdot \dfrac{\zeta^{\overline{\Delta}_{j'}}_{k_1}}{\zeta^{\overline{\Delta}_{j}}_{k_0}}
\end{aligned}
\end{equation}
where we have used the fact that the $n$-th band exhibits a change of IRREPs from $\overline{\Delta}(n,k_0)=\overline{\Delta}_j$ to $\overline{\Delta}(n,k_1)=\overline{\Delta}_{j'}$. Since $l_{101}$ defines the contour of a sixth slice $\mathcal{s}_6$ of a surface $\mathcal{s}$ wrapping the band crossing (\textit{i.e.} $\partial \mathcal{s}_6\equiv l_{101}$), we have
\begin{equation}
    \mathcal{s} = \bigcup\limits_{p=0}^{5} (C_6)^p \mathcal{s}_6\,.
\end{equation}
Then, invoking Stokes' theorem yields
\begin{equation}
\label{eq_chern2berry}
    \mathrm{e}^{-\text{i}2\pi c[\mathcal{s}]/6} = \mathrm{e}^{-\text{i}\gamma^{(n)}_B[l_{101}]} \,, 
\end{equation}
where $c[\mathcal{s}]$ is the Chern number over the closed surface $\mathcal{s}$, (\textit{i.e.} the chirality of the Weyl node wrapped by $\mathcal{s}$), from which we deduce
\begin{equation}
     \dfrac{\pi}{3} c[\mathcal{s}] =   \gamma^{(n)}_B[l_{101}] \mod 2\pi \,.   
\end{equation}
Upon further utilization of Eq.\,(\ref{eq_berry_6}), we ultimately arrive at
\begin{equation}
\label{eq_chern_6}
\left\{
\begin{aligned}
     c[\mathcal{s}] &=   \dfrac{3}{\pi} \gamma^{(n)}_B[l_{101}] \mod 6 
    = c^{(j,j')} \mod 6 \,,\\
    c^{(j,j')}&\equiv \dfrac{3\text{i}}{\pi}  \ln\left( \dfrac{\zeta^{\overline{\Delta}_{j'}}_{k_1}}{\zeta^{\overline{\Delta}_{j}}_{k_0}} \right)  \,.
\end{aligned}\right.
\end{equation}
We list in Table \ref{table_chern} the chiralities $c^{(j,j')}$ for all possible pairs of IRREPs $(\overline{\Delta}_{j},\overline{\Delta}_{j'})=(\overline{\Delta}(n,k_0),\overline{\Delta}(n,k_1))=(\overline{\Delta}(n+1,k_1),\overline{\Delta}(n+1,k_0))$, noting that the permutation $(j,j')\mapsto (j',j)$ gives a reversed chirality, \textit{i.e.} $c^{(j,j')} = -c^{(j',j)}$.

It is worth highlighting that our approach is not restricted to nonsymmorphic screw symmetries. In the case of a symmorphic hexagonal system, the above relations still hold with the only difference that the symmetry representations in the basis of the Bloch eigenstates, $\{\vert\psi_{nk}\rangle\}_n$, and of the cell-periodic Bloch eigenstates, $\{\vert u_{nk}\rangle\}_n$, match, \textit{i.e.} $\zeta^{\overline{\Delta}}_{k}=\chi^{\overline{\Delta}}_{k}$ in a symmorphic system. We discuss the extra constraints that the nonsymmorphic symmetry imposes on the band structure at the end of this section. In the following, we first consider the Weyl nodes on the first half of the $\Delta$-line, and then deduce the Weyl nodes on the second half of the $\Delta$-line imposed by the $\{C_{2z}\vert\boldsymbol{\tau}\}T$ symmetry using the fact that both pure rotation and time-reversal preserve chirality, or equivalently in this context, by $C_{2y}$ (or $\{C_{2x}\vert\boldsymbol{\tau}\}$) symmetry (see the discussion below on lowering the symmetry). 

We begin by corroborating the necessary presence of the Weyl nodes in ReO$_3$ indicated by the spinful IRREPs. As we will argue in Sec.\,\ref{sec_BBC}, only the energy gap at the Fermi level with a filling number $\nu=2$ is observable from the projected surface spectra. The energy gaps at the Fermi level with other filling numbers are hidden by the pockets formed by the Fermi surface. Here we thus focus on the Weyl nodes on the $\Delta$-line with $\nu=2$. For each Weyl node, we define an appropriate pair of momenta $(k_0,k_1)$ that directly surround the node, as in the above discussion. Starting with the Weyl node closest to the $\Gamma$ point shown in Fig.\,\ref{fig:band}(d), it is associated to the IRREPs' configuration $\overline{\Delta}(3,k_1)=\overline{\Delta}(2,k_0)=\overline{\Delta}_{10}$ and $\overline{\Delta}(2,k_1)=\overline{\Delta}(3,k_0)=\overline{\Delta}_8$, which gives $c^{(10,8)}=-2$ leading to a chirality $c = -2 \mod 6$. The next Weyl node with $\nu=2$ on the $\Delta$-line has $c^{(11,9)}=-2$, leading to a chirality $c=-2\mod 6$. The last Weyl node with $\nu=2$ on the first half of the $\Delta$-line has $c^{(9,8)}=+1$, leading to the chirality $c=+1\mod 6$. This Weyl node configuration must then be reproduced, in reversed order, on the second half of the $\Delta$-line by $\{C_{2z}\vert\boldsymbol{\tau}\}T$ symmetry or, equivalently, by $C_{2y}$ (or $\{C_{2x}\vert\boldsymbol{\tau}\}$) symmetry. 

We interestingly note that a lowering of the symmetry of the system from the point group $\mathsf{D}_6$ to $\mathsf{C}_6$ (\textit{i.e.} from $\mathsf{P6_322}=SG182$ to $\mathsf{P6_3}=SG173$) while preserving time-reversal symmetry, does not affect the topological configuration. Inversely, breaking time-reversal while preserving the point group $\mathsf{D}_6$ (\textit{i.e.} from the type II Shubnikov magnetic space group $\mathsf{P6_3221^\prime}=MSG182.180$ to type I $\mathsf{P6_322}=MSG182.179$~\cite{Bilbao,MSGtables}) would also preserve the Weyl node configuration.

So far, the chiralities were only defined modulo $6$, such that, for example, a chirality of $-2$ cannot be distinguished from $+4$ only from the IRREPs. Unsurprisingly, using the maximally localized Wannier model (see the details in Sec.\,\ref{sec_comp}), we find numerically that the Weyl nodes always realize the chirality with the lowest allowed absolute Chern number (see the discussion in Sec.\,\ref{sec_GC}). 
We thus conclude that the $\Delta$-line hosts a total chirality of $c_{\Delta}=(-2-2+1)\times 2=-6$, where the factor $2$ comes from the chirality-preserving $\{C_{2z}\vert\boldsymbol{\tau}\}T$ image of each node. Crucially, these Weyl nodes must be compensated by Weyl nodes located away from the $\Delta$-line to comply with the Nielsen-Ninomiya theorem~\cite{NIELSEN1981219} of global cancellation of Weyl charges over the whole BZ (see the discussion in Sec.\,\ref{sec_GC}). 

We end this subsection with the additional constraints imposed by the screw symmetry on the band structure. We directly see from Table \ref{table_C6_3} that the $k$-dependence of the IRREPs characters, through $\kappa = \mathrm{e}^{-\text{i}k\pi}$, implies the permutations of IRREPs under a shift by a full reciprocal vector along the screw axis. 

More precisely, given an $n$-th band associated with the IRREP $\overline{\Delta}_{j_n}$, such that the Bloch eigenstate $\vert \psi_{n},k \rangle $ has the symmetry eigenvalue $\chi^{\overline{\Delta}_{j_n}}_{k}$ at $k$, the same $n$-th band at the shifted momentum $k+1$ corresponds to the Bloch eigenstate $\vert \psi_{n}, k+1 \rangle$ with the symmetry eigenvalue $\chi^{\overline{\Delta}_{j_n'}}_{k+1}$. Then, the symmetry eigenvalue of $\vert \psi_{n}, k+1 \rangle $ must be equal to the symmetry eigenvalue of $\vert \psi_{n}, k \rangle $ by virtue of the commutation of the Bloch Hamiltonian with the translational representation in reciprocal space (see Appendix \ref{ap_wannier_basis}), \textit{i.e.}
\begin{equation}
\label{eq_shifted}
    \begin{aligned}
        \chi^{\overline{\Delta}_{j_n'}}_{k+1} &= -\chi^{\overline{\Delta}_{j_n'}}_{k} \stackrel{!}{=} \chi^{\overline{\Delta}_{j_n}}_{k}\,,\; j_n,j_n'\in\{7,\dots,12\}\,,
    \end{aligned}
\end{equation}
where the minus sign in the first equality comes from the factor $\kappa_{k=1}=\mathrm{e}^{\text{i} \pi} =-1$. The above equation defines the pair of IRREPs $(\overline{\Delta}_{j_n}\,,\overline{\Delta}_{j'_n})_{(j_n'\neq j_n)}$ that are associated by a reciprocal translation. 

If, instead of sticking to the $n$-th band, we follow the Bloch eigenstate $\vert \psi_n, k \rangle$ via a parallel transport to the shifted momentum $k+1$, \textit{i.e.} writing $\psi^{\parallel}$ as the parallel transported Bloch state,
\begin{equation}
 \vert \psi^{\parallel},k\rangle = \vert \psi_n,k\rangle \mapsto  \vert \psi^{\parallel},k+1\rangle = \vert \psi_m,k+1\rangle\,,
\end{equation}
which must hold for some $m\neq n$, and
we then obtain that the symmetry eigenvalue transforms as
\begin{equation}
    \chi^{\overline{\Delta}_{j_n}}_{k} \mapsto  \chi^{\overline{\Delta}_{j_n}}_{k+1} = - \chi^{\overline{\Delta}_{j_n}}_{k} \stackrel{!}{=} \chi^{\overline{\Delta}_{j_m}}_{k} \,,
\end{equation}
since the Bloch eigenstate $\vert \psi_m,k+1\rangle$ has the symmetry eigenvalue $\chi^{\overline{\Delta}_{j_m}}_{k}$. We emphasize that since we necessarily have $j_m\neq j_n$ in the above equation, the symmetry eigenvalue of the parallel transported Bloch eigenstate must be changed after a full reciprocal lattice translation, indeed $\overline{\Delta}_{j_m}\neq \overline{\Delta}_{j_n}$ for $j_m\neq j_n$. The above equation again defines the pair of IRREPs $(\overline{\Delta}_{j_n},\overline{\Delta}_{j_m})$ that are associated by a full reciprocal lattice translation. By consistency with Eq.\,(\ref{eq_shifted}), we furthermore conclude that
\begin{equation}
    \overline{\Delta}_{j_n'} = \overline{\Delta}_{j_m} \,,
\end{equation}
or 
\begin{equation}
   j_n' = j_m \,,
\end{equation}
with $j_m\in\{7,\dots,12\}\setminus\{j_n\}$.

\begin{figure}
\begin{tabular}{l}
\includegraphics[width=0.8\linewidth]{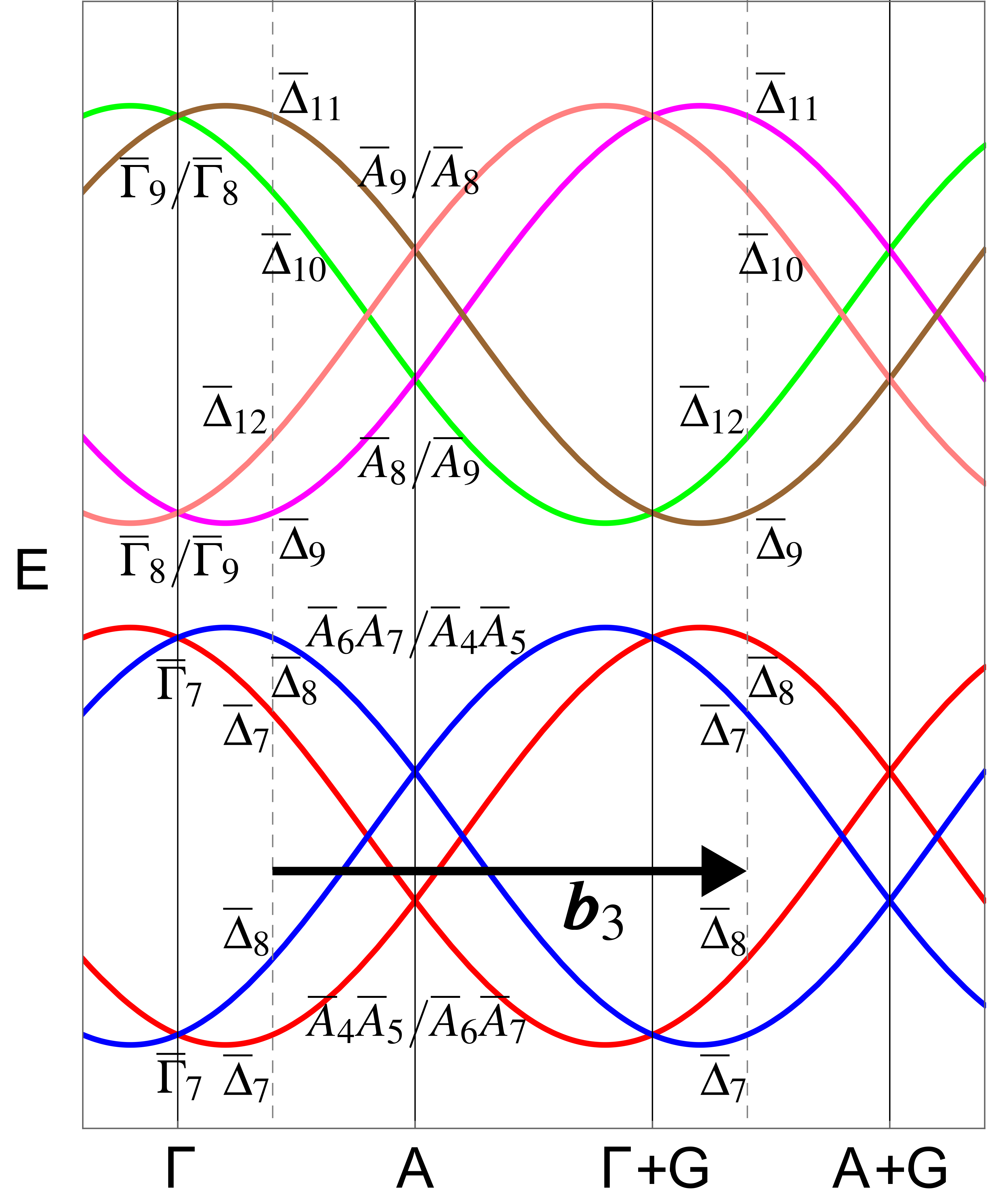} 
\end{tabular}
\caption{Symmetry enforced connectivity of bands along the $6_3$-screw axis. While the energy-ordered states preserve their symmetry eigenvalue under a shift by a full reciprocal lattice vector $\boldsymbol{b}_3$, \textit{i.e.} from $\vert \psi_{n}^{\overline{\Delta}_j},k\rangle $ to $\vert \psi_{n}^{\overline{\Delta}_j},k+1\rangle $, on the contrary, parallel transported states (marked by colors) must change their symmetry eigenvalue, \textit{e.g.} let us choose $\vert \psi^{\parallel},k\rangle= \vert \psi^{\overline{\Delta}_9}_5,k\rangle$ which becomes $\vert \psi^{\parallel},k+1\rangle= \vert \psi^{\overline{\Delta}_{11}}_8,k+1\rangle$. The compatibility relations between the IRREPs at the $\Gamma$ and A points and those on the $\Delta$-line can be found in Ref.~\cite{Bilbao}. Note the relative freedom in the choice of IRREPs at the $\Gamma$ and A points, denoted as $\overline{\text{X}}_j/\overline{\text{X}}_{j'}$ for $\text{X}=\Gamma\text{ and A}$.}
\label{fig_band_con}
\end{figure}

A direct inspection gives the following associated IRREPs' pairs $(\overline{\Delta}_7,\overline{\Delta}_8)$, $(\overline{\Delta}_9,\overline{\Delta}_{12})$ and $(\overline{\Delta}_{10},\overline{\Delta}_{11})$ for space group $\mathsf{P}6_322$. This, together with the compatibility relations between the IRREPs on the $\Delta$-line and the IRREPs at the $\Gamma$ and A points~\cite{Bilbao}, we conclude that every band must be part of a minimum block of four connected bands, with two types of blocks shown in Fig.\,\ref{fig_band_con} (note the relative freedom in the choice of IRREPs at the $\Gamma$ and A points, denoted as $\overline{\text{X}}_j/\overline{\text{X}}_{j'}$ for $\text{X}=\Gamma,\text{A}$). While there is some relative freedom in the choice of IRREPs at the high-symmetry points $\Gamma$ and A, the compatibility relations between these and the IRREPs on the $\Delta$-line leave only the two following types of connected-band-blocks, the group $\{\overline{\Delta}_7,\overline{\Delta}_7,\overline{\Delta}_8,\overline{\Delta}_8\}$ and the other group $\{\overline{\Delta}_9,\overline{\Delta}_{10},\overline{\Delta}_{11},\overline{\Delta}_{12}\}$. 


\subsection{Weyl nodes on other high-symmetry regions}\label{sec_other_WP}

The previous analysis can be directly transferred to the characterization of the Weyl nodes located on the other rotation-invariant lines. These are the two inequivalent $\text{P}\equiv\overline{\text{KH}}$-lines and the three inequivalent $\text{U}\equiv\overline{\text{ML}}$-lines, to which we add the two K points. We again only consider the Weyl nodes with the filling number $\nu=2$.

Starting with the bands crossing at the $\text{K}$ point at the Fermi level, we find an accidental degeneracy of a two-dimensional IRREP, $\overline{\text{K}}_6(2)$, with a one-dimensional IRREP, $\overline{\text{K}}_5(1)$. Given the compatibility relations from K to the $\mathsf{C}_3$-symmetric vertical P-line~\cite{Bilbao}, \textit{i.e.}
\begin{equation}
\left\{
\begin{aligned}
    \overline{\text{K}}_5(1)&\mapsto \overline{\text{P}}_4(1)\,,\\
    \overline{\text{K}}_6(2)&\mapsto \overline{\text{P}}_5(1)\oplus\overline{\text{P}}_6(1) \,,
\end{aligned}\right.
\end{equation}
the bands crossing at the $\text{K}$ point are characterized by their 3$\equiv$$\{C_3\vert \boldsymbol{0}\}$=$\{C_6\vert\boldsymbol{\tau}\}^2$-symmetry eigenvalues on the P-line~\cite{Bilbao}, \textit{i.e.} 
\begin{equation}
\label{eq_C3_characters}
\left\{
\begin{aligned}
    \chi^{\overline{\text{P}}_4}(3) &=\zeta^{\overline{\text{P}}_4}(3) =  -1 \,,\\
    \chi^{\overline{\text{P}}_5}(3) &= \zeta^{\overline{\text{P}}_5}(3) = \mathrm{e}^{-\text{i}\pi/3} \,,\\
    \chi^{\overline{\text{P}}_6}(3) &= \zeta^{\overline{\text{P}}_6}(3)= \mathrm{e}^{\text{i}\pi/3} \,.
\end{aligned}\right.
\end{equation}
In the direct vicinity below the $\text{K}$ point, we find the ordered 1D-IRREPs $(\overline{\text{P}}_5,\overline{\text{P}}_4)$, that become $(\overline{\text{P}}_6,\overline{\text{P}}_4)$ in the direct vicinity above the $\text{K}$ point. Applying the same method as in Sec.\,\ref{sec_theory_B}, but now for $C_3$ rotation symmetry [\textit{i.e.} using the symmetry eigenvalues presented in Eq.\,(\ref{eq_C3_characters})], we find the symmetry-indicated chirality of the Weyl node at the $\text{K}$ point to be 
\begin{equation}
\label{eq_c1K}
\begin{aligned}
    c[\mathcal{s}_{\text{K}}] = \dfrac{3\text{i}}{2\pi} \ln \left(\dfrac{\zeta^{\overline{\text{P}}_6}_{k_1}(3)}{\zeta^{\overline{\text{P}}_5}_{k_0}(3)}\right) \mod 3 = -1\mod 3\,,
\end{aligned}
\end{equation}
where we have chosen two momenta directly below and above the K point as $\boldsymbol{k}_0=(\boldsymbol{b}_1+\boldsymbol{b}_2)/3+k_0 \boldsymbol{b}_3$ and $\boldsymbol{k}_1=(\boldsymbol{b}_1+\boldsymbol{b}_2)/3+k_1 \boldsymbol{b}_3$, with $-k_0=k_1\gtrapprox 0 $.

We proceed with the Weyl node located on the P-line between K and H, and its image under $C_{2z}T$. Given the compatibility relation~\cite{Bilbao} 
\begin{equation}
    \overline{\text{H}}_6(2) \mapsto 
    \overline{\text{P}}_5(1) \oplus \overline{\text{P}}_6(1)\,,
\end{equation}
we find the symmetry-indicated chirality 
\begin{equation}
\begin{aligned}
    c[\mathcal{s}_{\text{P}}] = \dfrac{3\text{i}}{2\pi} \ln \left(\dfrac{
    \zeta^{\overline{\text{P}}_6}_{k_1}(3)}{\zeta^{\overline{\text{P}}_5}_{k_0}(3)}\right) \mod 3 = -1\mod 3\,,
\end{aligned}
\end{equation}
where again $[k_0,k_1]$ frames the Weyl node on the first half of the P-line. By $C_{2z}T$ symmetry, the Weyl node on the other half has the same chirality.    

We finally consider the Weyl node located on the first half of the $\text{U}$-line. The relevant compatibility relations are~\cite{Bilbao} 
\begin{equation}
\left\{
\begin{aligned}
    \overline{\text{M}}_5(2) &\mapsto \overline{\text{U}}_3(1)\oplus \overline{\text{U}}_4(1)\,,\\
    \overline{\text{L}}_3 \overline{\text{L}}_5(2)&\mapsto 
    2\overline{\text{U}}_4(1)
    \,.
\end{aligned}\right.
\end{equation}
Again using the method of Sec.\,\ref{sec_theory_B}, but now for the screw symmetry $\{C_{2z}\vert\boldsymbol{\tau}\}$$=$$\{C_{6z}\vert\boldsymbol{\tau}\}^3$, we find the symmetry-indicated chirality
\begin{equation}
\begin{aligned}
    c[\mathcal{s}_{\text{U}}] = \dfrac{2\text{i}}{2\pi} \ln \left(\dfrac{
    \zeta^{\overline{\text{U}}_{4}}_{k_1}(2_1)}{\zeta^{\overline{\text{U}}_{3}}_{k_0}(2_1)}\right) \mod 2 = 1\mod 2\,,
\end{aligned}
\end{equation}
where we have used the symmetry eigenvalues~\cite{Bilbao} $-\zeta^{\overline{\text{U}}_{3}}_{k_1}(2_1) = \zeta^{\overline{\text{U}}_{4}}_{k_1}(2_1) = \text{i}$.

\subsection{Principles of low chirality and global consistency}\label{sec_GC}

So far, we have derived the $C_\eta$-rotation indicated chirality (or Chern number) of Weyl nodes that is only defined modulo $\eta$, for $\eta=6,3,2$ \textcolor{black}{(depending on the rotation symmetry of the little co-group at the high-symmetry region), such that the strictly symmetry-constrained chiralities of Weyl points are not determined univocally.} In the context of the Wannier lattice model, realizing high chiralities generically requires that the symmetry-allowed long-range hopping parameters be dominant compared to symmetry-allowed short-range hopping parameters. Such a situation is highly unlikely given the general tendency of exponential decay of the hopping parameters with increasing hopping distance. \textcolor{black}{This pleads to a heuristic principle that among all the symmetry-allowed chiralities, the one with relatively low absolute value is more likely to be realized in the band structure of a real material system.} Applying this principle to the Weyl nodes indicated by symmetry, we would assign the following charges
\begin{equation}
\left\{
\begin{aligned}
    c_{\Delta} & =c_{\text{min}}[\mathcal{s}_{\Delta}] = -3 \,,\\ 
    c_{\text{P}} & = c_{\text{min}}[\mathcal{s}_{\text{P}}] = -1\,,\\
    c_{\text{U}} &= c_{\text{min}}[\mathcal{s}_{\text{U}}] \in \{+1,-1\}\,,\\
    c_{\text{K}} & = c_{\text{min}}[\mathcal{s}_{\text{K}}] \in \{-1,+2\}\,,
\end{aligned}\right.
\end{equation}
\textcolor{black}{where the sign of $c_{\text{U}}$ remains undecided using the symmetry data only, and we leave the possibility of the second lowest chirality at the K point given the higher symmetry constraints (corresponding to the point group $\mathsf{D}_3$) on the short-ranged hopping parameters.}

\textcolor{black}{Counting the multiplicity of Weyl nodes under the chirality preserving $C_{2z}T$ symmetry (\textit{i.e.} every Weyl node above the plane $k_z=0$ must have an identical image underneath it) and taking the multiplicity of the high-symmetry regions of the BZ into account, \textit{i.e.} there are two K-points, two P-lines, and three U-lines, we arrive at the consistency equation of global cancellation of chiralities (dictated by the Nielsen-Ninomiya theorem~\cite{NIELSEN1981219})
\begin{equation}
\label{eq_global_cancellation}
2 c_{\Delta} + (2\times 2) c_{\text{P}}+ (2\times 3) c_{\text{U}} + 2 c_{\text{K}} =0 \,.
\end{equation}
One readily finds that the only chiralities compatible with the global cancellation are 
\begin{equation}
    c_{\text{U}} = +1 \,,\;
    c_{\text{K}} = +2\,,
\end{equation}
so that the triply degenerate Weyl node at the K point must be of high chirality as well. We have verified these symmetry-based values through direct computation from an accurate symmetrized Wannier model, see Fig.\,\ref{berry} and Table\.~\ref{tab:net_C}.}

\begin{figure}
\centering
\includegraphics[width=1.0\linewidth]{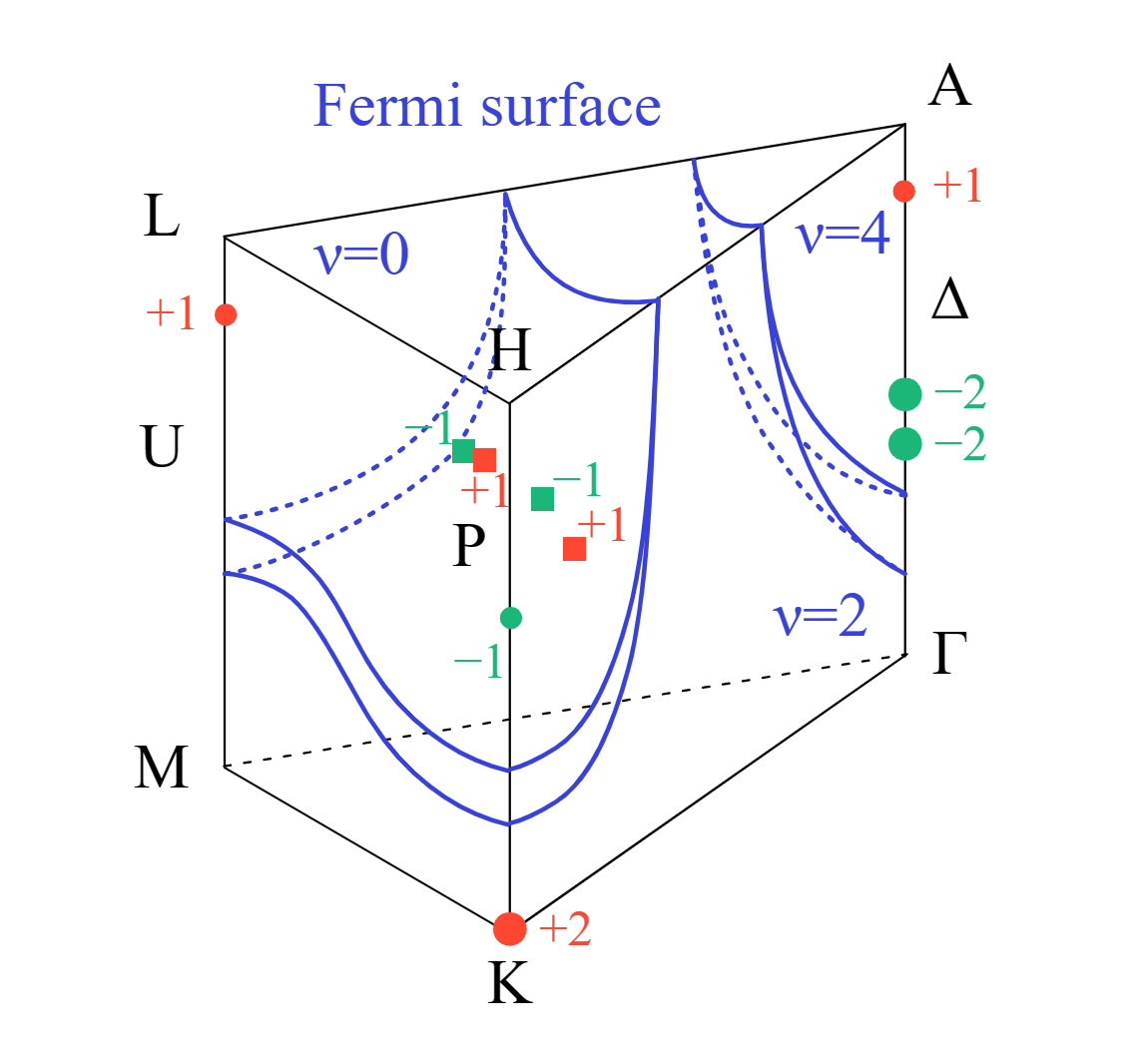}
\caption{Schematic Fermi surface sheets (blue line) separating the regions with distinct filling numbers $\nu=0,2,4$ (see definition in Sec.\,\ref{Sec: theory}) within the irreducible BZ of the hexagonal lattice. The Fermi surface sheets wrap all the symmetry-indicated (circle) and the non-symmetry-indicated (square) Weyl nodes with $\nu=2$, except the one located at the K point.}
\label{fig_FS}
\end{figure}

We render in Fig.\,\ref{fig_FS} the schematic configuration of the symmetry-indicated Weyl nodes and their chiralities as well as the Fermi surface over the fundamental domain of the BZ.

We conclude this theory part by noting that our method, except for the minimal global connectivity discussed at the end of Sec.\,\ref{sec_theory_B}, can be directly transferred to the context of symmorphic space groups. 

\section{Non-symmetry-indicated Weyl nodes and Berry curvature integration}
\label{BCvsIRREPS}

There remain two groups of Weyl nodes with the filling number $\nu=2$, neither of which is indicated by symmetry. We have evaluated their chiralities numerically from the maximally localized Wannier model and rendered them also in Fig.\,\ref{fig_FS}. First, there is one pair of Weyl nodes with opposite chiralities on the half vertical $C_2T$-invariant plane $\overline{\Gamma\text{A}\text{H}\text{K}}$. That means there are twelve such pairs in total, given the $\{C_{2z}\vert\boldsymbol{\tau}\}T$ images and the sixfold multiplicity of non-equivalent $\overline{\Gamma\text{A}\text{H}\text{K}}$ vertical planes. Second, there is a pair of Weyl nodes, again of opposite chiralities, on each half vertical $C_2T$-invariant plane $\overline{\text{K}\text{M}\text{L}\text{H}}$, which again means a total of twelve such pairs, given the $\{C_{2z}\vert\boldsymbol{\tau}\}T$ images and the multiplicity of the high-symmetry plane. Note that these non-symmetry-indicated Weyl nodes do not affect the global cancellation condition of Eq.\,(\ref{eq_global_cancellation}). We furthermore show in Sec.\,\ref{sec_BBC} that they are not visible in the surface spectra either.  

We next compare the symmetry-based method demonstrated in Sec.\,\ref{Sec: theory} with the traditional method involving the numerical integration of the Berry curvature, where the Berry curvature is often obtained from either first-principles Kohn-Sham orbitals or fitted Wannier orbitals.

First of all, it is worth noting that Kohn-Sham (spinful) orbitals in DFT, which are Bloch eigenfunctions of a Bloch Hamiltonian, are typically implemented in a plane wave basis coupled to bare $1/2$-spinors. While our symmetry arguments are completely general, in the sense that they do not depend on the method with which the Bloch eigenstates of the system are computed, we refer to a specific choice of basis functions for the sake of the argument. It is, in particular, convenient to expand the Kohn-Sham orbitals in a basis of localized Wannier functions representing atomic-like degrees of freedom that couple sublattice sites, electronic orbitals, and $1/2$-spins together. As we show in detail in Appendix \ref{ap_wannier_basis}, the Wannier-basis expansion of Bloch eigenstates allows us to readily derive the characters of all spinful IRREPs (equivalently, the symmetry eigenvalues of Bloch eigenstates) from the {\it diagonal} symmetry action on a properly chosen linear combination of microscopic degrees of freedom (and this constitutes the final result of Appendix \ref{sec_reps_6}). 

Moreover, we emphasize that while the numerical wannierization of the Kohn-Sham orbitals can be performed (and has been performed in this work), we do not need to do it explicitly for the argument in  Sec.\,\ref{Sec: theory} to be valid. Our conclusions are independent of whether or not we have access to numerically accurate Wannier-based Kohn-Sham orbitals. 

In terms of practical significance, the most notable advantage of using IRREPs indication to determine the chirality of Weyl nodes is the elimination of the need to construct the Wannier model. In fact, in the case of ReO$_3$ we have observed that even with a Wannier model that accurately reproduces a subset of the first principle bands, the Berry curvature integration for the Weyl nodes located on the $\Delta$-line is not completely robust. This is particularly evident when two Weyl nodes are very close to each other, with one of them having a chirality $|c|>1$. In principle, one can always calculate the chirality of a Weyl node by building a sphere that includes the Weyl node and excludes all others. However, when the two Weyl nodes are very close to each other, it results in a cumbersome process to find the appropriate size of the sphere enclosing the Weyl node of interest. Also, this often leads to a very small sphere, thus the numerical noise of the Berry curvature integration becomes considerable. By contrast, using the method based on the IRREPs, one can circumvent the above issue. We therefore propose that the IRREPs indication method may be more suitable for automated high-throughput identification of the chirality of symmetry-enforced Weyl nodes.

\section{Complementary ``real'' topologies}\label{sec_Euler}

\textcolor{black}{The Chern topology, at the origin of the chiralities of Weyl points, is the fundamental topology of complex vector bundles and is thus the fundamental topology of the Bloch eigenvectors of a complex Bloch Hamiltonian matrix (obtained in the basis of Bloch states, see Eq.\,(\ref{eq_bloch_Hamiltonian_matrix}) of Appendix \ref{ap_wannier_basis}). Yet, there are high-symmetry regions that satisfy a ``reality condition'', namely a region of the Brillouin zone over which there exists a change of basis that is constant for the whole region and that makes the Bloch Hamiltonian matrix real. The existence of such a region is protected by symmetry (see below) and the corresponding Bloch eigenvectors are made real under the change of basis such that their fundamental topology is captured by the topology of real vector bundles (note that real vector bundles necessarily have a trivial Chern topology). In this section, we study the patch Euler class of Weyl nodes formed by the crossing of two bands only \cite{bouhon2019nonabelian}, and the non-Abelian frame charges of nodes formed by the crossing of more than two bands \cite{wu2018,jiang2021observation}.}

\subsection{Reality condition of $C_2T$-invariant planes}

\textcolor{black}{The reality condition is satisfied whenever the high-symmetry region is invariant under an antiunitary symmetry that squares to identity, and can be understood as a consequence of the Takagi factorization of symmetric unitary matrices\,\cite{bouhon2019nonabelian,chen2021manipulation}. It turns out that any $C_2T$ symmetry constitutes such a symmetry and this is true whether we consider the symmorphic $C_2$ symmetries 
\begin{equation}
    C_{2y}\,,\;C_{2,(10)}\equiv C_{3z}^{-1}C_{2y}\,,\;C_{2,(01)}\equiv C_{3z}C_{2y},    
\end{equation}
or the nonsymmorphic $C_2$ symmetries of the present system
\begin{equation}
    \{C_{2z}\vert \boldsymbol{\tau}\}\,,\;
    \{C_{2x}\vert \boldsymbol{\tau}\}\,,\;
    \{C_{2,(12)}\vert \boldsymbol{\tau}\}\,,\; 
    \{C_{2,(21)}\vert \boldsymbol{\tau}\} \,,
\end{equation}
where $C_{2,(12)}\equiv C_{6z}C_{2y}$ and $C_{2,(21)}\equiv C_{6z}^{-1}C_{2y}$. Then, given a $C_{2\perp}$-axis and a plane perpendicular to it we define a point $(k_{\parallel 1},k_{\parallel 2},k_{\perp})$, where $(k_{\parallel 1},k_{\parallel 2})$ are the components in a basis within the plane and $k_{\perp}$ is the coordinate along the $C_{2\perp}$-axis. The action of $C_{2\perp} T$ on the momentum is thus $(k_{\parallel 1},k_{\parallel 2},k_{\perp}) \mapsto  (k_{\parallel 1},k_{\parallel 2},-k_{\perp})$, that is a mirror image with respect to the plane perpendicular to the $C_{2\perp}$-axis. We refer to this plane as the ``$C_{2\perp}T$-invariant plane''. We then conclude that any point of the $C_{2\perp}T$-invariant plane satisfies the reality condition.} 

\textcolor{black}{While we count seven $C_2T$-type symmetries for the space group $\mathsf{P6_322}$ (one for each $C_2$ symmetry listed above), they are not all independent. Indeed, e.g. the three generators $C_{6z}$, $C_{2y}$ and $T$ give rise to the whole (magnetic point) group under composition. In the following, it is enough to consider the ``real'' topological invariants protected by the three symmetries $\{C_{2z}\vert\boldsymbol{\tau}\}T$, $C_{2y}T$, and $\{C_{2x}\vert\boldsymbol{\tau}\}T$, which we write $C_{2z}T$, $C_{2y}T$, and $C_{2x}T$ in this section for brevity.}

\subsection{Patch Euler class of Weyl points}\label{sec_patch_euler}

Within any $C_2T$-invariant plane the reality condition is satisfied (\textit{i.e.} there exists a change of basis that makes the Bloch Hamiltonian real) and the Euler class of a Weyl node formed by two bands, say the bands $E_{n}(\boldsymbol{k})$ and $E_{n+1}(\boldsymbol{k})$, can be evaluated over a patch $\mathcal{D}$ of the BZ containing the Weyl node, such that there is no band crossing with the other adjacent bands over the whole patch $\mathcal{D}$. Mathematically, this can be expressed as 
\begin{equation}
    \xi[\mathcal{D}] = \dfrac{1}{2\pi}\left(\int_{\mathcal{D}}  \mathrm{Eu} -\oint_{\partial\mathcal{D}} a \right)\,,
\end{equation}
where the Euler connection is obtained from the Pfaffian of the two-band Berry connection, \textit{i.e.} $a = \text{Pf}\,\mathcal{A} $ with $\mathcal{A} = (\langle u_n,\boldsymbol{k}\vert~\langle u_{n+1},\boldsymbol{k}\vert)^{\top} \mathrm{d}(\vert u_n,\boldsymbol{k}\rangle~\vert u_{n+1},\boldsymbol{k}\rangle)$ evaluated from the real Bloch eigenvectors, and the Euler form is $\mathrm{Eu} = \mathrm{d}a = \mathrm{Pf}\,\mathrm{d}\mathcal{A} $ ~\cite{bouhon2019nonabelian}. The value of the patch Euler class indicates {\it half} the number of stable nodes, $N_{\mathrm{node}}[\mathcal{D}]\in\mathbb{Z}$, formed by the two-band subspace and located within the patch $\mathcal{D}$ (chosen such that no crossing with other bands happens over $\mathcal{D}$), \textit{i.e.} $\vert \xi[\mathcal{D}]\vert = \dfrac{1}{2} N_{\mathrm{node}}[\mathcal{D}]$. This Euler class-indicated number of nodes is topologically stable under the condition that the energy gaps above and below the two-band subspace containing the nodes are preserved. \textcolor{black}{While the sign of the patch Euler class of a Weyl point is completely independent of the sign of its Chern number (\textit{i.e.} chirality), the parity of the two invariants must be compatible, i.e.
\begin{equation}
      2\xi[\mathcal{D}]  \mod 2 = c[\mathcal{s}]  \mod 2 \,.
\end{equation}}
For example, a simple Weyl node with chirality $c=\pm1$, must exhibit a half-integer patch Euler class $\xi\in\{\pm1/2,\pm3/2,\dots\}$ \textcolor{black}{(there is no relation between the signs of $\xi$ and $c$)}, while a double Weyl node with chirality $c=\pm2$ must exhibit an integer patch Euler class $\xi\in\{0,\pm1,\dots\}$ \textcolor{black}{(again, no relation between the signs)}.  

The Weyl nodes on the vertical $\Delta$-line are characterized by two independent patch Euler classes, $(\xi_{C_{2x}T},\xi_{C_{2y}T})$, that are defined within two independent vertical $C_2T$-invariant planes (there are six $C_2T$-invariant planes crossing on the $\Delta$-line, but they split into two groups of three planes that are mapped onto each other under the action of the $C_{6z}$ symmetry), \textit{i.e.} the plane $\overline{\Gamma\text{A}\text{H}\text{K}}$ for $C_{2x}T$ and $\overline{\Gamma\text{A}\text{L}\text{M}}$ for $C_{2y}T$. The simple Weyl node on the $\Delta$-line, with $c=+ 1\mod 6$, is pinned on the intersecting vertical $C_{2}T$-invariant planes since it is its own image under the $C_{2}T$ symmetries. This topological pinning is captured by the (non-zero) half-integer value of both patch Euler classes, \textit{i.e.} $\vert\xi_{C_{2x}T}\vert$, $\vert\xi_{C_{2y}T}\vert \in (\mathbb{Z} + 1/2)$. 

The same applies to the simple Weyl nodes on the U-lines \textcolor{black}{protected by both $C_{2y}T$ and $C_{2x}T$}, while the Weyl nodes on the P-line acquire a single (independent) patch Euler class \textcolor{black}{protected by $C_{2x}T$}. Clearly, the pinning of these simple Weyl nodes to the vertical rotation-invariant axes is also required by the rotation symmetries themselves, \textit{i.e.} $C_{6z}$ for the $\Delta$-line, $C_{2z}$ for the U-line, and $C_{3z}$ for the P-line (\textcolor{black}{\textit{i.e.} these nodes are their own images under the rotation symmetries}).   

Considering next the high-chirality Weyl nodes of the $\Delta$-line, with $c=-2\mod 6$, they must each give rise to a pair of integer patch Euler classes, \textit{i.e.} $\vert\xi_{C_{2x}T}\vert$, $\vert\xi_{C_{2y}T}\vert \in \mathbb{Z}$.

\textcolor{black}{We find that the symmetry data alone (\textit{i.e.} the IRREPs) does not constrain the precise value of the Euler invariants beyond their parity such that they must be computed numerically from the Bloch eigenstates. We have done this here using the Bloch eigenvectors of a symmetrized Wannier model that reproduces accurately the band structure around the Energy level. A direct computation yields trivial Euler classes. This is in agreement with Sec.\,\ref{sec_strain}} where we study the $C_6$-symmetry breaking (and thus $C_{3z}$-symmetry breaking) induced by shear stress. Indeed, the splitting and trajectory of the high-chirality Weyl nodes of the $\Delta$-line induced by the symmetry breaking imply that both patch Euler classes vanish.

\subsection{\textcolor{black}{Multiple quaternion charges of the triply degenerate Weyl node}}\label{sec_real_top}

\textcolor{black}{More intriguing is the topology of the triply degenerate Weyl node at the K point. Indeed, beyond the high-chirality of $c=+2$ obtained above, it also supports a ``real'' topology characterization since it sits at the intersection of the $C_{2x}T$-invariant plane, and its two $C_2T$-invariant images under $C_{3z}$ symmetry, and the horizontal $C_{2z}T$-invariant plane. As mentioned above, from a symmetry point of view, the triply degenerate Weyl node is formed by the ``accidental'' superposition of the two-dimensional IRREP $\overline{\text{K}}_6$ with the one-dimensional IRREP $\overline{\text{K}}_5$. While it is accidental from a symmetry point of view, the triply degenerate Weyl node exhibits a certain robustness as seen, for example, upon breaking $C_{6z}$ symmetry in Sec.\,\ref{sec_strain}. Crucially, the resulting three-band crossing does not permit an evaluation of the topology through the patch Euler class (which is restricted to two-band subspaces only). We can instead characterize the triply degenerate Weyl node through its non-Abelian frame charge \cite{Wu1273}. More precisely, we consider the non-Abelian frame charge of the \textit{partial frame} made of the three Bloch eigenvectors forming the triply degenerate Weyl node. This leads to an effective three-band quaternion homotopy invariant $q\in \mathbb{Q} = \{+1,\pm i, \pm j,\pm k,-1\}$ computed over a loop within one $C_2T$-invariant plane and encircling the node \cite{bouhon2019nonabelian}. We denote the two independent quaternion charges, one for each independent $C_2T$-invariant plane containing the K point, by $(q_{C_{2z}T},q_{C_{2x}T})$. The remaining quaternion charges $(q_{C_{2,(12)}T},q_{C_{2,(21)}T})$, corresponding to the two remaining $C_2T$-invariant planes containing the K point, must be equal to $q_{C_{2x}T}$ by $C_{3z}$ symmetry. A direct computation using the Bloch eigenvectors of the symmetrized Wannier model yields two nontrivial quaternion charges for the 3-band partial frame:
\begin{equation}
    (q_{C_{2z}T},q_{C_{2x}T})=(-1,-1) \,, 
\end{equation}
and, by $C_{3z}$ symmetry:
\begin{equation}
     q_{C_{2,(12)}T} = q_{C_{2,(21)}T}= -1 \,.
\end{equation}
}
\begin{figure}
\centering
\includegraphics[width=0.475\linewidth]{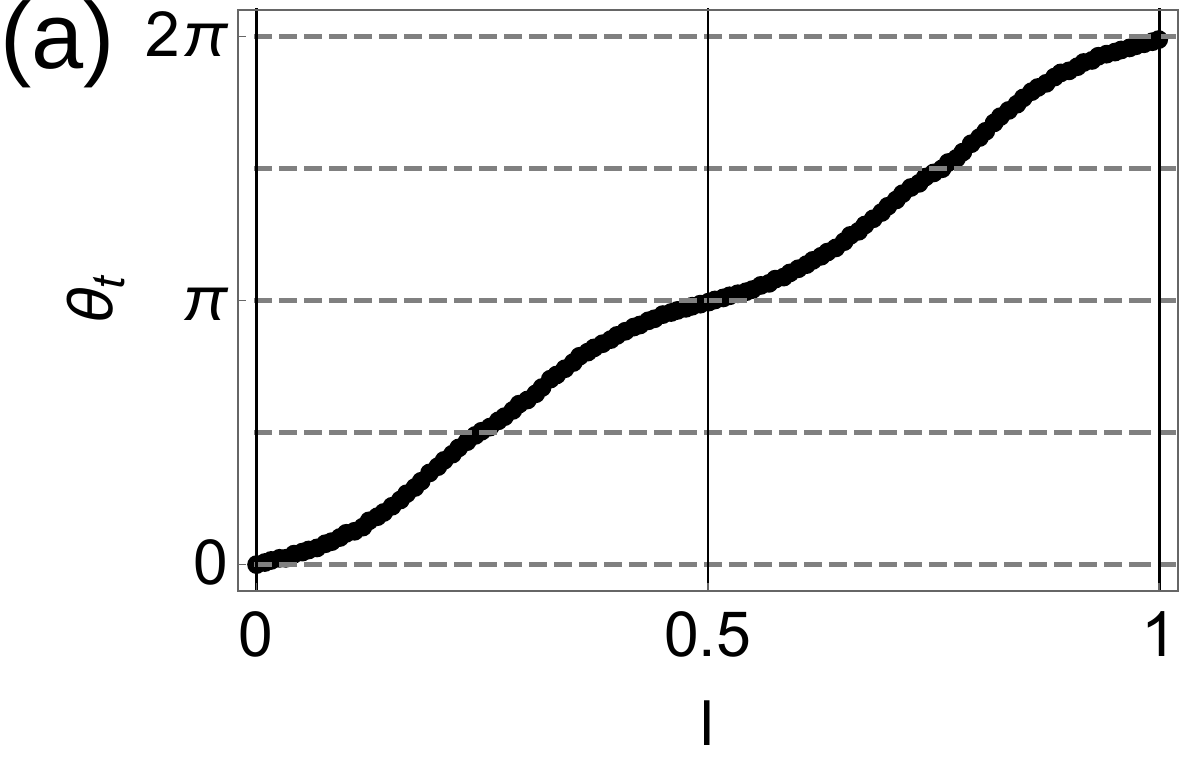}
\quad
\includegraphics[width=0.475\linewidth]{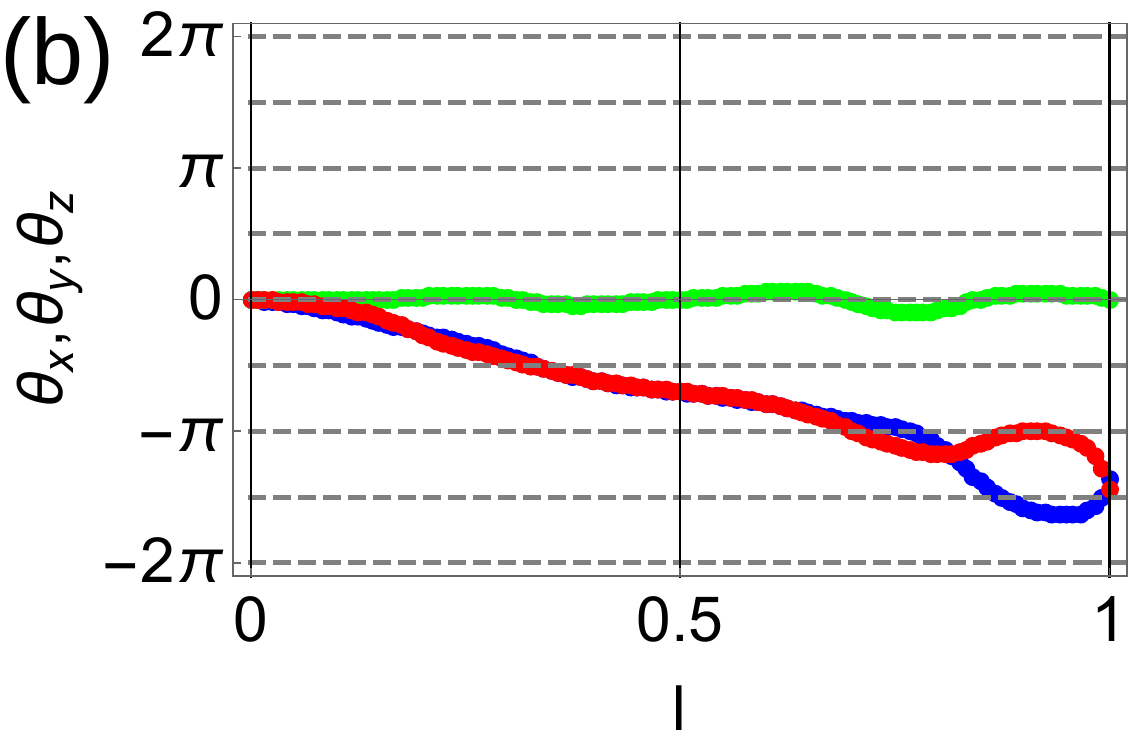}
\includegraphics[width=0.475\linewidth]{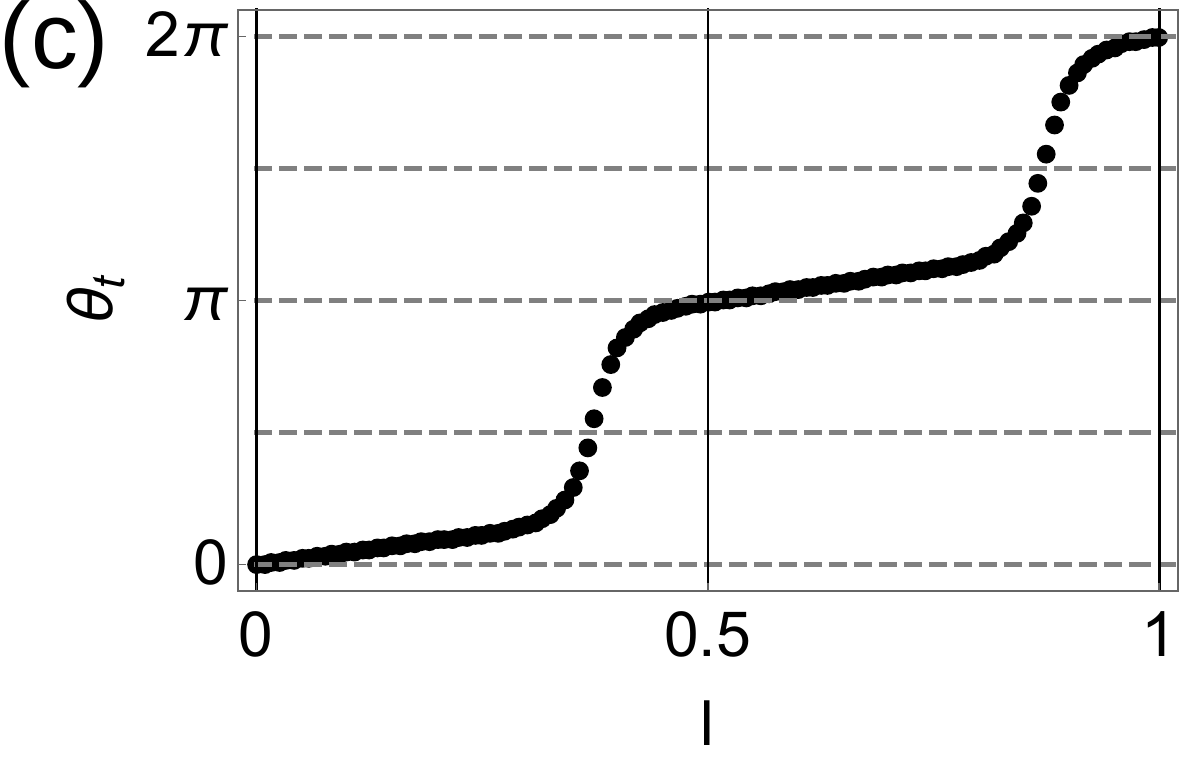}
\quad
\includegraphics[width=0.475\linewidth]{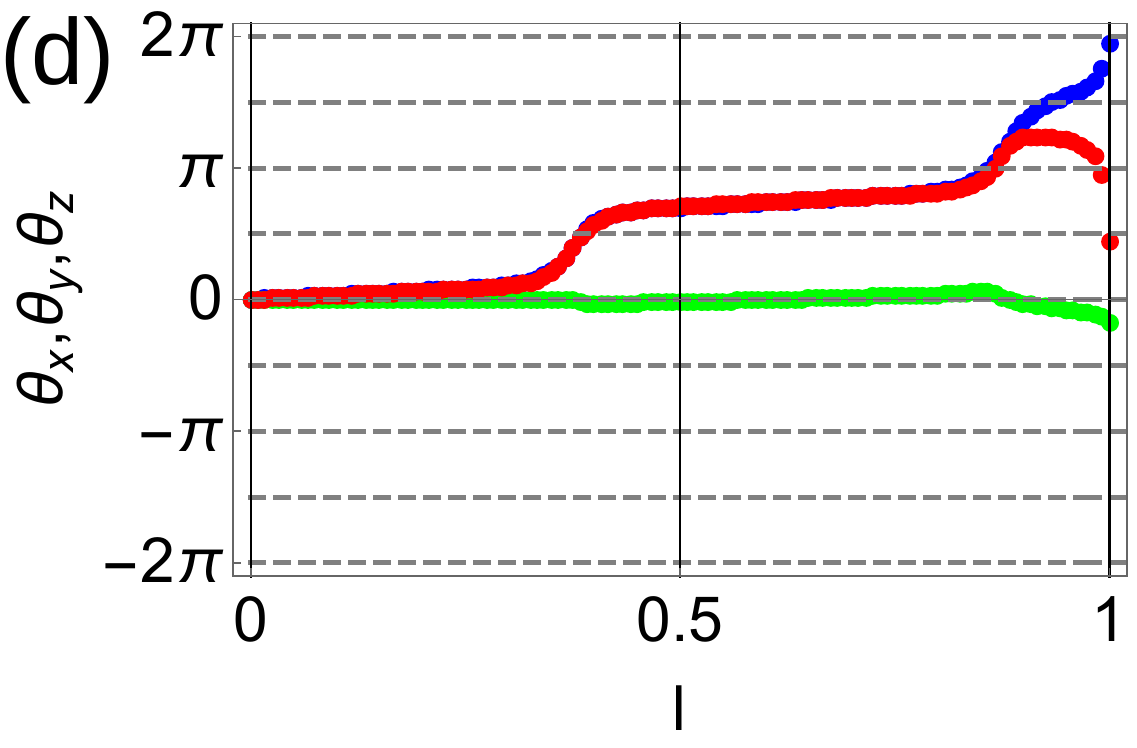}
\caption{\textcolor{black}{Accumulated total frame-rotation angle $\theta_t$ (black) of the three-band partial frame, along with its $\mathsf{so(3)}$ Lie algebra angle components $\{\theta_x,\theta_y,\theta_z\}$ (red, green and blue), calculated by following a loop $l$ encircling the triply degenerate Weyl node at the K point within (a-b) the horizontal $C_{2z}T$-invariant plane and (c-d) within the vertical $C_{2x}T$-invariant plane.}}
\label{fig_angle}
\end{figure}
\textcolor{black}{We show in Fig.\,\ref{fig_angle} the computed accumulated total frame rotation-angle of the partial frame, $\theta_t$, and its three $\mathsf{so}(3)$ Lie algebra angle components, $\{\theta_x,\theta_y,\theta_z\}$, along a loop encircling the triply degenerate Weyl node at the K point within the horizontal $C_{2z}T$-invariant plane in Figs.\,(a-b) and within the vertical $C_{2x}T$-invariant plane in Figs.\,(c-d). The total frame-rotation angle is obtained from the Lie algebra components through $\theta_t = \sqrt{\theta_x^2+\theta_y^2+\theta_z^2}$. We remind that a winding by $2\pi$ gives the quaternion charge $-1$ \cite{Wu1273}, which is compatible (minimally) with two nodes in the same energy gap and with the same non-Abelian frame charge (\textit{i.e.} $\pm i$, $\pm j$, or $\pm k$, giving the resultant quaternion charge $-1=i^2=j^2=k^2$), which, if there would be a way to split the triply degenerate Weyl node, would lead to a two-band node with the patch Euler class $\vert\xi \vert = 1$ \cite{bouhon2019nonabelian,jiang2021observation}. Then, similarly to the patch Euler class, each nontrivial quaternion charge implies that the triply degenerate Weyl node is pinned within the corresponding $C_2 T$-invariant plane. More precisely, the four nontrivial quaternion charges (with respect to the horizontal and the three vertical $C_2 T$-invariant planes) imply that, minimally, a double Weyl node is pinned at the K point, with a quadratic dispersion along each $C_2 T$-invariant plane, as long as the symmetries of the system are preserved, (in contrast to a simple Weyl node at the K point with a linear dispersion). Given the robust ``accidental'' triple degeneracy at the K point, the multiple nontrivial quaternion charges actually indicate the pinning of the triply degenerate Weyl node at the K point, since it is the only point intersecting all the $C_2 T$-invariant planes.}

\textcolor{black}{We highlight that this is the first reported Weyl node with nontrivial quaternion charges. In Sec.\,\ref{sec_strain}, we will discuss the manifestation of the above nontrivial quaternion charges, \textit{i.e. }robustness of the triply degenerate Weyl node upon the breaking of the hexagonal rotation symmetries while two $C_2T$ symmetries are preserved.} \textcolor{black}{The nontrivial non-Abelian topological charges imply that the capability of band nodes to pairwise annihilate depends on the trajectory used to bring them together in reciprocal space. Hence, not only do the non-Abelian charges hold storable information, but the information also is directly readable through the observation of the stability of nodes.}

\section{Bulk-boundary correspondence}\label{sec_BBC}

From the topological configuration of the bulk band structure, represented schematically in Fig.\,\ref{fig_FS}, where we show all the Weyl nodes realized at the filling number $\nu=2$ with their chiralities, we can predict the topology of the surface spectrum for different surface orientations according to the bulk-surface correspondence. 

To make these predictions, first, it is important to locate the Fermi surface sheets, schematically represented in Fig.\,\ref{fig_FS}. Any two regions of the BZ with distinct filling numbers (\textit{i.e.} numbers of occupied energy bands) must be separated by a number of Fermi surface sheets that match the filling number difference. For example, the A point ($\nu_{\text{A}} = 4$) and the $\Gamma$ point ($\nu_{\Gamma}=2$) are separated by  $2=\nu_{\text{A}}-\nu_{\Gamma}$ Fermi surface sheets. Similarly, there are $4=\nu_{\text{A}}-\nu_{\text{H}}=\nu_{\text{A}}-\nu_{\text{L}}$ Fermi surface sheets between the points A and H, and between the points A and L. There are $2=\nu_{\text{K}}-\nu_{\text{H}}=\nu_{\text{L}}-\nu_{\text{M}}$ Fermi surface sheets between the points K and H, and between the points M and L. The resulting Fermi surface is constituted of two sheets centered at the A point wrapping all the Weyl nodes with $\nu=2$ on the $\Delta$-line, and two sheets centered on the $\overline{\text{HL}}$-line wrapping the remaining Weyl nodes with $\nu=2$ away from the $k_z=0$ plane. There remain the Weyl nodes at the K points forming two disconnected point-like components of the Fermi surface. Second, it is worth noting that energy gaps of regions of the BZ wrapped by the Fermi surface may not be visible once projected on reduced surface BZs. Here and in Sec.\,\ref{sec_surface}, we consider the $[110]$ and $[001]$ surfaces \footnote{Defined here in terms of their Miller indices, \textit{i.e.} through the components of a normal vector in units of the primitive vector of the Bravais lattice.}. 

\begin{figure*}
\begin{tabular}{ccc}
\includegraphics[width=0.3\linewidth]{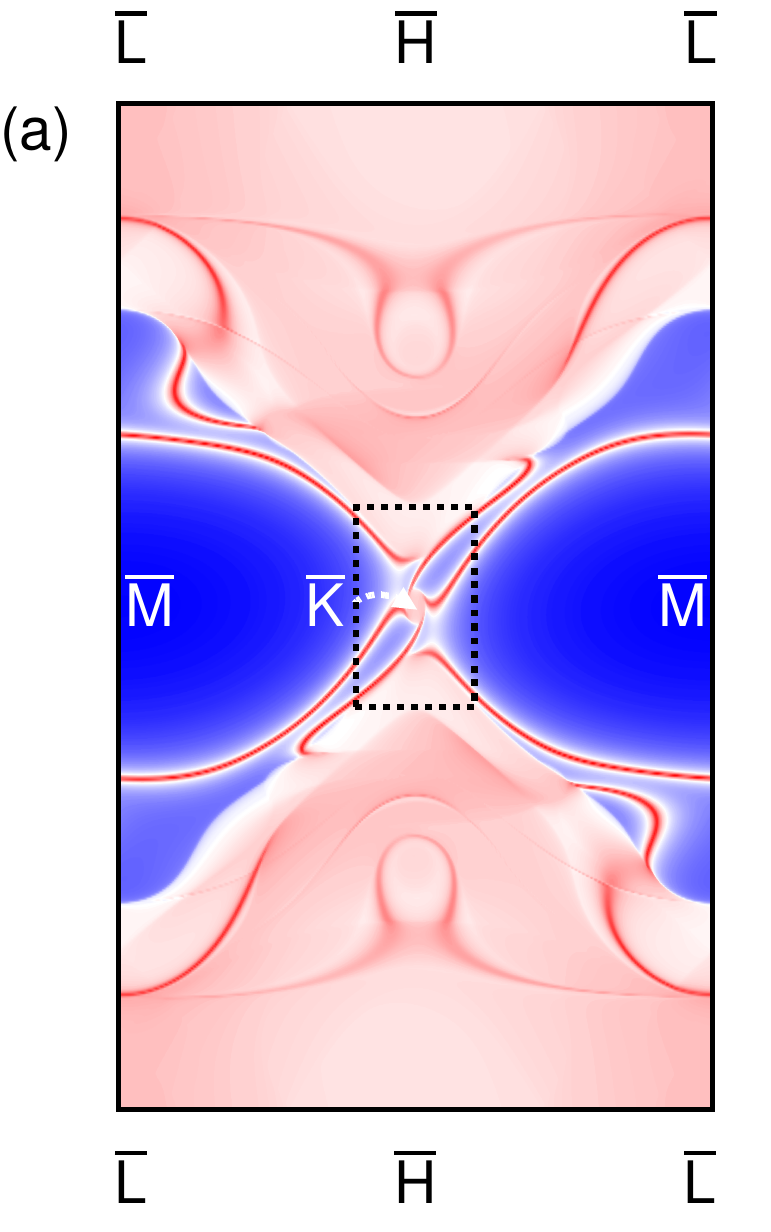} 
\includegraphics[width=0.3\linewidth]{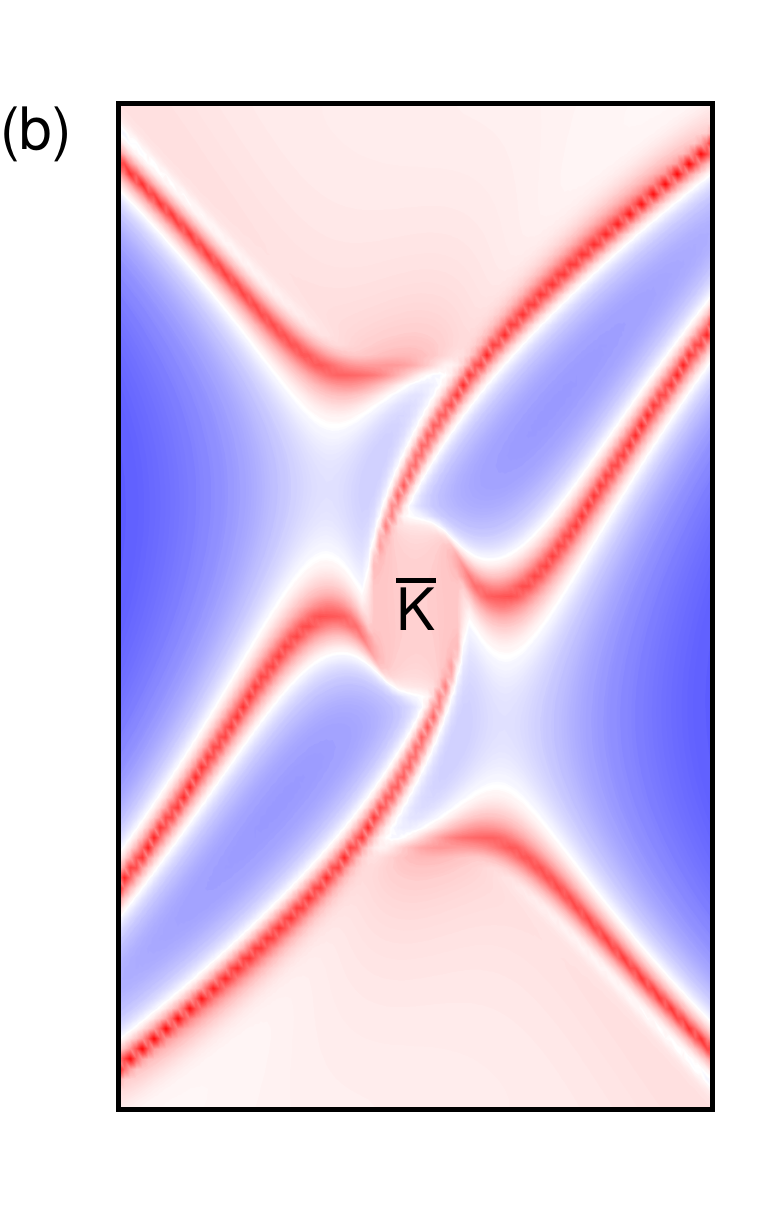} 
\includegraphics[width=0.3\linewidth]{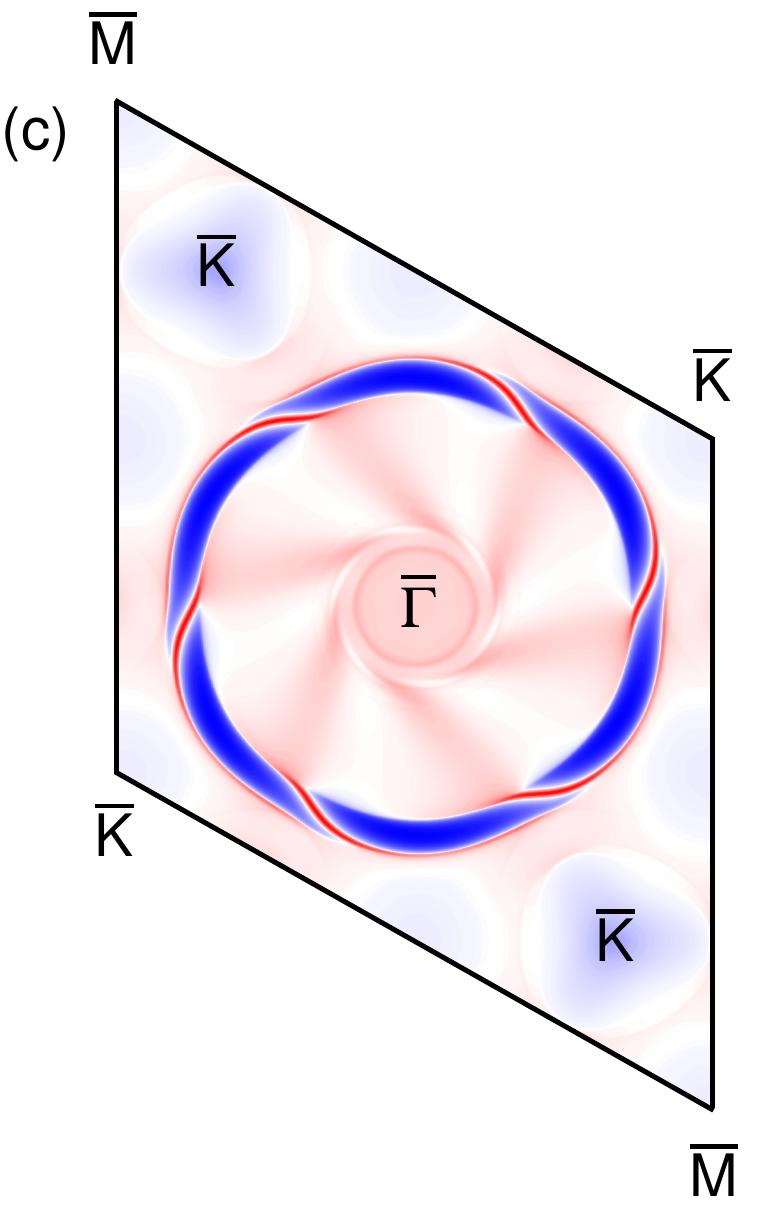}
\end{tabular}
\caption{$\boldsymbol{k}$-dependent local density of states at the Fermi level on the (a) [110] (the detail near the $\overline{\text{K}}$ point indicated by the dashed box is zoomed in (b)) and (c) [001] surfaces, where red (blue) color represents the high (low) density of states.}
\label{surface}
\end{figure*}

The $[110]$ surface corresponds to a projection of the bulk spectrum along one of the three $\overline{\text{KM}}$-lines, such that there is only a single connected region of the surface BZ with an energy gap, spanning a window of finite $\vert k_z\vert <1$. It has the filling number $\nu=2$. Since the two inequivalent K points are projected to a single point of the surface BZ, the latter acts as the source of the chirality of $+4$. Correspondingly, the total chirality of all the Weyl nodes with $\nu=2$ hidden by the Fermi surface is $-4$. We thus predict four Fermi arcs connecting the projected K points to the projected Fermi surface. This agrees with the numerical result given in Fig.\,\ref{surface}(a-b). 

The $[001]$ surface corresponds to a projection of the bulk spectrum along the $\Delta$-line, such that the points $\Gamma$ and A are projected onto the same point $\overline{\Gamma}$. The projected Fermi surface defines two gapless regions, one disc centered on the $\overline{\Gamma}$ point, and one strip covering the projected $\overline{\text{KM}}$-lines. There is thus a single connected gapped region with the filling number $\nu=2$ defining an annulus centered on the $\overline{\Gamma}$ point. The gapless disc hides the six Weyl nodes with $\nu=2$ on the $\Delta$-line with a total chirality of $-6$, and the gapless trip covering the BZ boundary hides the remaining Weyl nodes with a total chirality of $+6$. We thus predict that the gapped annulus must be crossed by six Fermi arcs. This agrees with the numerical result given in Fig.\,\ref{surface}(c). 

\section{Surface states from first principles}\label{sec_surface}

To validate the prediction of the symmetry-indicated bulk-boundary correspondence made in Sec.\,\ref{sec_BBC}, we numerically simulate the surface states of ReO$_3$ on the [110] and [001] surfaces. The simulation result can simultaneously serve as guidance for future experimental observation, especially for ARPES experiments as the chirality of a Weyl node can be directly measured by counting the number of Fermi arcs emanating from its corresponding surface projections.

On the [110] surface, the bulk high symmetry line $\overline{\text{K} \Gamma \text{K}}$ and the bulk Weyl node pinned on the line, is projected onto the center of the surface BZ, that is the $\overline{\text{K}}$ point shown in Fig.\,\ref{surface}(a). The surface states emanate from the $\overline{\text{K}}$ point and eventually merge with the surface projection of bulk states above and below the $\overline{\text{K}}$ point. It is worth noting that the surface states exhibit unusually long Fermi arcs, which cross the boundary of the surface BZ and span the entire surface BZ horizontally. This bears some resemblance to the long Fermi arcs of the CoSi materials family with space group $\mathsf{P}2_13$~\cite{tang2017, chang2017, rao2019, li2019, takane2019} but the Fermi arcs found here are even longer because the Fermi arcs do not need to terminate at the corners of the surface BZ. \textcolor{black}{Figure\,\ref{surface}(b) further depicts the details of the surface states near the $\overline{\text{K}}$ point, where one can clearly see that a total of $4$ Fermi arcs emanate from the $\overline{\text{K}}$ point.}

On the [001] surface, the bulk high symmetry line $\overline{\text{A} \Gamma \text{A}}$ is projected onto the center of the surface BZ, that is the $\overline{\Gamma}$ point shown in Fig.\,\ref{surface}(c).  \textcolor{black}{As anticipated in Sec.\,\ref{sec_BBC},  we observe $6$ Fermi arcs emanating from the $\overline{\Gamma}$ point, forming a distinctive sixfold petal shape.} On the other hand, since the bulk Weyl node at the $\text{K}$ point cancels its chirality with other Weyl nodes on the $\text{P}$-line, there is no discernible Fermi arc emanating from the $\overline{\text{K}}$ point in this surface orientation.

\section{Effect of strain on the Weyl nodes}\label{sec_strain}

Finally, we study the effect of strain on the configuration of symmetry-indicated Weyl nodes. \textcolor{black}{This showcases the great advantage of identifying the relevant topologies for the distinct symmetry conditions and determining the value of the corresponding topological invariants. Indeed, the accumulated knowledge of symmetry IRREPs, chiralities, Euler classes, and quaternion charges taken together predicts the fate of the Weyl points' configuration upon breaking the symmetries of the system.}

Specifically, we consider shear stress $\delta$, deforming the hexagonal lattice to an orthorhombic lattice, \textit{i.e.}
\begin{equation}
\begin{bmatrix}
a_x & a_y & 0 \\
b_x & b_y & 0 \\     
0 & 0 & c
\end{bmatrix}
=
\begin{bmatrix}
a/2 & -\sqrt{3}a/2 - \delta & 0 \\
a/2 & \sqrt{3}a/2 + \delta & 0 \\   
0 & 0 & c
\end{bmatrix}\,,
\end{equation}
which breaks the $6_3$-screw symmetry (as well as $C_{3z}$) while preserving $\{C_{2z}\vert \boldsymbol{\tau}\}$ \textcolor{black}{and the horizontal $C_{2y}$ and $\{C_{2x}\vert \boldsymbol{\tau}\}$ rotation symmetries}. This choice not only allows us to observe under stress the splitting and trajectories of the double Weyl nodes on the $\Delta$-line, but also the stability of the triply degenerate Weyl nodes at the K point.


\begin{figure}
\begin{tabular}{ccc}
\includegraphics[width=1.0\linewidth]{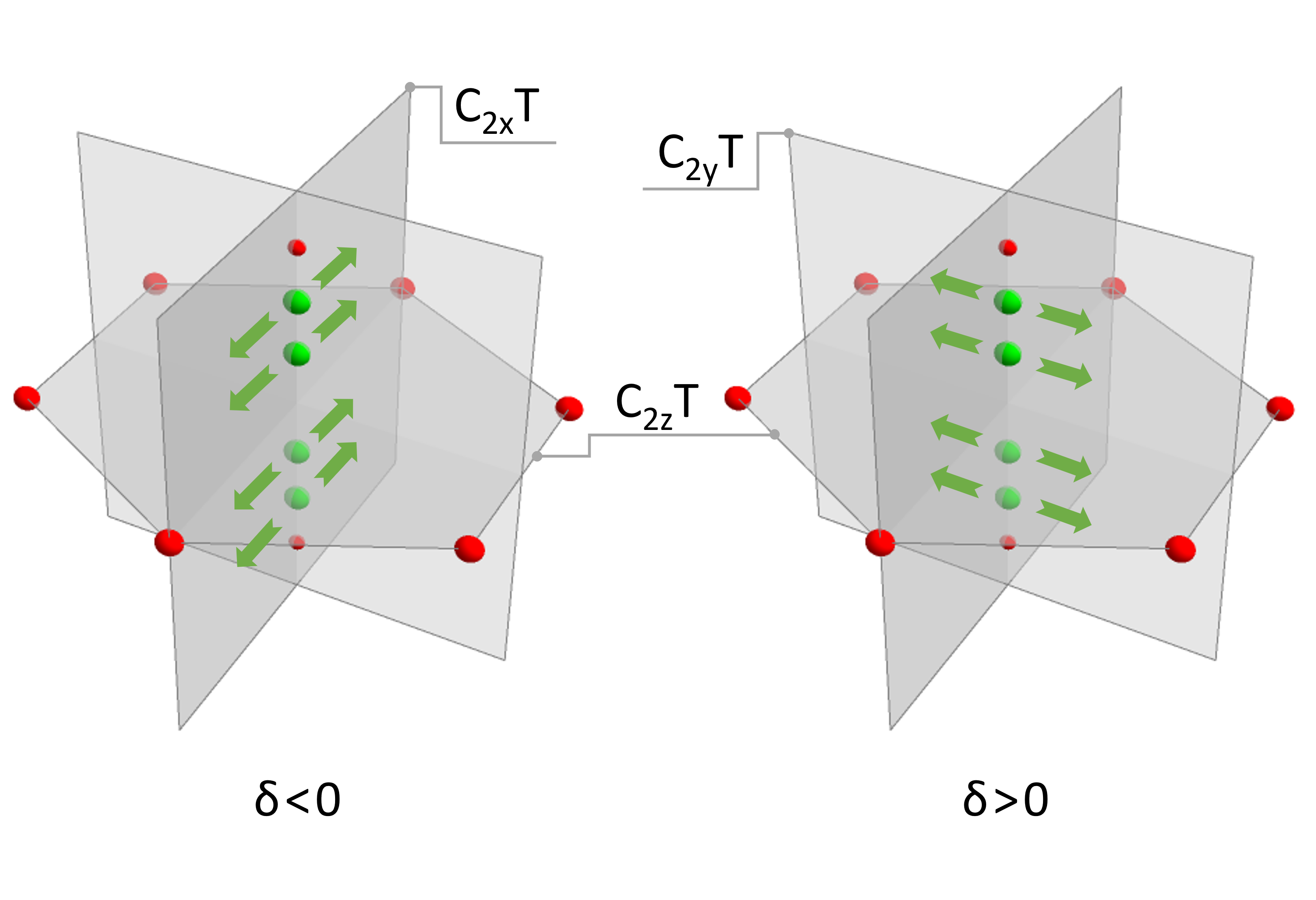} & 
\end{tabular}
\caption{Splitting and preserving of Weyl nodes on the $\Delta$-line and $\text{K}$ point due to positive (left panel) and negative (right panel) shear stress. The shaded planes represent the $C_{2}T$-invariant planes in reciprocal space, the size and color of the ball represent the chirality sign and absolute value of the Weyl node, and the arrows indicate the splitting direction~\cite{S3}.}
\label{weyl_stress}
\end{figure}

The shear stress changes the space group of ReO$_3$ from $\mathsf{P}6_322$ to $\mathsf{C}222_1$, \textcolor{black}{which means that only the $\{C_{2z}\vert\boldsymbol{\tau}\}$ and $\{C_{2x}\vert\boldsymbol{\tau}\}$ screw symmetries and the rotation symmetry $C_{2y}$ of the point group are preserved. Incidentally, the three $C_2T$-invariant planes that are perpendicular to the ${x}$-, ${y}$- and ${z}$-axis respectively are also preserved.} Figure~\ref{weyl_stress} visualizes these three $C_2T$-invariant planes and highlights the Weyl nodes on the high-symmetry $\Delta$-line and $\text{K}$ point with the filling number $\nu=2$.


For the Weyl nodes on the $\Delta$-line, we observe that their IRREPs, chiralities, and Euler classes directly determine their splitting behavior. \textcolor{black}{First, one predicts that the simple Weyl nodes (of chirality $+1$) are stable under the symmetry breaking, since their band branches belong to distinct $\{C_{2z}\vert\boldsymbol{\tau}\}$ symmetry eigenvalues (\textit{i.e.} they belong to different $C_{2z}$-IRREPs) and their patch Euler class are nonzero (\textit{i.e.} each node has $(\vert \xi_{C_{2x}T}\vert,\vert \xi_{C_{2y}T}\vert) = 1/2$). Then, one predicts that the Weyl nodes of chirality $-2$ split into two simple Weyl nodes of chirality $-1$ upon the symmetry breaking. Indeed, ($i$) the branches that form them belong to the same $C_{2z}$-IRREP, such that their crossings are not protected by $C_{2z}$ symmetry alone, then ($ii$) the $-2$ chirality is compatible with a splitting into two simple Weyl nodes that are images of each other under the preserved $C_{2z}$ symmetry, and finally ($iii$) their zero Euler class with respect to both $C_{2x}T$- and $C_{2y}T$-invariant planes implying that each pair of nodes are free to leave both planes.} 

It is worth highlighting that the shear stress preserves the vertical $\{C_{2x}\vert\boldsymbol{\tau}\}$ and $C_{2y}T$ symmetries, in addition to $\{C_{2z}\vert\boldsymbol{\tau}\}$ symmetry. This implies that the $C_2T$-protected patch Euler classes $(\vert \xi_{C_{2x}T}\vert,\vert\xi_{C_{2y}T}\vert)$ (defined in Sec.\,\ref{sec_Euler}) are preserved through the deformation of the system. Through the first-principle calculations, we show in Fig.\,\ref{weyl_stress} that the double Weyl nodes split alternatively within the $C_{2x}T$-invariant plane for $\delta>0$, and within the $C_{2y}T$-invariant plane for $\delta<0$. Since each patch Euler class indicates the topological pinning of Weyl nodes within the corresponding $C_2T$-invariant plane, the observation of the nodes split within both $C_2T$-invariant planes implies that both patch Euler classes vanish. \textcolor{black}{We conclude that the numerical findings in Fig.\,\ref{weyl_stress} precisely agree with the topological prediction.}

Interestingly, we also observe the robustness of the triple degeneracy at the $\text{K}$ point in a range of shear stress. \textcolor{black}{It is important first to note that from the symmetry data alone (\textit{i.e.} the IRREPs at the node), the triply degenerate Weyl node is only accidental even before breaking the symmetries, while the chirality of $-2$ permits its splitting into two Weyl points images of one another either under $\{C_{2x}\vert\boldsymbol{\tau}\}T$, if they remain on the horizontal $C_{2z}T$-invariant plane, or under $\{C_{2z}\vert\boldsymbol{\tau}\}T$, if they remain on the vertical $C_{2x}T$-invariant plane (and, in both cases, being each others image under $C_{2y}$). Hence, the two preserved independent non-Abelian frame charges $(q_{{C_{2z}}T},q_{{C_{2x}}T})$ of the triply degenerate Weyl node, both equal to the nontrivial quaternion charge $-1$, are crucial to predict the persistence of the nodes on the line at the intersection of the $C_{2x}T$- and $C_{2z}T$-invariant planes, independently of the way the hexagonal symmetries are broken. It turns out that the triply degenerate Weyl node holding the double quaternion is even more robust than being only confined to the line where the two $C_2T$-invariant planes intersect.}

\section{Conclusions}
\label{sec_conclusion}
To summarize, we identify ReO$_3$ as an ideal Weyl semimetal, hosting a series of high-chirality Weyl nodes close to the Fermi level. These findings rest on a detailed symmetry analysis of the spinful IRREPs structure as well as matching first-principles calculation results. We find that the present hexagonal screw symmetry plays a crucial role in the assignment of chiralities to the symmetry-indicated Weyl nodes, while the $C_2T$ symmetries and the associated patch Euler classes and \textcolor{black}{non-Abelian frame charges} complement the topological landscape of the electronic structure of the system. Apart from these local charge investigations, we complete the full characterization by accounting for all Weyl nodes and demanding overall charge neutrality, thereby giving a full account of all Weyl nodes relevant at the Fermi level, including those indicated by symmetry arguments as well as accidental nodes that are not constrained by rotation eigenvalues. \textcolor{black}{Notably, we identify the first high-chirality Weyl node sitting exactly at the Fermi level holding multiple nontrivial quaternion charges of $-1$.} We corroborate these bulk analyses with a surface perspective that provides very clear Fermi arc signatures of both types of high-chirality Weyl nodes at the Fermi level, providing an accessible route towards observing this physics in ARPES experiments. \textcolor{black}{Additionally, these high chirality Weyl nodes are expected to amplify unique phenomena observed in conventional Weyl semimetals. For instance, the stronger Berry curvature fields associated with high-chirality nodes can enhance effects such as anomalous Hall conductivity, orbital magnetization, and chiral anomaly-induced transport phenomena such as negative magnetoresistance.} \textcolor{black}{Finally, we have studied the effect of breaking the hexagonal rotation symmetries, while preserving the $C_2T$ symmetries, showcasing the predictive power of the complementary knowledge of the ``complex'' topologies, the symmetry indicated chiralities, together with the ``real'' topologies, Euler classes and quaternions.}

\begin{acknowledgments}
We thank Wojciech J. Jankowski for valuable insights and fruitful discussions. S.C. and B.M. acknowledge funding from EPSRC [EP/V062654/1]. S.C. also acknowledges financial support from the Cambridge Trust and from the Winton Programme for the Physics of Sustainability. R.-J.S. acknowledges funding from an EPSRC New Investigator Award (EP/W00187X/1), from the Winton Programme for the Physics of Sustainability, and from Trinity College Cambridge. B.M. also acknowledges support from a UKRI Future Leaders Fellowship (MR/V023926/1), from the Gianna Angelopoulos Programme for Science, Technology, and Innovation, and from the Winton Programme for the Physics of Sustainability. A.B. was partially funded by a Marie-Curie fellowship, grant no. 101025315 and acknowledges financial support from the Swedish Research Council (Vetenskapsrådet) grant no. 2021-04681. The calculations in this work have been performed using resources provided by the Cambridge Tier-2 system (operated by the University of Cambridge Research Computing Service and funded by EPSRC [EP/P020259/1]), as well as by the UK Materials and Molecular Modelling Hub (partially funded by EPSRC [EP/P020194]), Thomas, and by the UK National Supercomputing Service, ARCHER. Access to Thomas and ARCHER was obtained \textit{via} the UKCP consortium and funded by EPSRC [EP/P022561/1]).  
\end{acknowledgments}

\bibliography{references}

\appendix

\begin{widetext}
\section{Symmetry representation in the basis of the Bloch eigenstates}

We here briefly review the basic relations underlying the unitary representation of a nonsymmorphic symmetry in terms of the electronic Bloch eigenstates. While the group theory results do not depend on the method of computation of the Bloch eigenstates, we assume that these have the typical structure of Kohn-Sham orbitals obtained from first-principles methods and that an expansion in atomic-like Wannier functions is known, \textit{e.g.} through maximally localized wannierization~\cite{wannier90}. The discussion here is general and we illustrate it for the special case of the screw symmetry $6_3$ of the space group $\mathsf{P}6_322$ (see also Sec.\,\ref{Sec: theory} for more details on the space group). 

In the following, we consider a 3D crystal system characterized by the primitive Bravais vectors $\{\boldsymbol{a}_1,\boldsymbol{a}_2,\boldsymbol{a}_3\}$ and the dual primitive reciprocal lattice vectors,
\begin{equation}
    \begin{aligned}
        \boldsymbol{b}_1 &= 2\pi \dfrac{\boldsymbol{a}_2\times \boldsymbol{a}_3}{\boldsymbol{a}_1\cdot \boldsymbol{a}_2\times \boldsymbol{a}_3} \,,\\
        \boldsymbol{b}_2 &= 2\pi \dfrac{\boldsymbol{a}_3\times \boldsymbol{a}_1}{\boldsymbol{a}_1\cdot \boldsymbol{a}_2\times \boldsymbol{a}_3} \,,\\
        \boldsymbol{b}_3 &= 2\pi \dfrac{\boldsymbol{a}_1\times \boldsymbol{a}_2}{\boldsymbol{a}_1\cdot \boldsymbol{a}_2\times \boldsymbol{a}_3} \,.
    \end{aligned}
\end{equation}
We then define the first BZ
\begin{equation}
    \text{BZ}=\left\{
        \boldsymbol{k} = k_1 \boldsymbol{b}_1 + k_2 \boldsymbol{b}_2 + k_3 \boldsymbol{b}_3 \;\big\vert \;
        (k_1,k_2,k_3) \in [0,1]^3
    \right\} \,.
\end{equation}
We write generic vectors of the Bravais lattice and of the reciprocal lattice as
\begin{equation}
    \begin{aligned}
        \boldsymbol{R} & = m_1 \boldsymbol{a}_1 +m_2 \boldsymbol{a}_2 +m_3 \boldsymbol{a}_3  \,,\\
        \boldsymbol{G} & = m_1 \boldsymbol{b}_1 + m_2 \boldsymbol{b}_2 + m_3 \boldsymbol{b}_3  \,.
    \end{aligned}
\end{equation}

\subsection{Bloch eigenfunctions, Bloch eigenstates, and Bloch eigenvectors and their associated symmetry representations}\label{ap_wannier_basis}

We start here assuming we are given the spin-dependent Kohn-Sham orbitals that are solutions of the Kohn-Sham equation, \textit{i.e.} a Schr{\"o}dinger differential operator including periodic potentials and spin-orbit coupling $\mathfrak{h}(\boldsymbol{x})$. Due to the discrete translation symmetries of the Kohn-Sham Hamiltonian, the Kohn-Sham orbitals are Bloch {\it eigenfunctions}
\begin{equation}
\left\{
\begin{aligned}
    \mathfrak{H}(\boldsymbol{x}) \,\psi_{n\boldsymbol{k}}(\boldsymbol{x}) &= E_{n}(\boldsymbol{k})\,\psi_{n\boldsymbol{k}}(\boldsymbol{x})\,,\\
    \widetilde{\mathfrak{H}}_{\boldsymbol{k}}(\boldsymbol{x}) \,u_{n\boldsymbol{k}}(\boldsymbol{x}) &= E_{n}(\boldsymbol{k})\,u_{n\boldsymbol{k}}(\boldsymbol{x}) \,,\\
    \widetilde{\mathfrak{H}}_{\boldsymbol{k}}(\boldsymbol{x})&= \mathrm{e}^{-\text{i}\boldsymbol{k}\cdot \boldsymbol{x}} \mathfrak{H}(\boldsymbol{x})\mathrm{e}^{\text{i}\boldsymbol{k}\cdot \boldsymbol{x}}\,,
\end{aligned}\right.
\end{equation}
where $u_{n\boldsymbol{k}}(\boldsymbol{x}) = \mathrm{e}^{-\text{i}\boldsymbol{k}\cdot \boldsymbol{x}} \psi_{n,\boldsymbol{k}}(\boldsymbol{x})$ is the cell-periodic part of the Bloch eigenfunctions, \textit{i.e.} $u_{n\boldsymbol{k}}(\boldsymbol{x}+\boldsymbol{R})=u_{n\boldsymbol{k}}(\boldsymbol{x})$ for any Bravais lattice vector $\boldsymbol{R}$. 

The Kohn-Sham orbitals are typically written in a basis of plane waves and bare $1/2$-spinors, \textit{i.e.}
\begin{equation}
\label{eq_KS_eigenfunctions}
\begin{aligned}
    \psi_{n\boldsymbol{k}}(\boldsymbol{x}) & = \sum\limits_{\boldsymbol{G},\sigma} 
 \mathrm{e}^{\text{i} (\boldsymbol{k}+\boldsymbol{G})\cdot \boldsymbol{x}} \,\chi_{\sigma}\,c_{n\boldsymbol{k}}(\boldsymbol{G},\sigma) \,,\\
 & =
 \sum\limits_{\boldsymbol{G}} 
 \mathrm{e}^{\text{i} (\boldsymbol{k}+\boldsymbol{G})\cdot \boldsymbol{x}} \left(\chi_{\uparrow}\,c_{n\boldsymbol{k}}(\boldsymbol{G},\uparrow) + 
 \chi_{\downarrow}\,
 c_{n\boldsymbol{k}}(\boldsymbol{G},\downarrow) \right)\,,
\end{aligned}
\end{equation}
where $\boldsymbol{G}$ is a reciprocal lattice vector and the bare $1/2$-spinors are defined with respect to the chosen quantization axis $z$ such that $\sigma_z(\chi_{\uparrow},\chi_{\downarrow})  = (\chi_{\uparrow},-\chi_{\downarrow})$.

Let us now assume that we know the decomposition of the Bloch Kohn-Sham orbitals in the basis of Bloch-L{\"owdin} states and in the basis of atomic-like Wannier states, \textit{i.e.}
\begin{equation}
\label{eq_Bloch_wannier}
\left\{
\begin{aligned}
    \vert \psi_{n},\boldsymbol{k} \rangle &=  
    \sum\limits_{\alpha,\sigma} \vert \varphi_{\alpha\sigma},\boldsymbol{k}\rangle \; \left[\mathcal{U}_{n}(\boldsymbol{k})\right]_{\alpha\sigma} \,,\\
    \vert \varphi_{\alpha\sigma}, \boldsymbol{k} \rangle &=   \dfrac{1}{\sqrt{N_{\alpha\sigma}}}
    \sum\limits_{\boldsymbol{R}} \mathrm{e}^{\text{i}\boldsymbol{k}\cdot (\boldsymbol{R}+\boldsymbol{r}_{\alpha\sigma})} \vert w_{\alpha\sigma} , \boldsymbol{R} + \boldsymbol{r}_{\alpha\sigma}  \rangle    \,,
\end{aligned}\right.
\end{equation}
where $\boldsymbol{R}$ locates the centers of the unit cells of the Bravais lattice and $\boldsymbol{r}_{\alpha\sigma}$ locates the orbital $\alpha$ with spin $\sigma$ within the unit cell (\textit{i.e.} it identifies its sublattice degree of freedom), and $N_{\alpha\sigma}$ is the number of such orbitalo-spinor degrees of freedom in the whole system. (Formally, we start from a finite lattice with periodic boundary conditions and then take the limit to infinity, that is the Born-von Karman boundary conditions.) We note that the Bloch-L{\"o}wdin basis $\vert \varphi_{\alpha\sigma}, \boldsymbol{k} \rangle$ is defined in the Zak-gauge, that is including the phase factor $\mathrm{e}^{\text{i}\boldsymbol{k}\cdot \boldsymbol{r}_{\alpha\sigma}}$, which is the appropriate gauge to compute the Zak phases (on top of using the periodic gauge for the Bloch {\it eigenvectors}, see below). We recover the Bloch eigenfunctions simply through the position representation
\begin{equation}
     \langle \boldsymbol{x} \vert \psi_{n}, \boldsymbol{k} \rangle = \psi_{n\boldsymbol{k}} (\boldsymbol{x})   = \sum\limits_{\alpha,\sigma,\boldsymbol{R}} \dfrac{\mathrm{e}^{\text{i}\boldsymbol{k}\cdot (\boldsymbol{R}+\boldsymbol{r}_{\alpha\sigma})}}{\sqrt{N}_{\alpha\sigma}} \left[\mathcal{U}_n(\boldsymbol{k})\right]_{\alpha\sigma} 
       w_{\alpha\sigma} (\boldsymbol{x}- \boldsymbol{R} - \boldsymbol{r}_{\alpha\sigma} )  \,,
\end{equation}
with the Wannier function $\langle \boldsymbol{x}\vert w_{\alpha\sigma} , \boldsymbol{R} + \boldsymbol{r}_{\alpha\sigma}  \rangle = w_{\alpha\sigma} (\boldsymbol{x}- \boldsymbol{R} - \boldsymbol{r}_{\alpha\sigma} )  $. We now seek the equation for which the coefficients $[\mathcal{U}_{n}(\boldsymbol{k})]_{\alpha\sigma}$ are the eigen-solutions. For this, we first evaluate the Kohn-Sham differential operator in the basis of Wannier states, obtained through the projection on the Wannier basis
\begin{equation}
\left\{
\begin{aligned}
    t_{\alpha\sigma,\beta\sigma'}(\boldsymbol{R}_i-\boldsymbol{R}_j) &= \langle w_{\alpha\sigma} ,\boldsymbol{R}_i + \boldsymbol{r}_{\alpha\sigma}
    \vert \hat{\mathfrak{H}}\vert  w_{\beta\sigma'} ,\boldsymbol{R}_j + \boldsymbol{r}_{\beta\sigma'} \rangle \,,\\
    &= \int d^3\boldsymbol{x} \; \langle w_{\alpha\sigma} ,\boldsymbol{R}_i + \boldsymbol{r}_{\alpha\sigma}
    \vert \boldsymbol{x}\rangle \;
    \mathfrak{H}(\boldsymbol{x})\;
    \langle \boldsymbol{x} \vert 
    w_{\beta\sigma'} ,\boldsymbol{R}_j + \boldsymbol{r}_{\beta\sigma'} \rangle
    \,,\\
    &= \int d^3\boldsymbol{x} \; w_{\alpha\sigma}^* (\boldsymbol{x}-\boldsymbol{R}_i - \boldsymbol{r}_{\alpha\sigma}
    ) \;
    \mathfrak{H}(\boldsymbol{x})\;
    w_{\beta\sigma'} (\boldsymbol{x}-\boldsymbol{R}_j - \boldsymbol{r}_{\beta\sigma'} )
    \,,\\
    \langle \boldsymbol{x}'\vert \hat{\mathfrak{H}} \vert \boldsymbol{x} \rangle &= \mathfrak{H}(\boldsymbol{x}) \delta(\boldsymbol{x}'-\boldsymbol{x}) \,,
\end{aligned}\right.
\end{equation}
as
\begin{equation}
    \hat{\mathfrak{H}} =\sum\limits_{ij,\alpha\beta,\sigma\sigma'} \vert w_{\alpha\sigma} ,\boldsymbol{R}_i + \boldsymbol{r}_{\alpha\sigma} \rangle  \; t_{\alpha\sigma,\beta\sigma'}(\boldsymbol{R}_i-\boldsymbol{R}_j)\; \langle w_{\beta\sigma'} ,\boldsymbol{R}_j + \boldsymbol{r}_{\beta\sigma'} \vert \,.
\end{equation}
We emphasize that this form is explicitly invariant under translation by a Bravais lattice vector. Indeed, given the action of a translation by a Bravais vector on the Wannier function, which is formally defined as a {\it function-space operator} (here taken in the {\it active} convention)~\cite{BradCrack}
\begin{equation}
\begin{aligned}
    ^{\{E\vert \widetilde{\boldsymbol{R}}\}}\langle \boldsymbol{x}\vert w_{\alpha\sigma}, \boldsymbol{R} +\boldsymbol{r}_{\alpha\sigma} \rangle &= \langle \boldsymbol{x}-\widetilde{\boldsymbol{R}}\vert w_{\alpha\sigma}, \boldsymbol{R} +\boldsymbol{r}_{\alpha\sigma} \rangle\\
    &= w_{\alpha\sigma}(\boldsymbol{x} - \widetilde{\boldsymbol{R}} - \boldsymbol{R} -\boldsymbol{r}_{\alpha\sigma})\,,\\
    &= \langle\boldsymbol{x} \vert w_{\alpha\sigma},\boldsymbol{R}+\boldsymbol{r}_{\alpha\sigma} + \widetilde{\boldsymbol{R}} \rangle  \,,
\end{aligned}
\end{equation}
or, when transferred as an action on the Wannier states,
\begin{equation}
    ^{\{E\vert \widetilde{\boldsymbol{R}}\}}\vert w_{\alpha\sigma}, \boldsymbol{R} +\boldsymbol{r}_{\alpha\sigma} \rangle =\vert w_{\alpha\sigma},\boldsymbol{R}+\boldsymbol{r}_{\alpha\sigma} + \widetilde{\boldsymbol{R}} \rangle\,,
\end{equation}
where $E$ is the identity element of the crystallographic point group, we have
\begin{equation}
    \begin{aligned}
        ^{\{E\vert \widetilde{\boldsymbol{R}}\}}\hat{\mathfrak{H}} &= \sum\limits_{ij,\alpha\beta,\sigma\sigma'} \vert w_{\alpha\sigma} ,\boldsymbol{R}_i + \boldsymbol{r}_{\alpha\sigma} + \widetilde{\boldsymbol{R}} \rangle  \; t_{\alpha\sigma,\beta\sigma'}(\boldsymbol{R}_i-\boldsymbol{R}_j)\; \langle w_{\beta\sigma'} ,\boldsymbol{R}_j + \boldsymbol{r}_{\beta\sigma'} + \widetilde{\boldsymbol{R}}\vert \,,\\
        &= \sum\limits_{i'j',\alpha\beta,\sigma\sigma'} \vert w_{\alpha\sigma} ,\boldsymbol{R}'_i + \boldsymbol{r}_{\alpha\sigma}  \rangle  \; t_{\alpha\sigma,\beta\sigma'}(\boldsymbol{R}'_i- \widetilde{\boldsymbol{R}}-\boldsymbol{R}'_j+\widetilde{\boldsymbol{R}})\; \langle w_{\beta\sigma'} ,\boldsymbol{R}'_j + \boldsymbol{r}_{\beta\sigma'} \vert \,,\\
        &= \hat{\mathfrak{H}} \,.
    \end{aligned}
\end{equation}
Then, converting the Wannier functions into the Bloch-orbital basis via the inverse Fourier transform
\begin{equation}
    \vert w_{\alpha\sigma}, \boldsymbol{R}+\boldsymbol{r}_{\alpha\sigma} \rangle = \dfrac{1}{\sqrt{N_{\alpha\sigma}}} \sum\limits_{\boldsymbol{k}} \mathrm{e}^{-\text{i} \boldsymbol{k}\cdot (\boldsymbol{R}+\boldsymbol{r}_{\alpha\sigma})} \vert \varphi_{\alpha\sigma},\boldsymbol{k} \rangle\,,
\end{equation}
we rewrite the Kohn-Sham operator in the Bloch-L{\"o}wdin basis
\begin{equation}
\label{eq_bloch_Hamiltonian_matrix}
\left\{
\begin{aligned}
    \hat{\mathfrak{H}} 
    &= \sum\limits_{\boldsymbol{k},\alpha,\sigma,\sigma'} \vert \varphi_{\alpha\sigma} ,\boldsymbol{k} \rangle  \; H_{\alpha\sigma,\beta\sigma'} (\boldsymbol{k}) \; \langle \varphi_{\beta\sigma'} ,\boldsymbol{k}\vert \,,\\
    H_{\alpha\sigma,\beta\sigma'} (\boldsymbol{k}) &= \langle \varphi_{\alpha\sigma} ,\boldsymbol{k} \vert \hat{\mathfrak{H}} \vert \varphi_{\beta\sigma'} ,\boldsymbol{k} \rangle\,,\\
    &= 
    \sum\limits_{j} \mathrm{e}^{\text{i} \boldsymbol{k}\cdot (\boldsymbol{R}_j+\boldsymbol{r}_{\beta\sigma'}-\boldsymbol{0}-\boldsymbol{r}_{\alpha\sigma})}\; t_{\alpha\sigma,\beta\sigma'}(\boldsymbol{0}-\boldsymbol{R}_j) \,,\\
    &\equiv\left[H(\boldsymbol{k})\right]_{\alpha\sigma,\beta\sigma'}\,,
\end{aligned}\right.
\end{equation}
where the elements $H_{\alpha\sigma,\beta\sigma'} (\boldsymbol{k})$ define the matrix Bloch Hamiltonian $H(\boldsymbol{k})\in \mathbb{C}^N \times \mathbb{C}^N$, with $N$ the total number of microscopic degrees of freedom, within a single unit cell (\textit{i.e.} sublattice sites, electronic orbitals, and spins), and represented by the atomic-like Wannier functions. 

Finally, writing the eigenvalue equation for the Bloch {\it eigenstates}
\begin{equation}
\label{eq_eigenvalue_equation}
    \hat{\mathfrak{H}} \vert \psi_{n},\boldsymbol{k}\rangle = \vert \psi_{n},\boldsymbol{k}\rangle\;E_n(\boldsymbol{k})\,,
\end{equation}
we find that the coefficients $\left\{[\mathcal{U}_n(\boldsymbol{k})]_{\alpha\sigma}\right\}_{\alpha\sigma}$ in Eq.\,(\ref{eq_Bloch_wannier}) are solutions of the algebraic equation
\begin{equation}
    H(\boldsymbol{k}) \;\mathcal{U}_n(\boldsymbol{k}) = \mathcal{U}_n(\boldsymbol{k})\;E_n(\boldsymbol{k})\,.
\end{equation}
We call $\mathcal{U}_{n}(\boldsymbol{k})$ the Bloch {\it eigenvectors} of the matrix Bloch Hamiltonian. Writing the equations for all $n=1,2,\dots,N$ in a matrix form, we have
\begin{equation}
\label{eq_eigenvalue_matrix}
\left\{    
\begin{aligned}
    H(\boldsymbol{k})
    \cdot \mathcal{U}(\boldsymbol{k}) &= \mathcal{U}(\boldsymbol{k})\cdot \mathcal{E}(\boldsymbol{k})\,,\\
    \mathcal{E}(\boldsymbol{k}) &= \text{diag}(E_1(\boldsymbol{k}),\dots,E_N(\boldsymbol{k})) \,,
\end{aligned}
\right.
\end{equation}
where we chose to label the energy eigenvalues from the bottom, \textit{i.e.} $E_{n}(\boldsymbol{k}) \leq E_{n+1}(\boldsymbol{k})$ for all $0<n<N-1$.

\subsubsection{Periodic gauge}

The Kohn-Sham Bloch operator must be invariant under reciprocal lattice translations, \textit{i.e.}
\begin{equation}
\begin{aligned}
    ^{\{E\vert \boldsymbol{G}\}}\hat{\mathfrak{H}} 
    &= \sum\limits_{\boldsymbol{k}}\vert \boldsymbol{\varphi},\boldsymbol{k}+\boldsymbol{G}\rangle \cdot H(\boldsymbol{k}+\boldsymbol{G}) \cdot \langle\boldsymbol{\varphi},\boldsymbol{k}+\boldsymbol{G}\vert \,,\\
    &= \sum\limits_{\boldsymbol{k}} \vert \boldsymbol{\varphi},\boldsymbol{k}\rangle \cdot V(\boldsymbol{G})\cdot H(\boldsymbol{k}+\boldsymbol{G}) \cdot V^{\dagger}(\boldsymbol{G})\cdot \langle\boldsymbol{\varphi},\boldsymbol{k}\vert \,,\\
    &\stackrel{!}{=} \hat{\mathfrak{H}}\,,
\end{aligned}
\end{equation}
with the matrix representing the reciprocal translation 
\begin{equation}
    V(\boldsymbol{G}) = 
    \text{diag}\left(
        \mathrm{e}^{\text{i} \boldsymbol{G}\cdot \boldsymbol{r}_{A\uparrow}},\mathrm{e}^{\text{i} \boldsymbol{G}\cdot \boldsymbol{r}_{A\downarrow}},
        \mathrm{e}^{\text{i} \boldsymbol{G}\cdot \boldsymbol{r}_{B\uparrow}},\mathrm{e}^{\text{i} \boldsymbol{G}\cdot \boldsymbol{r}_{B\downarrow}},\dots
    \right)\,,
\end{equation}
leading to the symmetry constraint on the matrix Bloch Hamiltonian 
\begin{equation}
   H(\boldsymbol{k}+\boldsymbol{G}) = V^{\dagger}(\boldsymbol{G})\cdot H(\boldsymbol{k}) \cdot V(\boldsymbol{G})\,.
\end{equation}
This implies that the energy eigenvalues are the same at $\boldsymbol{k}$ and at the shifted quasi-momentum. Indeed, by the invariant of eigenvalues under similitude transformations, we have
\begin{equation}
\begin{aligned}
    \mathcal{E}(\boldsymbol{k}) &= \mathcal{U}^{\dagger}(\boldsymbol{k})\cdot H(\boldsymbol{k})\cdot \mathcal{U}(\boldsymbol{k}) \\
    &= \left[\mathcal{U}^{\dagger}(\boldsymbol{k}) \cdot 
    V(\boldsymbol{G})\right] \cdot H(\boldsymbol{k}+\boldsymbol{G})\cdot
    \left[V^{\dagger}(\boldsymbol{G})\cdot \mathcal{U}(\boldsymbol{k})\right] \\
    &= \mathcal{U}^{\dagger}(\boldsymbol{k}+\boldsymbol{G}) \cdot H(\boldsymbol{k}+\boldsymbol{G})\cdot
    \mathcal{U}(\boldsymbol{k}+\boldsymbol{G}) \\
    &= \mathcal{E}(\boldsymbol{k}+\boldsymbol{G})\,,
\end{aligned}
\end{equation}
from which we also conclude the identity of the shifted Bloch eigenvectors modulo a gauge phase factor, \textit{i.e.} 
\begin{equation}
    \mathcal{U}_n(\boldsymbol{k}+\boldsymbol{G}) = V^{\dagger}(\boldsymbol{G})\cdot\mathcal{U}_n(\boldsymbol{k})\, \mathrm{e}^{\text{i} \phi_n(\boldsymbol{k}) }        \,.
\end{equation}
We always chose the {\it periodic gauge} by setting $\phi_n(\boldsymbol{k})=0$ for all $n$, which implies the periodicity of the Bloch eigenstates
\begin{equation}
    \vert \psi_n, \boldsymbol{k}+\boldsymbol{G} \rangle =  \vert \psi_n, \boldsymbol{k} \rangle \cdot 
    \left[\mathcal{U}^{\dagger}(\boldsymbol{k}) \cdot V(\boldsymbol{G})\cdot \mathcal{U}_n(\boldsymbol{k}+\boldsymbol{G})\right] =  \vert \psi_n, \boldsymbol{k} \rangle\,.
\end{equation}

\subsubsection{Symmetry representation}

In the following, we use the rotation matrix by the angle $\theta_g$ of the point group symmetry $g$ around the $z$-axis
\begin{equation}
    D_g = \mathrm{e}^{\theta_g L_z} = \left(
        \begin{array}{ccc}
            \cos\theta_g & -\sin\theta_g & 0\\
            \sin\theta_g & \cos\theta_g & 0\\
             0 & 0 & 1
        \end{array}
    \right)\,,
\end{equation}
acting on the three-component quasi-momentum as $D_g\cdot \boldsymbol{k}$, and the $\mathsf{SU}(2)$ representation of the same rotation in the basis of the bare $1/2$-spinors~\cite{BrinkSatchler}
\begin{equation}
\label{eq_halfspin_rep}
\left\{
\begin{aligned}
    {^{g}}\left(\chi_{\uparrow}~\chi_{\downarrow}\right) &= \left(\chi_{\uparrow}~\chi_{\downarrow}\right)\cdot \mathcal{D}^{1/2}(g)\,,\\
    \mathcal{D}^{1/2}(g) & = \mathrm{e}^{-\text{i} \theta_{g} \frac{\sigma_z}{2}}  \,.
\end{aligned}\right.
\end{equation}
These can be readily generalized for all types of point group symmetries.

We find the action of the symmetry on the Wannier on the position operator (again formally given by a {\it function-space operator}, still in the {\it active} convention~\cite{BradCrack}) reads~\cite{hourglass}
\begin{equation}
\begin{aligned}
    ^{\{g\vert \boldsymbol{\tau}\}}\langle \boldsymbol{x}\vert w_{\alpha\sigma}, \boldsymbol{R} +\boldsymbol{r}_{\alpha\sigma} \rangle &= \langle D_g^{-1}(\boldsymbol{x}-\boldsymbol{\tau})\vert w_{\alpha\sigma}, \boldsymbol{R} +\boldsymbol{r}_{\alpha\sigma} \rangle\\
    &=  w_{\alpha\sigma}(D_g^{-1}(\boldsymbol{x}-\boldsymbol{\tau}) - \boldsymbol{R} -\boldsymbol{r}_{\alpha\sigma} ) \\
    &= w_{\alpha\sigma}\left(D_g^{-1}[\boldsymbol{x} - (D_g\boldsymbol{R} +D_g\boldsymbol{r}_{\alpha\sigma} + \boldsymbol{\tau}] \right) \\
    &= \sum\limits_{\beta,\sigma'}w_{\beta\sigma'}(\boldsymbol{x} - (D_g\boldsymbol{R} +D_g\boldsymbol{r}_{\alpha\sigma} +\boldsymbol{\tau}) ) \, \hat{U}_{\beta\sigma',\alpha\sigma}(g) \\
    &=  \sum\limits_{\beta\sigma'} w_{\beta\sigma'}(\boldsymbol{x} - \widetilde{\boldsymbol{R}} -\boldsymbol{r}_{\beta\sigma'} ) \, \hat{U}_{\beta\sigma',\alpha\sigma}(g) \\
    &= \sum\limits_{\beta\sigma'} \langle \boldsymbol{x}\vert w_{\beta\sigma'}, \widetilde{\boldsymbol{R}} +\boldsymbol{r}_{\beta\sigma'} \rangle
     \, \hat{U}_{\beta\sigma',\alpha\sigma}(g) \,,
\end{aligned}
\end{equation}
where we have set
\begin{equation}
\label{eq_bravais_rel}
    \widetilde{\boldsymbol{R}} +\boldsymbol{r}_{\beta\sigma'} = D_g\boldsymbol{R} +D_g\boldsymbol{r}_{\alpha\sigma} +\boldsymbol{\tau} \,,
\end{equation}
that points to a sub-lattice site $\boldsymbol{r}_{\beta\sigma'}$ in the unit cell $\widetilde{\boldsymbol{R}}$ occupied by the orbital $\beta$ and the spin $\sigma'$, and where the unitary matrix $\hat{U}(g)$ represents the symmetry action on the microscopic fermionic degrees of freedom combining sub-lattice sites, electronic orbitals, and spins. Concretely, the Wannier functions are obtained as linear combinations of the basis functions of the IRREPs associated to the microscopic degrees of freedom (thus mixing bosonic and fermionic IRREPs), with product coefficients (the Clebsch-Gordan coefficients) that are obtained through the reduction of a product of IRREPs in a direct sum of IRREPs~\cite{Koster}. See Appendix \ref{sec_reps_6} for a simple concrete example. 

Transferring the above symmetry action on Wannier {\it wave-functions} to a symmetry action on the Wannier {\it states}, we have
\begin{equation}
    ^{\{g\vert \boldsymbol{\tau}\}}\vert w_{\alpha\sigma}, \boldsymbol{R} +\boldsymbol{r}_{\alpha\sigma} \rangle = \sum\limits_{\beta\sigma'}\vert w_{\beta\sigma'},  \widetilde{\boldsymbol{R}} +\boldsymbol{r}_{\beta\sigma'}\rangle \, \hat{U}_{\beta\sigma',\alpha\sigma}(g) \,,
\end{equation}
which implies the following representation of $\{g\vert\boldsymbol{\tau}\}$ in the Bloch-L{\"o}wdin basis 
\begin{equation}
\begin{aligned}
    ^{\{g\vert\boldsymbol{\tau}\}}\vert \varphi_{\alpha\sigma}, \boldsymbol{k} \rangle &= \sum\limits_{\boldsymbol{R}} \mathrm{e}^{\text{i} \boldsymbol{k}\cdot (\boldsymbol{R}+\boldsymbol{r}_{\alpha\sigma})} \; {^{\{g\vert \boldsymbol{\tau}\}}}\vert w_{\alpha\sigma}, \boldsymbol{R} +\boldsymbol{r}_{\alpha\sigma} \rangle \,,\\
    &= \sum\limits_{\widetilde{\boldsymbol{R}},\beta,\sigma'} \mathrm{e}^{\text{i} D_g\boldsymbol{k}\cdot (\widetilde{\boldsymbol{R}} +\boldsymbol{r}_{\beta\sigma'}-\boldsymbol{\tau})} \; \vert w_{\beta\sigma'},  \widetilde{\boldsymbol{R}} +\boldsymbol{r}_{\beta\sigma'}\rangle \, \hat{U}_{\beta\sigma',\alpha\sigma}(g) \,,\\
    &=
    \sum\limits_{\beta,\sigma'}\vert \varphi_{\beta\sigma'}, D_g\boldsymbol{k} \rangle \;\hat{U}_{\beta\sigma',\alpha\sigma}(g)  
    \; \mathrm{e}^{-\text{i} \boldsymbol{k}\cdot D_g^{-1}\boldsymbol{\tau}} \,,
\end{aligned}
\end{equation}
where we have used $\boldsymbol{R} +\boldsymbol{r}_{\alpha\sigma}  = D_g^{-1}\widetilde{\boldsymbol{R}} +D_g^{-1}\boldsymbol{r}_{\beta\sigma'}-D_g^{-1}\boldsymbol{\tau}$ [Eq.\,(\ref{eq_bravais_rel})]. We then get the representation in the Bloch eigenstates
\begin{equation}
\label{eq_sym_rep_BE}
    \begin{aligned}
    \mathcal{S}^{\boldsymbol{k}}_{mn}(\{g\vert\boldsymbol{\tau}\}) &\equiv \langle \psi_{m}, D_g\boldsymbol{k} \vert^{\{g\vert\boldsymbol{\tau}\}}\vert  \psi_{n}, \boldsymbol{k} \rangle \,,\\
    & = 
    \sum\limits_{\alpha\beta,\sigma\sigma'} \left[\mathcal{U}^{\dagger}_{m}(D_g\boldsymbol{k})\right]_{\beta\sigma'}
    \hat{U}_{\beta\sigma',\alpha\sigma}(g)  
    \; 
    \left[\mathcal{U}_{n}(\boldsymbol{k})\right]_{\alpha\sigma}\;
    \mathrm{e}^{-\text{i} \boldsymbol{k}\cdot D_g^{-1}\boldsymbol{\tau}} \,.
\end{aligned}
\end{equation}

In the case the above representation is evaluated at a {\it high-symmetry point} $\boldsymbol{k}^*$ of the Brillouin zone for which $g$ is an element of the associated little co-group $\overline{G}^{\boldsymbol{k}^*}$ (see \cite{BradCrack}), then the following equation holds
\begin{equation}
      D_g\boldsymbol{k}^*  = \boldsymbol{k}^* + \boldsymbol{G}_{g,\boldsymbol{k}^*}     \,,
\end{equation}
with $\boldsymbol{G}_{g,\boldsymbol{k}^*}$ a vector of the reciprocal Bravais lattice. It is then crucial to evaluate the representation by using the periodic gauge (see above), that is
\begin{equation}
\label{eq_sym_rep_BE_general}
    \begin{aligned}
    \mathcal{S}^{\boldsymbol{k}^*}_{mn}(\{g\vert\boldsymbol{\tau}\}) 
    & = 
    \sum\limits_{\alpha\beta,\sigma\sigma'} \left[\mathcal{U}^{\dagger}_{m}(\boldsymbol{k}^*+\boldsymbol{G}_{g,\boldsymbol{k}^*})\right]_{\beta\sigma'}
    \hat{U}_{\beta\sigma',\alpha\sigma}(g)  
    \; 
    \left[\mathcal{U}_{n}(\boldsymbol{k}^*)\right]_{\alpha\sigma}\;
    \mathrm{e}^{-\text{i} \boldsymbol{k}^*\cdot D_g^{-1}\boldsymbol{\tau}} \\
    &= \sum\limits_{\alpha\beta\gamma,\sigma\sigma'\sigma''} \left[\mathcal{U}^{\dagger}_{m}(\boldsymbol{k}^*)\right]_{\gamma\sigma''}
     V_{\gamma\sigma'',\beta\sigma'}(\boldsymbol{G}_{g,\boldsymbol{k}^*}) \hat{U}_{\beta\sigma',\alpha\sigma}(g)  
    \; 
    \left[\mathcal{U}_{n}(\boldsymbol{k}^*)\right]_{\alpha\sigma}\;
    \mathrm{e}^{-\text{i} \boldsymbol{k}^*\cdot D_g^{-1}\boldsymbol{\tau}}\\
    &= \left[\mathcal{U}(\boldsymbol{k}^*)^{\dagger}\cdot V(\boldsymbol{G}_{g,\boldsymbol{k}^*})\cdot
    \hat{U}(g) \cdot \mathcal{U}(\boldsymbol{k}^*)
    \right]_{mn} \, \mathrm{e}^{-\text{i} \boldsymbol{k}^*\cdot D_g^{-1}\boldsymbol{\tau}}\,,
\end{aligned}
\end{equation}
which removes the relative gauge indeterminacy between the Bloch eigenvectors at $\boldsymbol{k}^*$ and $D_g\boldsymbol{k}^*$. It is this gauge-invariant quantity that defines the IRREP of $\overline{G}^{\boldsymbol{k}^*}$ associated with the $\{m,n\}$-energy levels (bands) at the high-symmetry point $\boldsymbol{k}^*$. 

The invariance of the Hamiltonian operator, namely 
\begin{equation}
\begin{aligned}
    ^{\{g\vert\boldsymbol{\tau}\}}\hat{\mathfrak{H}}   &=  \sum\limits_{\boldsymbol{k}} \vert \boldsymbol{\varphi} ,D_g\boldsymbol{k} \rangle \cdot  \hat{U}(g) \cdot H (\boldsymbol{k}) \cdot \hat{U}^{\dagger}(g) \cdot \langle \boldsymbol{\varphi} ,D_g\boldsymbol{k}\vert \,,\\
    &=  \sum\limits_{D_g^{-1}\boldsymbol{k}} \vert \boldsymbol{\varphi} ,\boldsymbol{k} \rangle \cdot  \hat{U}(g) \cdot H (D_g^{-1}\boldsymbol{k}) \cdot \hat{U}^{\dagger}(g) \cdot \langle \boldsymbol{\varphi} ,\boldsymbol{k}\vert \,,\\
    &\stackrel{!}{=} \hat{\mathfrak{H}} \,,
\end{aligned}
\end{equation}
implies the following symmetry condition on the matrix Bloch Hamiltonian
\begin{equation}
    \hat{U}(g) \cdot H (\boldsymbol{k}) \cdot \hat{U}^{\dagger}(g) = H(D_g\boldsymbol{k})\,.
\end{equation}

Combining the eigenvalue equation Eq.\,(\ref{eq_eigenvalue_equation}) with the symmetry-invariance of $\hat{\mathfrak{H}}$, we find
\begin{equation}
    \begin{aligned}
        \langle \psi_{\widetilde{m}},D_g\boldsymbol{k}\vert\left(
        ^{\{g\vert\boldsymbol{\tau}\}} \hat{\mathfrak{H}} \;{^{\{g\vert\boldsymbol{\tau}\}}}\vert \psi_n,\boldsymbol{k}\rangle\right) &= 
        \langle \psi_{\widetilde{m}},D_g\boldsymbol{k}\vert
        \left({^{\{g\vert\boldsymbol{\tau}\}}}
        \vert \psi_n,\boldsymbol{k}\rangle \, E_n(\boldsymbol{k})\right) \,,\\
        \Leftrightarrow\quad\quad
        \langle \psi_{\widetilde{m}},D_g\boldsymbol{k}\vert\left(\sum\limits_m
         \hat{\mathfrak{H}} \;\vert \psi_m,D_g\boldsymbol{k}\rangle \, \mathcal{S}^{\boldsymbol{k}}_{mn}(\{g\vert \boldsymbol{\tau}\})\right) &= 
         \mathcal{S}^{\boldsymbol{k}}_{{\widetilde{m}}n}(\{g\vert \boldsymbol{\tau}\}) \, E_n(\boldsymbol{k}) \,,\\
         \Leftrightarrow\quad\quad
         E_{\widetilde{m}}(D_g\boldsymbol{k})\, \mathcal{S}^{\boldsymbol{k}}_{\widetilde{m}n}(\{g\vert \boldsymbol{\tau}\}) &= 
         \mathcal{S}^{\boldsymbol{k}}_{\widetilde{m}n}(\{g\vert \boldsymbol{\tau}\}) \, E_n(\boldsymbol{k}) \,,
    \end{aligned}
\end{equation}
implying  
\begin{equation}
     \left\{E_{\widetilde{m}}(D_g\boldsymbol{k})\right\}_{\widetilde{m}} = \left\{E_n(\boldsymbol{k})\right\}_{n}\,,
\end{equation}
such that the spectrum at $D_g\boldsymbol{k}$ is identical to the spectrum at $\boldsymbol{k}$.

Let us rewrite the above relation in a matrix form
\begin{equation}
    \left.S^{\boldsymbol{k}}(\{g\vert\boldsymbol{\tau}\})\right.^{-1} \cdot \mathcal{E}(D_g \boldsymbol{k}) \cdot \mathcal{S}^{\boldsymbol{k}}(\{g\vert\boldsymbol{\tau}\}) = \mathcal{E}(\boldsymbol{k})\,, 
\end{equation}
where $\mathcal{E}(\boldsymbol{k})$ is the diagonal matrix of ordered energy eigenvalues [Eq.\,(\ref{eq_eigenvalue_matrix})]. In case we choose a symmetry-invariant quasi-momentum, \textit{i.e.} $D_g\boldsymbol{k}^*=\boldsymbol{k}^*$, and assuming there is no degeneracy, \textit{i.e.} $E_m(\boldsymbol{k}^*)\neq E_n(\boldsymbol{k}^*)$ whenever $m\neq n$, we get
\begin{equation}
    \left[\mathcal{E}( \boldsymbol{k}^*),\mathcal{S}^{\boldsymbol{k}^*}(\{g\vert\boldsymbol{\tau}\})\right] = 0\,,
\end{equation}
which implies that $\mathcal{S}^{\boldsymbol{k}^*}(\{g\vert\boldsymbol{\tau}\})$ is diagonal. Therefore, 
\begin{equation}
\label{eq_eigenvalue_irrep}
\left\{
\begin{aligned}
 ^{\{g\vert\boldsymbol{\tau}\}}\vert  \psi_{n}, \boldsymbol{k}^* \rangle &= \vert \psi_{n}, \boldsymbol{k}^* \rangle \;\chi_{\boldsymbol{k}^*}^{\Gamma_{j_n}}(\{g\vert \boldsymbol{\tau}\}) \,,\\
 \chi_{\boldsymbol{k}^*}^{\Gamma_{j_n}}(\{g\vert \boldsymbol{\tau}\}) &\equiv \mathcal{S}^{\boldsymbol{k}^*}_{nn}(\{g\vert\boldsymbol{\tau}\})\,,
\end{aligned}\right.
\end{equation}
where $\chi_{\boldsymbol{k}^*}^{\Gamma_{j_n}}(\{g\vert \boldsymbol{\tau}\})$ is the character of the spinful IRREP $\Gamma_{j_n}$ to which the $n$-th Bloch eigenstate belongs (see below). 

\subsubsection{Feature of the symmetry representation in the Wannier-basis at a symmetry invariant quasi-momentum}\label{ap_wannierbasis_feature}

Let us rewrite explicitly the symmetry representation in the basis of Bloch eigenstates at a symmetry invariant quasi-momentum ($D_g\boldsymbol{k}^*=\boldsymbol{k}^*$),
\begin{equation}
\mathcal{S}^{\boldsymbol{k}^*}(\{g\vert\boldsymbol{\tau}\}) = \langle \boldsymbol{\psi}, \boldsymbol{k}^* \vert^{\{g\vert\boldsymbol{\tau}\}}\vert  \boldsymbol{\psi}, \boldsymbol{k}^* \rangle =  \mathcal{U}(\boldsymbol{k}^*)^{\dagger}\cdot \hat{U}(g) \cdot \mathcal{U}(\boldsymbol{k}^*) \, \mathrm{e}^{-\text{i} \boldsymbol{k}\cdot D_g^{-1}\boldsymbol{\tau}}\,. 
\end{equation}
It has the important feature to be diagonal in the basis of Bloch eigenvectors, \textit{i.e.} it only involves one Bloch eigenvector. As we show in Sec.\,\ref{sec_reps_6}, this leads to a direct relationship between the symmetry eigenvalues in the basis of Bloch eigenstates (equivalently, the characters of the spinful IRREPs) and the symmetry representation in the (microscopic) Wannier basis.

\subsubsection{Symmetry representation in the basis of cell-periodic Bloch eigenstates}\label{ap_cellper}

We now introduce the cell-periodic Bloch eigenstates
\begin{equation}
\begin{aligned}
    \vert u_{n}, \boldsymbol{k} \rangle &=  \mathrm{e}^{-\text{i}\boldsymbol{k}\cdot \boldsymbol{x}}
    \sum\limits_{\alpha,\sigma}\vert \varphi_{\alpha\sigma}, \boldsymbol{k} \rangle \; \left[\mathcal{U}_{n}(\boldsymbol{k})\right]_{\alpha\sigma} \,,\\
    &=\dfrac{1}{\sqrt{N_{\alpha\sigma}}}\sum\limits_{\boldsymbol{R},\alpha,\sigma}
    \mathrm{e}^{-\text{i}\boldsymbol{k}\cdot (\boldsymbol{x}-\boldsymbol{R}-\boldsymbol{r}_{\alpha\sigma})}
    \vert w_{\alpha\sigma}, \boldsymbol{R}+\boldsymbol{r}_{\alpha\sigma} \rangle \,  \left[\mathcal{U}_{n}(\boldsymbol{k})\right]_{\alpha\sigma}  \,,
\end{aligned}
\end{equation}
for which the symmetry acts as 
\begin{equation}
\begin{aligned}
    ^{\{g\vert\boldsymbol{\tau}\}}\vert u_n, \boldsymbol{k} \rangle &= 
    \mathrm{e}^{-\text{i}\boldsymbol{k}\cdot (D_g^{-1}\boldsymbol{x}-D_g^{-1}\boldsymbol{\tau})}
    \sum\limits_{\alpha,\sigma}
    {^{\{g\vert\boldsymbol{\tau}\}}} 
    \vert \varphi_{\alpha\sigma}, \boldsymbol{k} \rangle \; \left[\mathcal{U}_{n}(\boldsymbol{k})\right]_{\alpha\sigma} \,,\\
    &= \mathrm{e}^{-\text{i}D_g\boldsymbol{k}\cdot (\boldsymbol{x}-\boldsymbol{\tau})}
    \sum\limits_{\alpha\beta,\sigma\sigma'}
    \vert \varphi_{\beta\sigma'}, D_g\boldsymbol{k} \rangle \;\hat{U}_{\beta\sigma',\alpha\sigma}(g) \left[\mathcal{U}_{n}(\boldsymbol{k})\right]_{\alpha\sigma} \, \mathrm{e}^{-\text{i}D_g\boldsymbol{k}\cdot \boldsymbol{\tau}} \,,\\
    &= \mathrm{e}^{-\text{i}D_g\boldsymbol{k}\cdot \boldsymbol{x}}
    \sum\limits_{m,\alpha\beta,\sigma\sigma'}
    \vert \psi_m, D_g\boldsymbol{k} \rangle \;\left[\mathcal{U}_{m}(\boldsymbol{k})\right]^*_{\beta\sigma'}\hat{U}_{\beta\sigma',\alpha\sigma}(g) \left[\mathcal{U}_{n}(\boldsymbol{k})\right]_{\alpha\sigma}\,,\\
    &=     \sum\limits_{m,\alpha\beta,\sigma\sigma'}
    \vert u_m, D_g\boldsymbol{k} \rangle \;\left[\mathcal{U}^{\dagger}_{m}(\boldsymbol{k})\right]_{\beta\sigma'}\hat{U}_{\beta\sigma',\alpha\sigma}(g) \left[\mathcal{U}_{n}(\boldsymbol{k})\right]_{\alpha\sigma}\,.
\end{aligned}
\end{equation}
We thus arrive to the symmetry representation in the cell-periodic Bloch eigenstates
\begin{equation}
\label{eq_symm_rep_ap_1}
\begin{aligned}
    \mathcal{R}^{\boldsymbol{k}}_{mn}(\{g\vert\boldsymbol{\tau}\}) &\equiv \langle u_{m}, D_g\boldsymbol{k} \vert^{\{g\vert\boldsymbol{\tau}\}}\vert  u_{n}, \boldsymbol{k} \rangle \,,\\
    &= \sum\limits_{\alpha\beta,\sigma\sigma'}
    \left[\mathcal{U}^{\dagger}_{m}(D_g\boldsymbol{k})\right]_{\beta\sigma'}
    \hat{U}_{\beta\sigma',\alpha\sigma}(g)  
    \; 
    \left[\mathcal{U}_{n}(\boldsymbol{k})\right]_{\alpha\sigma} = \mathcal{S}^{\boldsymbol{k}}_{mn}(\{g\vert\boldsymbol{\tau}\}) \; \mathrm{e}^{\text{i} \boldsymbol{k}\cdot D_g^{-1}\boldsymbol{\tau}}  \,.
\end{aligned}
\end{equation}
We emphasize the absence of $\boldsymbol{k}$-dependent phase factor as compared to the symmetry representation in the Bloch eigenstates Eq.\,(\ref{eq_sym_rep_BE}). Crucially, the Wilson loop and other gauge invariant topological quantities are defined from the cell-periodic Bloch eigenstates. It is thus the latter symmetry representation that appears in the symmetry reduction of the Wilson loop of Sec.\,\ref{sec_theory_B}.  

It is convenient to rewrite Eq.\,(\ref{eq_symm_rep_ap_1}) as an equation that relates the Bloch eigenvectors of the symmetry transformed momentum $D_g\boldsymbol{k}$ to the Bloch eigenvectors at $\boldsymbol{k}$, namely in a matrix form
\begin{equation}
    \hat{U}(g) \cdot \mathcal{U}(\boldsymbol{k})  
    = \mathcal{U}(D_g\boldsymbol{k}) \cdot \mathcal{R}^{\boldsymbol{k}}(\{g\vert \boldsymbol{\tau}\})
    \,.
\end{equation}

Choosing an invariant quasi-momentum $\boldsymbol{k}^*$ (\textit{i.e.} $D_g\boldsymbol{k}^*=\boldsymbol{k}^*$), we find, similarly to Eq.\,(\ref{eq_eigenvalue_irrep}), that the representation matrix becomes diagonal, \textit{i.e.}
\begin{equation}
\left\{
\begin{aligned}
 ^{\{g\vert\boldsymbol{\tau}\}}\vert  u_{n}, \boldsymbol{k}^* \rangle &= \vert u_{n}, \boldsymbol{k}^* \rangle \;\zeta_{\boldsymbol{k}^*}^{\Gamma_{j_n}}(\{g\vert \boldsymbol{\tau}\}) \,,\\
 \zeta_{\boldsymbol{k}^*}^{\Gamma_{j_n}}(\{g\vert \boldsymbol{\tau}\}) &\equiv \mathcal{R}^{\boldsymbol{k}^*}_{nn}(\{g\vert\boldsymbol{\tau}\}) = \chi_{\boldsymbol{k}^*}^{\Gamma_{j_n}}(\{g\vert \boldsymbol{\tau}\}) \mathrm{e}^{\text{i} \boldsymbol{k}^*\cdot D_g^{-1}\boldsymbol{\tau}}\,,
\end{aligned}\right.
\end{equation}
with $\mathcal{R}^{\boldsymbol{k}^*}_{mn}(\{g\vert\boldsymbol{\tau}\})=0$ for $m\neq n$.

In the main text, we express the chirality of a Weyl node in terms of the flow of a Berry phase, such that the base loop sweeps a gapped surface wrapping the Weyl node. The Berry phase is obtained as the determinant of a Wilson loop. The Wilson loop is most robustly defined as a path-ordered product of band-projectors in the basis of cell-periodic Bloch eigenstates. For a single-band subspace, that is
\begin{equation}
\left\{
\begin{aligned}
    \mathcal{W}^{(n)}_l &= \langle u_n,\boldsymbol{k}_0 \vert \prod\limits_{\boldsymbol{k}}^{\boldsymbol{k}_0\leftarrow \boldsymbol{k}_0}\mathcal{P}(\boldsymbol{k}) \vert  u_n,\boldsymbol{k}_0 \rangle\,,\\
    \mathcal{P}(\boldsymbol{k})&= \vert u_n,\boldsymbol{k}\rangle \langle u_n,\boldsymbol{k} \vert \,,
\end{aligned}
\right.
\end{equation}
where $\boldsymbol{k}_0$ is the initial and final point of the oriented loop $l$.

\subsection{Spinful irreducible representations of $6_3$-screw symmetry}\label{sec_reps_6}

We here derive the spinful IRREPs of the $6_3$-screw symmetry directly from the symmetry representation in the basis of atomic-like Wannier functions. For this, we only need to assume that the rotation symmetry acts diagonally on the Wannier basis. More precisely, there always exists a choice of linear combinations of Wannier basis functions representing the ``atomic'' degrees of freedom (accounting for the sub-lattice sites, the electronic orbitals, and the spins) which makes the symmetry action diagonal. 

Let us consider a generic quasi-momentum located on a $C_6$-invariant axis while avoiding the high-symmetry points, say the $\overline{\Delta}$-line and avoiding the $\Gamma$ and A points, \textit{i.e.} 
\begin{equation}
    \boldsymbol{k}^{\Delta} = k \,\boldsymbol{b}_3\;,\quad k\in(0,\frac{1}{2})\cup (\frac{1}{2},1)\,,
\end{equation}
for which $D_6\boldsymbol{k}^{\Delta}=\boldsymbol{k}^{\Delta}$. Assuming that the atomic-like degrees of freedom, represented by localized Wannier functions, are diagonal under the screw symmetry, \textit{i.e.}
\begin{equation}
\begin{aligned}
    ^{\{g\vert \boldsymbol{\tau}\}}\vert w_{\alpha\sigma}, \boldsymbol{R} +\boldsymbol{r}_{\alpha\sigma} \rangle &= 
    \sum\limits_{\beta,\sigma'}\vert w_{\beta\sigma'},  \widetilde{\boldsymbol{R}} +\boldsymbol{r}_{\beta\sigma'}\rangle \, \mathrm{e}^{-\text{i} (\theta_{\sigma l_{\alpha}}+k \pi)} \delta_{\beta,\alpha}\delta_{\sigma,\sigma'}\,,\\
    &= \vert w_{\alpha\sigma},  \widetilde{\boldsymbol{R}} +\boldsymbol{r}_{\alpha\sigma}\rangle \, \mathrm{e}^{-\text{i} (\theta_{\sigma l_{\alpha}}+k \pi)}\,,
\end{aligned}
\end{equation}
with
\begin{equation}
\label{eq_irreps_formula_orbi_spin}
    \mathrm{e}^{-\text{i}\theta_{\sigma l_{\alpha}}} =  \mathrm{e}^{-s_{\sigma}\text{i} \frac{\pi}{6}}\chi_6^{l_{\alpha}} = \mathrm{e}^{\text{i} \frac{\pi}{6} (2l_{\alpha}-s_{\sigma})} \;,\quad l_{\alpha}\in\{0,1,2,3,4, 5\}\,,
\end{equation}
where $s_{\uparrow}=+1$ and $s_{\downarrow}=-1$, and $\chi_6^{l_{\alpha}} = \mathrm{e}^{\text{i} \frac{\pi}{3}l_{\alpha}}$. See Eq.\,(\ref{eq_halfspin_rep}) for the component of the phase coming from the action on the bare spin-$1/2$. (See also the beginning of Sec.\,\ref{sec_theory_A}.) 

We then find the symmetry representations in the basis of Bloch-eigenstates
\begin{equation}
\label{eq_similitude}
    \begin{aligned}
        \text{diag}\left(\chi^{\Gamma_{j_1}}_{k}(6_3),
        \dots,
        \chi^{\Gamma_{j_N}}_{k}(6_3)
        \right)&\equiv  \langle \boldsymbol{\psi}, k \vert^{\{C_6\vert\boldsymbol{b}_3/2\}}\vert  \boldsymbol{\psi}, k \rangle \,,\\
        &= \text{diag}\left(\mathcal{S}^{k}_{11}(\{C_6\vert\boldsymbol{b}_3/2\}),\dots, \mathcal{S}^{k}_{NN}(\{C_6\vert\boldsymbol{b}_3/2\}) \right)  \,,\\
    &= \mathcal{U}^{\dagger}(k)\cdot d^k(6_3)
    \cdot\mathcal{U}(k)  \;  \mathrm{e}^{-\text{i}k\pi}\,,
    \end{aligned}
\end{equation}
with the diagonal matrix
\begin{equation}
\begin{aligned}
    d^k(6_3) &= \text{diag}\left(
    \mathrm{e}^{-\text{i} \theta_{\uparrow l_{A}}}, \mathrm{e}^{-\text{i} \theta_{\downarrow l_{A}}},
    \mathrm{e}^{-\text{i} \theta_{\uparrow l_{B}}}, \mathrm{e}^{-\text{i} \theta_{\downarrow l_{B}}},\dots 
        \right) \,,\\
    &= \text{diag}\left(
    \mathrm{e}^{\text{i} \frac{\pi}{6} (2l_{A}-1)}, \mathrm{e}^{\text{i} \frac{\pi}{6} (2l_{A}+1)},
    \mathrm{e}^{\text{i} \frac{\pi}{6} (2l_{B}-1)}, \mathrm{e}^{\text{i} \frac{\pi}{6} (2l_{B}+1)},\dots 
        \right)\,, \;l_A,l_B,\dots \in \{0,1,\dots,5\}\,,  
\end{aligned}
\end{equation}
where the angles [Eq.\,(\ref{eq_irreps_formula_orbi_spin})] are obtained from the symmetry action on the microscopic degrees of freedom corresponding to the orbital $A$ with spin $+1/2$ and orbital $A$ with spin $-1/2$, then orbital $B$ with spin $+1/2$ and orbital $B$ with spin $-1/2$, and so forth. Given the similitude relation Eq.\,(\ref{eq_similitude}) between the symmetry representation in the Wannier basis and the symmetry representation in the Bloch-eigenstate basis that are both diagonal, we have  
\begin{equation}
        \left\{\chi_k^{\Gamma_{j_n}}(6_3)\right\}_{n=1,2,\dots,N} = 
        \left\{
            \mathrm{e}^{\text{i} \frac{\pi}{6} (2l_{\alpha}-1)-\text{i}k\pi}, \mathrm{e}^{\text{i} \frac{\pi}{6} (2l_{\alpha}+1)-\text{i}k\pi}
    \right\}_{\alpha=A,B,\dots} \,,
\end{equation}
\textit{i.e.} the two symmetry representations are equal as sets. We conclude that the character, $\chi_k^{\Gamma_{j_n}}(6_3)$, of any spinful irreducible representation, $\Gamma_{j_n}$, is readily given as a phase factor $\mathrm{e}^{\text{i} \frac{\pi}{6} (2l_{\alpha}-s_{\sigma})}$ for some combination of $s_{\sigma}\in\{\pm1\}$ and $l_{\alpha}\in\{0,1,2,3,4,5\}$. Going through the combinatorial set of all index-pairs $(l_{\alpha},s_{\sigma})$, we find the two-to-one relation
\begin{equation}
     \left\{
        (l_{\alpha},s_{\sigma})
     \right\}_{l_{\alpha}\in \mathbb{Z}_6,s_{\sigma}\in \{+1,-1\}} \mapsto
     \left\{
        \mathrm{e}^{-\text{i}\frac{\pi}{6}},
        \mathrm{e}^{\text{i}\frac{\pi}{6}},
        \mathrm{e}^{-\text{i}\frac{\pi}{2}},
        \mathrm{e}^{\text{i}\frac{\pi}{2}},
        \mathrm{e}^{-\text{i}\frac{5\pi}{6}},
        \mathrm{e}^{\text{i}\frac{5\pi}{6}}
    \right\}\mathrm{e}^{-\text{i}k\pi} \ni \chi_k^{\Gamma_{j_n}}(6_3) \,,
\end{equation}
from which we conclude that there only exists six inequivalent spinful irreducible representations. There is a one-to-one correspondence with the spinful irreducible representations of $C_6$ symmetry of point group $\mathsf{C}_6$ which we list in Table \ref{table_C6_alt_brad} below. We conclude that the symmetry representation in the basis of the microscopic degrees of freedom readily determines the irreducible representations and the symmetry eigenvalues in the basis of the Bloch eigenstates.
{\renewcommand{\arraystretch}{1.6}
\begin{table}[]
    \centering
\begin{tabular}{c|cccccc}
    \hline\hline
      $\mathsf{C}_6$ IRREPs from Ref.\,\cite{altmann2011tables} & $^{1}E_{1/2}$ & 
     $^{2}E_{1/2}$ & 
     $^{1}E_{3/2}$ & 
     $^{2}E_{3/2}$ &
     $^{1}E_{5/2}$ & 
     $^{2}E_{5/2}$  \\
     \hline
     $\mathsf{C}_6$ IRREPs from Ref.\,\cite{BradCrack} & 
     $\Gamma_8$ & 
     $\Gamma_7$ & 
     $\Gamma_{12}$ & 
     $\Gamma_{11}$ &
     $\Gamma_{9}$ & 
     $\Gamma_{10}$  \\
     \hline\hline
     $(l,s_{\sigma})$ & 
     $\begin{array}{c}(0,-1)\\
        (1,+1)
     \end{array}$   & 
     $\begin{array}{c}(0,1)\\
        (5,-1)
     \end{array}$ & 
     $\begin{array}{c}(4,-1)\\
        (5,+1)
     \end{array}$ & 
     $\begin{array}{c}(1,+1)\\
        (2,-1)
     \end{array}$ &
     $\begin{array}{c}(2,-1)\\
        (3,+1)
     \end{array}$ & 
     $\begin{array}{c}(4,+1)\\
        (3,-1)
     \end{array}$
     \\
     \hline
      $\chi_k^{\Gamma_j}(C_6) = \zeta^{\Gamma_j}(\{C_6\vert\boldsymbol{a}_3/2\}) $ & 
      $ \mathrm{e}^{\text{i}\frac{\pi}{6}}$ & 
      $ \mathrm{e}^{-\text{i}\frac{\pi}{6}}$ & 
      $ -\text{i} $ & 
      $ \text{i}$ & 
      $ \mathrm{e}^{\text{i} \frac{5\pi}{6}}$ & 
      $ \mathrm{e}^{-\text{i} \frac{5\pi}{6}}$ \\
         \hline\hline 
\end{tabular}
\caption{\label{table_C6_alt_brad}
Spinful irreducible representations of $C_6$ symmetry of point group $\mathsf{C}_6$ according to the notations in Ref.\,\cite{altmann2011tables} and in Ref.\,\cite{BradCrack}. See also Table \ref{table_C6_3} with the notation from the Bilbao Crystallographic Server~\cite{Bilbao}.}
\end{table}
}

Similarly to the above derivation, we find the symmetry eigenvalues in the basis of cell-periodic Bloch eigenstates
\begin{equation}
\label{eq_similitude_b}
    \begin{aligned}
        \text{diag}\left(\zeta^{\Gamma_{j_1}}_{k}(6_3),
        \dots,
        \zeta^{\Gamma_{j_N}}_{k}(6_3)
        \right)&= \text{diag}\left(\mathcal{R}^{k}_{11}(\{C_6\vert\boldsymbol{b}_3/2\}),\dots, \mathcal{R}^{k}_{NN}(\{C_6\vert\boldsymbol{b}_3/2\}) \right) = \langle \widetilde{\boldsymbol{\psi}}, k \vert^{\{C_6\vert\boldsymbol{b}_3/2\}}\vert  \widetilde{\boldsymbol{\psi}}, k \rangle \,,\\
    &= \mathcal{U}^{\dagger}(k)\cdot d^k(6_3)
    \cdot\mathcal{U}(k)  \,,
    \end{aligned}
\end{equation}
such that 
\begin{equation}
        \left\{\zeta_k^{\Gamma_{j_n}}(6_3)\right\}_{n=1,2,\dots,N} = 
        \left\{
            \mathrm{e}^{\text{i} \frac{\pi}{6} (2l_{\alpha}-1)}, \mathrm{e}^{\text{i} \frac{\pi}{6} (2l_{\alpha}+1)}
    \right\}_{\alpha=A,B,\dots} \,.
\end{equation}
These eigenvalues match with the characters of the spinful irreducible representations of symmetry $C_6$ in symmorphic point group $\mathsf{C}_6$ listed in Table \ref{table_C6_3} above.  

\end{widetext}

\end{document}


\title{Supplemental Material for\\ High-chirality \textcolor{black}{and multi-quaternion Weyl nodes} in hexagonal ReO$_3$}

\author{Siyu Chen}
\affiliation{TCM Group, Cavendish Laboratory, University of Cambridge,
J. J. Thomson Avenue, Cambridge CB3 0HE, United Kingdom}
\author{Robert-Jan Slager}
\affiliation{TCM Group, Cavendish Laboratory, University of Cambridge,
J. J. Thomson Avenue, Cambridge CB3 0HE, United Kingdom}
\author{Bartomeu Monserrat}
\affiliation{TCM Group, Cavendish Laboratory, University of Cambridge,
J. J. Thomson Avenue, Cambridge CB3 0HE, United Kingdom}
\affiliation{Department of Materials Science and Metallurgy, University of Cambridge, 27 Charles Babbage Road, Cambridge CB3 0FS, United Kingdom}
\author{Adrien Bouhon}
\affiliation{TCM Group, Cavendish Laboratory, University of Cambridge,
J. J. Thomson Avenue, Cambridge CB3 0HE, United Kingdom}
\affiliation{Nordita, Stockholm University and KTH Royal Institute of Technology, Hannes Alfv{\'e}ns v{\"a}g 12, SE-106 91 Stockholm, Sweden}

\maketitle

\section{Phonon dispersion of R\lowercase{e}O$_3$}

\begin{figure}[!h]
\includegraphics[width=0.7\linewidth]{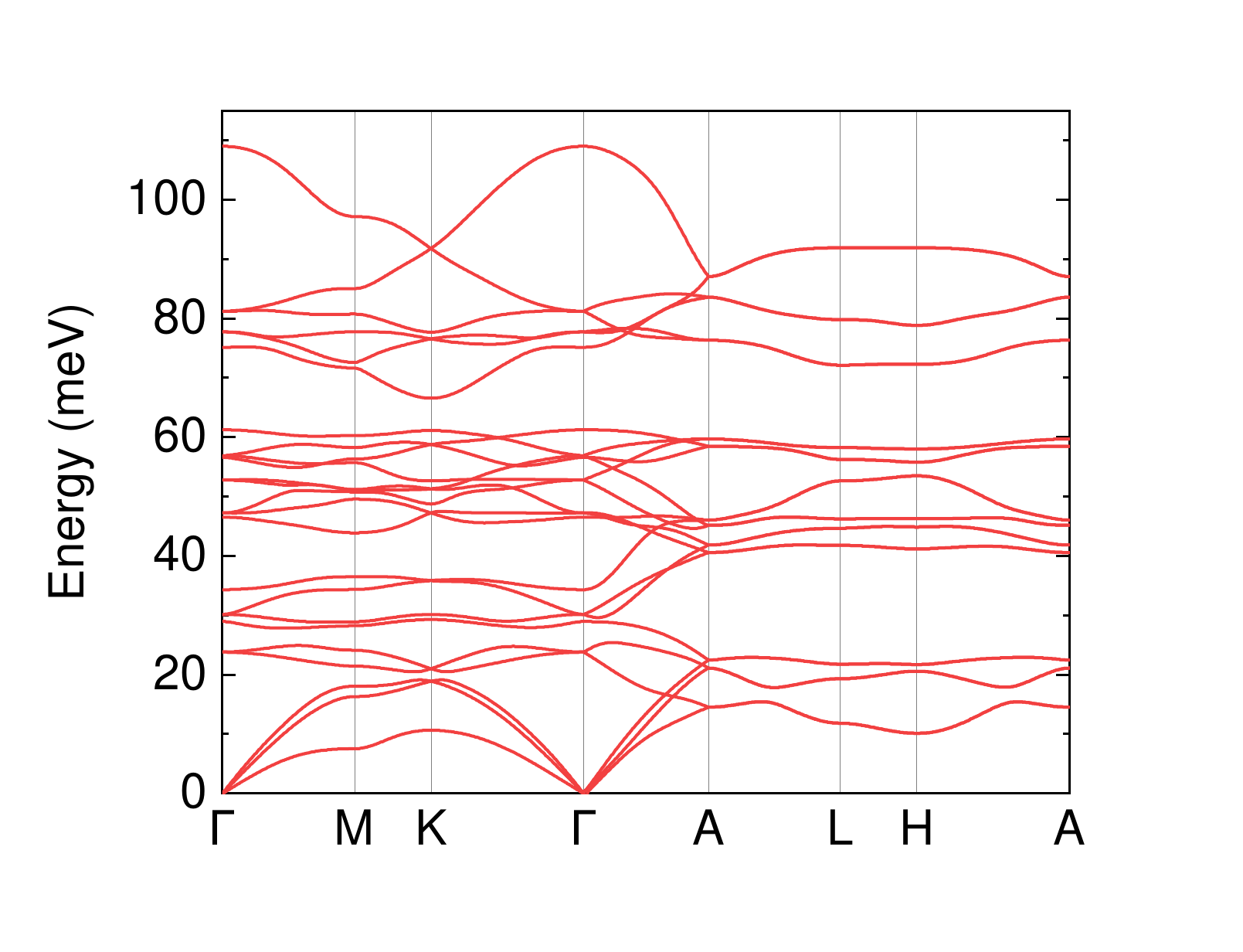}
\caption{Phonon dispersion of ReO$_3$.}
\label{fig:phonon}
\end{figure}

We calculate the phonon dispersion of ReO$_3$ using the finite displacement method in conjunction with nondiagonal supercells~\cite{Lloyd2015}, where a Farey $\boldsymbol{q}$-grid~\cite{chen2022} of order $3$ is adopted in order to guarantee that phonons on all high-symmetry points for the hexagonal lattice are directly computed. Before performing the phonon calculation, both the primitive cell and the internal atomic coordinates are optimized such that the pressure on the cell is converged to better than $10^{-2}$\,GPa and the force experienced by all atoms in the cell is lower than $10^{-4}$\,eVÅ$^{-1}$.  The phonon dispersions of ReO$_3$ along the high-symmetry path are shown in Fig.~\ref{fig:phonon}. The absence of imaginary frequencies is consistent with the experimental finding that hexagonal ReO$_3$ is stable under ambient conditions.

\section{DFT+U band structure}
ReO$_3$ is a transition metal oxide that may exhibit strong electronic correlations. To explore this possibility, we use the DFT$+U$ method~\cite{Liechtenstein1995, dudarev1998} to re-calculate the electronic structure. Figure~\ref{fig:band} compares the band structure of ReO$_3$ calculated using DFT and DFT+$U$, where we have empirically chosen the effective on-site Coulomb and exchange parameters to be $U=3$\,eV and $J=0.5$\, eV for the $d$-orbitals of Re. The inclusion of the Hubbard $U$ term renormalizes the bands compared to standard DFT, with the largest changes observed for energies $E>0.5$ eV. Importantly, there is no significant difference between the DFT and DFT$+U$ bands for $|E|<0.5$ eV, and applying a Hubbard $U$ correction does not modify the distribution of Weyl points discussed in the main text. This finding aligns with the fact that the bands near the Fermi level do not have a significant contribution from the Re $5d$-orbitals.

\begin{figure}[!h]
\begin{tabular}{cc}
\label{fig:band}
\includegraphics[width=0.8\linewidth]{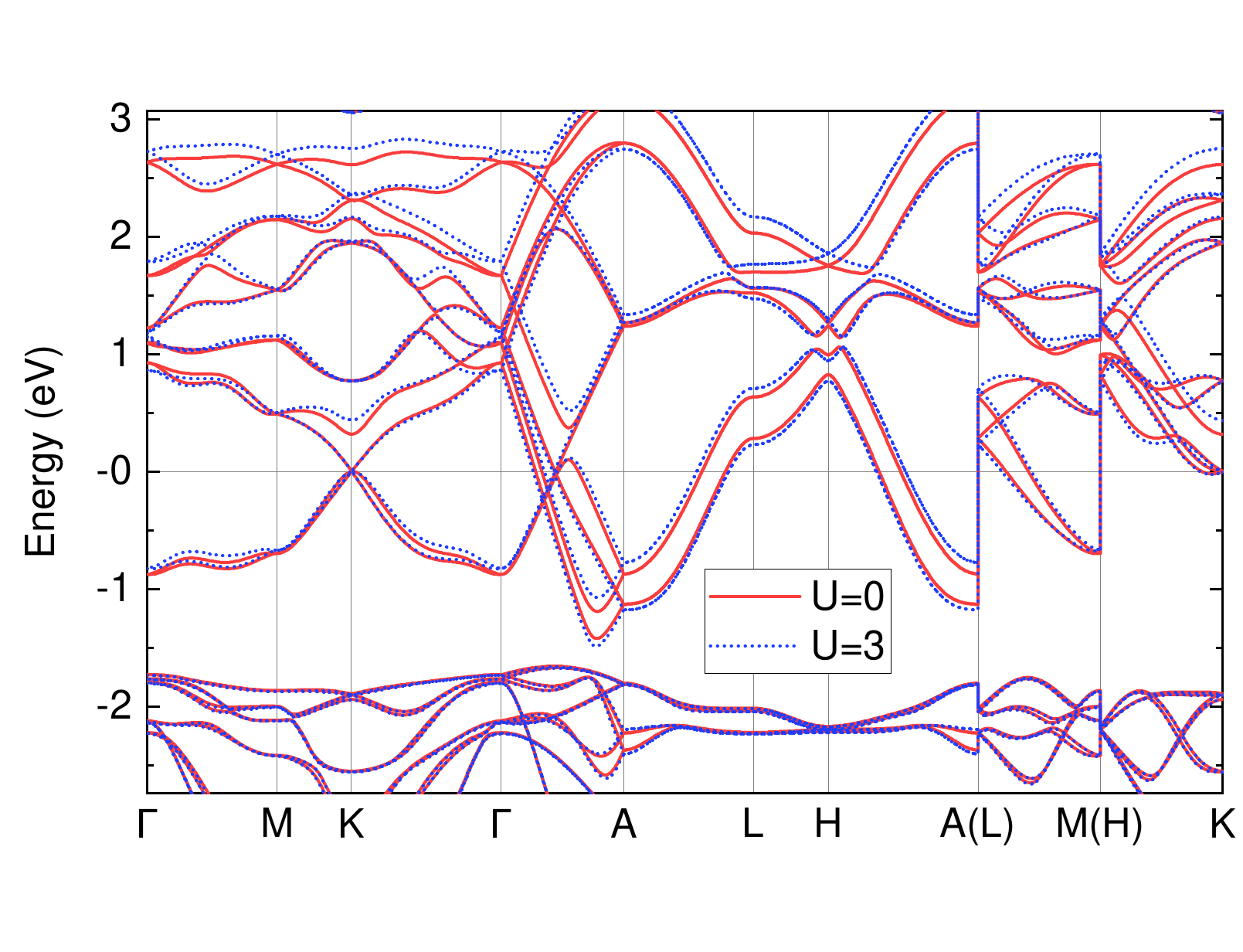}   
\end{tabular}
\caption{Band structure of ReO$_3$ with $U=0$ eV and $U=3$ eV.}
\label{fig:band}
\end{figure}

\section{Band structure under shear stress}

\begin{figure}[H]
\begin{tabular}{cc}
\label{fig:band}
\includegraphics[width=0.95\linewidth]{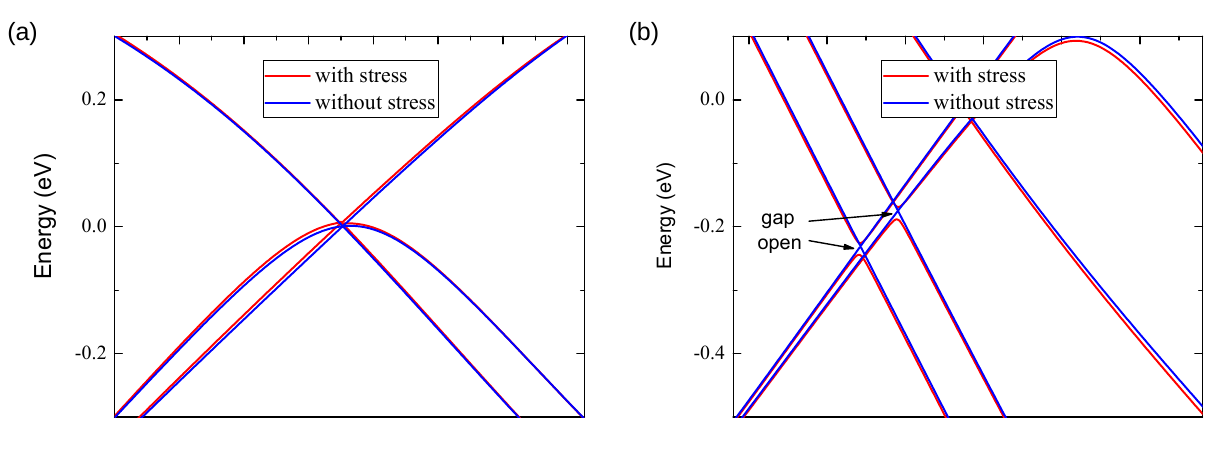}   
\end{tabular}
\caption{Comparison of band structures of ReO$_3$ (a) near the K point and (b) on the $\Delta$-line with and without shear stress.}
\label{fig:band_stress}
\end{figure}

As a supplement to the schematic diagram of Weyl nodes splitting in Fig.\,9 of the main text, Fig.\,~\ref{fig:band_stress}(a-b) compares the calculated band structures of ReO$_3$ near the K point and on the $\Delta$-line, both with and without shear stress. The preserved triple-degenerate point shown in Fig.\,~\ref{fig:band_stress}(a) corresponds to the robust Weyl node depicted by the red sphere at the K point in Fig.\,9. Meanwhile, the highlighted band gap opening in Fig.\,\ref{fig:band_stress}(b) reflects the Weyl node splitting depicted by the green spheres and arrows in Fig.\,9.

\bibliography{references}